\documentclass[12pt,notitlepage,a4paper]{article}

\pdfoutput=1

\usepackage{a4wide}
\usepackage{epsfig}
\usepackage{color,graphicx}

\usepackage{amssymb,amsmath}
\usepackage{jheppub}
\usepackage{adjustbox}
\usepackage{multirow}
\usepackage{makecell}
\newcommand{\be}{\begin{equation}}
\newcommand{\ee}{\end{equation}}
\newcommand{\bea}{\begin{eqnarray}}
\newcommand{\eea}{\end{eqnarray}}

\begin{document}

\begin{center}  

\vskip 2cm 

\centerline{\Large {\bf\boldmath $\mathcal{N}=5$ SCFTs and quaternionic reflection groups}}

\vskip 1cm

\renewcommand{\thefootnote}{\fnsymbol{footnote}}

   \centerline{
    {\large \bf Anirudh Deb${}^{a}$} \footnote{anirudh.deb@stonybrook.edu}, {\large \bf Gabi Zafrir${}^{a,b,c}$} \footnote{gabi.zafrir@oranim.ac.il}}
      
\vspace{1cm}
\centerline{{\it ${}^a$ C.~N.~Yang Institute for Theoretical Physics,  Stony Brook University, Stony Brook,}}
\centerline{{\it NY 11794-3840, USA}}
\centerline{{\it ${}^b$ Simons Center for Geometry and Physics, Stony Brook University, Stony Brook,}}
\centerline{{\it NY 11794-3840, USA}}
\centerline{{\it ${}^c$ Department of Mathematics and Physics, University of Haifa at Oranim,}}
\centerline{{\it Kiryat Tivon 3600600, Israel}}
\vspace{1cm}

\end{center}

\vskip 0.3 cm

\setcounter{footnote}{0}
\renewcommand{\thefootnote}{\arabic{footnote}}   
   
\begin{abstract}

It was previously noted that for 3d SCFTs with $\mathcal{N}\geq 6$ the moduli space has the form of $\mathbb{C}^{4r}/\Gamma$, where $\Gamma$ is a complex reflection group, at least following suitable gauging of finite symmetries. Here we argue that this observation can be extended also to 3d SCFTs with $\mathcal{N}\geq 5$ SUSY, where $\Gamma$ is now a quaternionic reflection group. To do this, we study the moduli space of the known 3d $\mathcal{N}=5$ SCFTs. For Lagrangian cases, the results for the moduli space are further checked using the superconformal index. 

\end{abstract}
 
 \newpage
 
\tableofcontents

\section{Introduction}
\label{sec:intro}

 Conformal field theories (CFTs) have been a rich playground for studying the interplay between quantum field theories and mathematics. These usually emerge at the starting and end point of RG flows\footnote{Generally speaking, the theory at the UV and IR fixed point should be Lorentz and scale invariant. This does not necessarily implies conformal invariance, though it is believed that every Lorentz and scale invariant interacting theory is also conformal invariant.}, and such describe the UV and IR dynamics of the theory. These also emerge at critical points and so describe the behavior at second order phase transitions. Due to their ample symmetry, CFTs are rather constrained, leading to the phenomena of duality and universality, where different theories share the same CFT as their IR fixed point, UV fixed point or critical point. 

CFTs can be further constrained if additional global symmetries exist. As such, if one insist that the fixed point or critical point exhibits certain global symmetries, the list of possible CFTs describing it should shrink considerably. One may then hope that with enough degree of symmetry only a manageable list of candidate CFTs would remain. This has lead to the interest in classifying CFTs, particularly with ample symmetry, as it can place significant constraints on RG flows preserving such symmetries. Additionally, this may also lead to insight on more general properties of quantum field theories, reveal interesting connections with mathematics, and motivate novel string theory constructions. Of particular interest are superconformal field theories (SCFTs), where supersymmetry is expected to place additional constraints, with particular emphasis on cases with close to maximal supersymmetry.

The natural place to start is with the case of maximal SUSY, corresponding to sixteen conserved supercharges. Indeed $\mathcal{N}=(2,0)$ SCFTs in 6d, which is the maximal dimension where SCFTs are expected to exist, are suspected to be classified by a choice of ADE group, see for instance \cite{Henningson:2004dh}. The moduli space of the theory is then $\mathbb{R}^{5r}/\Gamma$, where $\Gamma$ is the Weyl group of the ADE group. Dimensional reduction on a tori, when allowing for discrete symmetry twists, then leads to all known 4d $\mathcal{N}=4$ SCFTs, see for instance \cite{Tachikawa:2011ch}. These appear to be classifiable by a choice of Weyl group, if we identify SCFTs differing by marginal deformations or gauging discrete symmetries\footnote{Note that a choice of Weyl group is rougher than a choice of Lie group, as there are different Lie groups which share the same Weyl group. Here the existence of dualities relating different gauge groups is important.}. The moduli space of the theory (at least for some choice regarding the gauging of discrete symmetries) is then $\mathbb{R}^{6r}/\Gamma$, where $\Gamma$ is the Weyl group of the chosen Lie group.

The situation becomes more interesting for 3d $\mathcal{N}=8$ SCFTs. By dimensional reduction of 4d $\mathcal{N}=4$ SCFTs or 6d $\mathcal{N}=(2,0)$ SCFTs, we seem to again associate a 3d SCFT with a choice of Lie group. However, there exist 3d $\mathcal{N}=8$ SCFTs, the so-called BLG theories \cite{Bagger:2007jr,Gustavsson:2007vu}, which are not of this type. Notably, their moduli space is $\mathbb{R}^{8r}/D_k$, for $D_k$ the dihedral group of order $2k$. This is still a quotient, but not by a Weyl group, as for generic $k$ the dihedral groups are not Weyl groups. It was pointed out in \cite{Tachikawa:2019dvq}, that both the Weyl groups and the dihedral groups are a special case of a real reflection group. As such, a natural generalization of the classification scheme is that 3d $\mathcal{N}=8$ SCFTs are classified by a choice of real reflection group, and that their moduli space is then given by $\mathbb{R}^{8r}/\Gamma$, where $\Gamma$ is the chosen reflection group. Interestingly, the list of real reflection groups includes the Weyl groups, the dihedral groups and two exceptional cases called $H_3$ and $H_4$. Given that the SCFTs based on the two former cases exist, it was then suggested in \cite{Tachikawa:2019dvq} that new 3d $\mathcal{N}=8$ SCFTs associated with $H_3$ and $H_4$ exist, though these have yet to be discovered.

The next interesting case is that of twelve supercharges, particularly 4d $\mathcal{N}=3$ SCFTs and 3d $\mathcal{N}=6$ SCFTs. It was pointed out in \cite{Tachikawa:2019dvq}, that for all known cases the moduli space is again an orbifold, $\mathbb{C}^{3r}/\Gamma$ for 4d and $\mathbb{C}^{4r}/\Gamma$ for 3d, with $\Gamma$ a complex reflection group (again assuming a suitable choice of gauged discrete symmetries was made). This suggests a correspondence between complex reflection groups and SCFTs with twelve supercharges in 4d and 3d. Note that here multiple different SCFTs may have the same moduli space, hence this is merely a correspondence\footnote{Like in the maximal SUSY case, the 4d case is more restrained than the 3d one. First, only crystallographic complex reflection groups may appear. Second, the spectrum of known 4d $\mathcal{N}=3$ SCFTs appears to be consistent with a classification by crystallographic complex reflection groups, that is the choice of crystallographic complex reflection group uniquely determines the SCFT, again if we identify theories differing by discrete gauging. We refer the reader to \cite{Tachikawa:2019dvq} for the details.}. Like in the previous case, this correspondence suggests the existence of new 4d $\mathcal{N}=3$ SCFTs and 3d $\mathcal{N}=6$ SCFTs associated with certain complex reflection groups. Indeed, some of these have since been discovered \cite{Kaidi:2022lyo,Kaidi:2022uux}, though complex reflection groups with no known SCFT realizations still remain.

This leads us to the cases with less SUSY, the notable next case being theories with eight supercharges. Here significant effort has been divested for the development of classification schemes, with partial success. In the highest dimensional case, 6d, a classification of $\mathcal{N}=(1,0)$ SCFTs based on F-theory constructions has been put forward in \cite{Heckman:2013pva,Heckman:2015bfa,Bhardwaj:2018jgp} (see \cite{Heckman:2018jxk} for a review). Based on this classification, and using geometric methods and circle reduction of $\mathcal{N}=(1,0)$ SCFTs, a classification attempt for 5d $\mathcal{N}=1$ SCFTs has been ongoing, see \cite{Jefferson:2017ahm,Jefferson:2018irk,Hayashi:2018lyv,Apruzzi:2018nre,Closset:2018bjz,Apruzzi:2019vpe,Apruzzi:2019opn,Bhardwaj:2019jtr,Apruzzi:2019enx,Bhardwaj:2019xeg,Apruzzi:2019kgb,Bhardwaj:2020gyu} for a partial list of references. In 4d, a bottom up classification based mostly on the structure of the Coulomb branch, has been ongoing, with a proposed classification of 4d $\mathcal{N}=2$ SCFTs of ranks 1 and 2, see \cite{Argyres:2015ffa,Argyres:2015gha,Argyres:2016xua,Argyres:2016xmc,Argyres:2016yzz,Argyres:2018zay,Martone:2020nsy,Argyres:2020wmq,Martone:2021ixp,Argyres:2022lah,Argyres:2022puv,Argyres:2022fwy} for a partial list of references and also \cite{Argyres:2020nrr} for a short review. However, the situation in 3d seems hopeless at the moment as any 3d $\mathcal{N}=4$ gauge theory is expected to flow to a 3d $\mathcal{N}=4$ SCFT, so the list of possible 3d $\mathcal{N}=4$ SCFTs seems unwieldy, barring a significant proliferation of IR dualities. Furthermore, the moduli space in all these cases is significantly more complicated, made from two distinct moduli spaces, potentially with mixed branches, which are not necessarily orbifolds. As such, it seems that a simple classification scheme, or a correspondence, as the ones observed for higher SUSY cases will not be applicable here.

This leaves the case of 3d $\mathcal{N}=5$ SCFTs, associated with ten supercharges. These appear to obey many of the properties observed for higher SUSY SCFTs. Notably, the number of known ones is quite limited, see the review in \ref{sec:reviewN5}, and their moduli space consists of a single branch which is an orbifold of $\mathbb{C}^{4r}$. As such, we might hope that the above mentioned correspondence might extend also to these cases. Indeed, in the 4d case the moduli space was spanned by multiples of six real scalars, with the symmetry rotating them being the $SO(6) \equiv SU(4)$ R-symmetry of 4d $\mathcal{N}=4$ SCFTs. If we consider the coordinates as real, then the natural quotient is by a real reflection group. If we want to take a quotient by a complex reflection group, then it is more natural to think of the six real scalars as three complex scalars instead. Note that now the symmetry rotating them is $U(1)\times SU(3)$, which is the R-symmetry of 4d $\mathcal{N}=3$ SCFTs.

Similarly in 3d, the moduli space is spanned by multiples of eight real scalars, which we can alternatively think of as four complex scalars. When thought of as real scalars, we would naturally quotient by real reflection groups, and the symmetry rotating them will be $SO(8)$, which is the R-symmetry of 3d $\mathcal{N}=8$ SCFTs. Alternatively, when considered as four complex scalars, the natural quotient would be by complex reflection groups, and the symmetry rotating them will be $SU(4) \equiv SO(6)$, which is the R-symmetry of 3d $\mathcal{N}=6$ SCFTs\footnote{In addition to the $SU(4)$, there is also a $U(1)$. This is not an R-symmetry, but rather a global symmetry that appears to be present in any 3d $\mathcal{N}=6$ SCFT \cite{Bashkirov:2011fr}. We should also mention that the same reference also argued that every 3d $\mathcal{N}=7$ SCFT atomatically has $\mathcal{N}=8$ SUSY, and hence why we do not discuss these here.}. However, since the number of scalars is a multiple of four, we can also consider them as two quaternionic scalars. In that case the symmetry rotating them would be $USp(4) \equiv SO(5)$, which is the R-symmetry of 3d $\mathcal{N}=5$ SCFTs. Analogy with the above cases then leads us to speculate a similar relation between 3d $\mathcal{N}=5$ SCFTs and quaternionic reflection groups. The latter are a generalization of complex reflection groups to $\mathbb{H}^r$, considered and classified in \cite{Cohen:1980qrg}. We refer the reader to appendix \ref{sec:QRG} for more information on this type of discrete groups.

This leads us to propose that the moduli space of 3d $\mathcal{N}=5$ SCFTs, upon suitable gauging of discrete symmetries, is always of the form of $\mathbb{H}^{2r}/\Gamma$, with $\Gamma$ a quaternionic reflection group. This suggests a correspondence between 3d $\mathcal{N}=5$ SCFTs and quaternionic reflection groups. Like for 3d $\mathcal{N}=6$ SCFTs, we find that there can be multiple different $\mathcal{N}=5$ SCFTs with the same moduli space, so this does not amount to a classification of 3d $\mathcal{N}=5$ SCFTs in terms of quaternionic reflection groups. Also, we note that the current list of known 3d $\mathcal{N}=5$ SCFTs does not fully cover all known quaternionic reflection groups. This leaves a significant number of quaternionic reflection groups, particularly for low rank cases (acting on $\mathbb{H}^{r}$ for small $r$), for which there is no known $\mathcal{N}=5$ SCFT whose moduli space is a quotient by these groups. This may suggest the existence of many yet to be discovered $\mathcal{N}=5$ SCFTs.

Finally, we want to comment on the subtle issue of discrete gauging. Given a CFT with a discrete non-anomalous symmetry, we can consider gauging said symmetry. If the symmetry acts on the moduli space, then the moduli space after gauging would be the previous one further quotiented by the gauged symmetry. Furthermore, after said gauging we acquire a $d-2$ form symmetry, whose gauging would ungauge the gauged 0-form symmetry \cite{Gaiotto:2014kfa}. This would now change the moduli space back by removing the quotient by the gauged 0-form symmetry. As such, we see that when the moduli space is an orbifold, we can enlarge the quotiented group by gauging discrete 0-form symmetries and shrink it by gauging discrete 1-form symmetries (in 3d). It is possible to show that gauging discrete 0-form symmetries can produce 3d $\mathcal{N}=8$ and 4d $\mathcal{N}=4$ SCFTs whose moduli space is not a quotient by a real reflection group \cite{Tachikawa:2019dvq}, hence why we needed to enforce some conditions in regards to gauging discrete symmetries. The seemingly correct statement is that if we look at the variant with the smallest possible quotient group, then said quotient group would be a reflection group of the appropriate kind. Since we can shrink the group by gauging 1-form symmetries in 3d (and 2-form symmetries in 4d), such variant would have the property of having no non-anomalous 1-form symmetry (2-form symmetry in 4d)\footnote{In principle this should include both invertible and non-invertible 1-form symmetries, though here we shall mosly concentrate on the invertible ones.}. Other variants can then be generated by gauging 0-form discrete symmetries of this mother SCFT. As such the more accurate form of the correspondence we propose is that every 3d $\mathcal{N}=5$ SCFT can be generated by gauging a 0-form discrete symmetry of a mother SCFT, in the sense defined previously, whose moduli space is a quotient by a quaternionic reflection group.

To test our proposal, we examine the moduli space of all known 3d $\mathcal{N}=5$ SCFTs. Here, we need also consider the moduli space of different global variants, notably the ones with no (invertible) 1-form symmetries. While much is known about the moduli spaces of 3d $\mathcal{N}=5$ SCFTs, there are still many cases where it was not explicitly studied. Our work determines the moduli spaces also in these cases, which includes exotic $\mathcal{N}=5$ SCFTs as well as variants of orthosymplectic ABJ theories. Here we mostly considered Lagrangian theories, for which Lagrangian methods can be employed to study the moduli space. We further use the superconformal index to determine and test our expectations for the moduli space in certain cases, like various exotic 3d $\mathcal{N}=5$ SCFTs, as well as variants of orthosymplectic ABJ theories for certain low-rank cases. We indeed find that in all cases the moduli space is a quotient by a quaternionic reflection group, as expected from our proposal. This provides some support for our conjecture. There are a few caveats that we do not tackle here, and hope to consider in future work, notably: the potential presence of non-invertible 1-form symmetries, non-Lagrangian $\mathcal{N}=5$ SCFTs where we don't study the various variants, and undertaking a complete classification of Lagrangian $\mathcal{N}=5$ SCFTs.

The paper is organized as follows. In section \ref{sec:reviewN5} we review the known $\mathcal{N}=5$ theories and what is the expected moduli space of each theory. In section \ref{sec:SCI} we discuss the superconformal index in a limit which gives us the Hilbert Series of the moduli space of vacua. We match this Hilbert Series with the one expected form a quotient by the appropriate quaternionic reflection group. Finally we conclude and discuss some future directions in \ref{sec:concl}. In the appendices we collect some useful facts and information to perform the calculations. Appendix \ref{sec:QRG} reviews quaternionic reflection groups. Some more details about the computation of the superconformal index is given in appendix \ref{app:SCIdetails}. Appendix \ref{sec:exdet}, discusses one example in detail. Appendix \ref{app:HSQRG} discusses how to obtain Hilbert series by computing invariants of a given quaternionic reflection group and also contains tables displaying invariants for select reflection groups. Appendix \ref{app:CSLEVcons} discusses various details regarding the enhancement to $\mathcal{N}=5$ supersymmetry. Finally, appendix \ref{app:MS} provides a detailed discussion on the moduli spaces of certain $\mathcal{N}=5$ SCFTs.

\section{Review of known \texorpdfstring{$\mathcal{N}=5$}{} SCFTs}
\label{sec:reviewN5}

Here we shall review the known $\mathcal{N}=5$ SCFTs, see table \ref{N=5table} for a summary (see also \cite{Bergshoeff:2008bh}). There are several known ways of engineering $\mathcal{N}=5$ SCFTs. One way is through Lagrangian methods. Following the work of \cite{Gaiotto:2008sd}, it is known that such theories are related to Lie superalgebras, where the gauge content is made from the bosonic generators and the flavor content is made from the fermionic generators. The list of theories engineered in this way is then equivalent to the simple Lie superalgebras. The main list that appears as physical theories include the $SU(M|N)$, $PSU(N|N)$, $OSp(M|2N)$ and the three exceptionals: $D(2|1;\alpha)$, $F(4)$ and $G(3)$ \footnote{There are also the strange superalgebras, denoted by $P$ and $Q$, but there does not appear to be an $\mathcal{N}=5$ SCFT associated with these.}. The case of $SU(M|N)$ and $PSU(N|N)$ actually have $\mathcal{N}\geq 6$ supersymmetry. The other cases, $OSp(M|2N)$ and the three exceptionals, are $\mathcal{N}=5$ SCFTs\footnote{In certain cases, these have enhanced $\mathcal{N}=6$ SUSY, like the $OSp(2|2N)$ case. See \cite{Tachikawa:2019dvq} for more details on such cases.}. It should be noted that there might be additional manifestly $\mathcal{N}=5$ SCFTs besides this list. Indeed for the case of $\mathcal{N}=6$ SCFTs, a classification of such Lagrangian cases exists in \cite{Schnabl:2008wj}. It actually coincide with the list suggested by simple Lie superalgebras, up to the addition of $U(1)$ gauge groups. It would be interesting to extend such analysis also to the case of $\mathcal{N}=5$ SCFTs, though we shall not carry this here.

\begin{table}[htbp]
\begin{center}
\resizebox{\textwidth}{!}{
	\begin{tabular}{|c||c|c|c||c|}
		\hline
		$\mathcal{N}=5$ SCFT & Field theory realization & String theory realization & Moduli space  \\
		\hline
Unitary ABJ theories & $U(N+x)_k\times U(N)_{-k}$ & M$2$-branes probing & $Sym^N(\mathbb{C}^4/\mathbb{Z}_k)=$ \\
($\mathcal{N}=6$) & & $\mathbb{C}^4/\mathbb{Z}_k$ singularity  & $\mathbb{C}^{4N}/G_{N}(\mathbb{Z}_k,\mathbb{Z}_k)$ \\
		\hline
	SCFTs based on & $[SU(N)_k\times SU(N)_{-k}]/\mathbb{Z}_N$ & Equivalent to an & $\mathbb{C}^{4N}/G_{N}(\mathbb{Z}_k,\mathbb{I})$ \\
	$PSU(N|N)$ ($\mathcal{N}=6$) & & ABJM theory & \\
	 & & see \cite{Tachikawa:2019dvq,Bergman:2020ifi} & \\
		\hline
	Orthosymplectic ABJ & $O(2N+x)_{2k}\times USp(2N)_{-k}$ & M$2$-branes probing & $Sym^N(\mathbb{C}^4/\hat{D}_k)=$ \\
	theories & $O(2N+0/1)_{2k}\times USp(2N+2x)_{-k}$ & $\mathbb{C}^4/\hat{D}_k$ singularity & $\mathbb{C}^{4N}/G_{N}(\hat{D}_k,\hat{D}_k)$ \\
		\hline
		SCFTs based on & $SU(2)_{k_1}\times SU(2)_{k_2} \times SU(2)_{k_3}$  & & $\mathbb{C}^{4}/\hat{D}_{\frac{I_0}{2}}$ \\
		$D(2|1;\alpha)$ &  $k^{-1}_1 + k^{-1}_2 + k^{-1}_3 = 0$ &  & (see \ref{MSD}) \\
		\hline
		SCFTs based on $F(4)$ & $SU(2)_{2k}\times Spin(7)_{-3k}$ &  & $\mathbb{C}^{4}/\hat{D}_{6k}$ \\
         & & & ($\mathbb{C}^{4}/\hat{D}_{2}$ for $k=1$) \\
		\hline
		SCFTs based on $G(3)$ & $SU(2)_{3k}\times (G_2)_{-4k}$ &  & $\mathbb{C}^{4}/\hat{D}_{6k}$ \\
		\hline
		branes on  &  & M$2$-branes probing & $Sym^N(\mathbb{C}^4/\hat{E}_k)=$ \\
		$E$-type singularities &  & $\mathbb{C}^4/\hat{E}_k$ singularity & $\mathbb{C}^{4N}/G_N(\hat{E}_k,\hat{E}_k)$ \\
		\hline
	\end{tabular}}
\end{center}
\caption{A list of known $\mathcal{N}=5$ SCFTs, together with their field theory and string theory realizations. Also written is their moduli space known from either existing literature or the work of this paper. Not all theories have both realizations, so some fields are empty. Additionally, some cases have more than one realization in these classes, and there are equivalences between some of them. Also, many cases possess $1$-form symmetries that can be gauged to change the moduli space. Finally, some cases have higher supersymmetry. This is specified if the entire family has it, but not if it is only present for special cases (like the $k=1,2$ cases for ABJM or the $O(2)$ case of the orthosymplectic ABJ theories). Here we are mainly interested in purely $\mathcal{N}=5$ SCFTs, for which more information about the moduli space and the different variants related by gauging discrete symmetries can be found in the rest of this paper. We will not touch $\mathcal{N}\geq 6$ SCFTs in detail, and refer the reader to \cite{Tachikawa:2019dvq} for a similar analysis for these theories. Some known $\mathcal{N}\geq 6$ SCFTs are not listed in this table and we refer the reader to \cite{Schnabl:2008wj,Tachikawa:2019dvq} for a more complete listing of these.} 
\label{N=5table}
\end{table}

Out of the above listed examples, the theories based on the "classical" variants have been well studied. Specifically, the ones based on $SU(M|N)$ are known as the ABJM and ABJ models \cite{Aharony:2008ug,Aharony:2008gk}. They have gauge group $U(M)_k\times U(N)_{-k}$, with the subscript denoting the Chern-Simons level, and matter content being two bifundamental hypermultiplets (the case of $M=N$ is the ABJM model \cite{Aharony:2008ug}, while the unequal rank cases are usually referred to as the ABJ model \cite{Aharony:2008gk}). The case based on $PSU(N|N)$ has gauge group $SU(N)_k\times SU(N)_{-k}$ and again two bifundamental hypermultiplets. The case based on $OSp(M|2N)$ is refereed to as the ABJ model (orthosymplectic version to differentiate it from the unitary version) \cite{Aharony:2008gk,Hosomichi:2008jb}, has gauge group $O(M)_{2k}\times USp(2N)_{-k}$ and a bifundamental hyper.

The ones based on the exceptional superconformal algebras are significantly less studied. Out of these, the most well studied is the one based on $D(2|1;\alpha)$, see for instance \cite{Assel:2022row}. Its matter content consists of gauge group $SU(2)_{k_1}\times SU(2)_{k_2}\times SU(2)_{k_3}$ obeying $k^{-1}_1 + k^{-1}_2 + k^{-1}_3 =0$, and a trifundamental hyper. The other two $\mathcal{N}=5$ SCFTs remain, to our knowledge, relatively unstudied. We do know that their gauge groups are $SU(2)\times Spin(7)$ for $F(4)$ and $SU(2)\times G_2$ for $G(3)$. The flavor content can also be read from the supercharges, and consists of a bifundamental hyper for $G(3)$ and a hyper in the $({\bf 2},{\bf 8})$ for $F(4)$. The Chern-Simons levels can be read from the condition of SUSY enhancement in \cite{Gaiotto:2008sd} (see also \cite{Hosomichi:2008jb,Schnabl:2008wj}), and we find these to be: $k_{SU(2)} = 3k , k_{G_2} = -4k$ for $G(3)$ and $k_{SU(2)} = 2k , k_{Spin(7)} = -3k$ for $F(4)$ (see appendix \ref{app:CSLEVcons} for the computation), with $k$ an integer.   

Another way of engineering SCFTs is using string theory. Here again there are various methods, where we shall merely concentrate on brane systems. Other methods also exist, but to our knowledge, do not lead to additional $\mathcal{N}=5$ SCFTs, besides the ones mentioned here. A notable way to generate $\mathcal{N}=5$ SCFTs is as the low-energy theory living on M$2$-branes probing a $\mathbb{C}^4/\Gamma$ singularity, where $\Gamma$ is a discrete subgroup of $SU(2)$. These obey an ADE classification. For the A-type case, which is the discrete group $\mathbb{Z}_k$, the resulting theories are $\mathcal{N}=6$ SCFTs (further enhanced to $\mathcal{N}=8$ for $k=1,2$), while for the D and E-type cases one gets purely $\mathcal{N}=5$ SCFTs (save for a handful of cases with more supersymmetry). The D-type cases are equivalent to the orthosymplectic ABJ type theories that also have a Lagrangian construction, but the E-type theories are new. To our knowledge, there is no Lagrangian construction realizing these 3d SCFTs.

Of special interest to us here is the moduli space of such theories. For the $O(M)_{2k}\times USp(2N)_{-k}$ theories the moduli space is known to be the $N$-symmetric product of $\mathbb{C}^4/\hat{D}_k$, with $\hat{D}_k$ the binary dihedral group (for $M>N$ and similarly in the opposite case). This follows from the brane construction, and can also be understood from the field theory. Similarly, for M$2$-branes probing a $\mathbb{C}^4/E$-type singularities, the moduli space should be the symmetric product of $\mathbb{C}^4/\hat{E}_k$, where $\hat{E}_6 = \hat{T}$ , $\hat{E}_7 = \hat{O}$, $\hat{E}_8 = \hat{I}$ are the binary tetrahedral, octahedral and icosahedral groups, respectively. These moduli spaces can also be described as a quotient by a quaternionic reflection group: $\mathbb{C}^{4N}/G_N(\hat{D}_k,\hat{D}_k)$, in the $\hat{D}_k$ case and as $\mathbb{C}^{4N}/G_N(\hat{E}_k,\hat{E}_k)$, in the $\hat{E}_k$ case (see appendix \ref{sec:QRG} for a review on quaternionic reflection groups). As such, these are consistent with our conjecture. An interesting question then is what is the moduli space for other global variants, particularly the ones without higher form symmetries. Our conjecture suggests that these should be described by a quotient by a quaternionic reflection group. We explore this issue in this paper, particularly for the $O(M)_{2k}\times USp(2N)_{-k}$ theories that have a Lagrangian description. It would be interesting to preform a similar analysis for the E-type cases.

Finally, there are the three exceptional cases, on which less is known. The moduli space for the $G(3)$ and $F(4)$ cases are not known to our knowledge, and we determine them here from field theory and test the results using the superconformal index. The derivation of the moduli space from field theory is quite technical so we defer it to appendix \ref{app:MS}, and shall only state the results here. We find that in both cases the moduli space is of the form $\mathbb{C}^4/\hat{D}_{6k}$, with $k$ related to the CS level of the $SU(2)$ gauge group by: $k^{G(3)}_{SU(2)} = 3k$ , $k^{F(4)}_{SU(2)} = 2k$, with the exception of the $k=1$ case of $F(4)$, where the moduli space is enlarged to $\mathbb{C}^4/\hat{D}_2$. Similarly, using the superconformal index, we find that the moduli space of the $D(2|1;\alpha)$ case to be of type $\mathbb{C}^4/\hat{D}_{\frac{I_0}{2}}$, and we refer the reader to section \ref{MSD} for the details. As such we see that also for the exceptional cases, the moduli space conforms to our expectations. We still need to check the variants given by gauging 1-form symmetries, which we shall consider next.

So far we have been concerned with $\mathcal{N}=5$ SCFTs that have manifest $\mathcal{N}=5$ SUSY. However, it is possible that there are $3d$ theories that manifest less SUSY and flow to $\mathcal{N}=5$ SCFTs. An example of such a theory was given in \cite{Garozzo:2019ejm,Beratto:2020qyk,Gang:2021hrd}, through an $\mathcal{N}=3$ theory that supposedly flows to an $\mathcal{N}=5$ SCFT. Said SCFT appears to have no moduli space, and so again seem to conform to our exceptions. 

\subsection{Expectations for the moduli space}
\label{sec:expmodspace}
We have seen that the moduli space of all known $\mathcal{N}=5$ SCFTs is of the form $\mathbb{C}^{4N}/G$ for $G$ some quaternionic reflection group. However, here we have checked only one global variant, that usually has 1-form symmetries. To truly justify our claim we should consider other global variants, particularly the ones with no 1-form symmetries\footnote{Here we shall only consider invertible 1-form symmetries.}. We shall next detail the various 1-form symmetries and our expectations for the moduli space of other variants for the purely $\mathcal{N}=5$ SCFTs with a Lagrangian description. As previously summarized, this includes the cases based on the superalgebras $OSp(M|2N)$, $D(2|1;\alpha)$, $G(3)$ and $F(4)$. Out of these, the SCFTs based on $G(3)$ have no 1-form symmetries. This leaves the other three cases. We shall begin with the $OSp(M|2N)$ cases, which are the most intricate, and then turn to the two exceptional cases.

\subsubsection*{$OSp(M|2N)$}  

Here we again consider the $OSp(M|2N)$ case, where the associated $\mathcal{N}=5$ SCFT can be described by an $O(M)_{2k}\times USp(2N)_{-k}$ gauge theory. As we have seen previously, the moduli space in this case is given by $\mathbb{C}^{4L}/G_L(\hat{D}_{k},\hat{D}_{k})$, where $L=N$ or $\lfloor \frac{M}{2} \rfloor$, depending on which is smaller. We expect the moduli space of a variant related by gauging a 1-form symmetry to be given by $\mathbb{C}^{4L}/G_Q$, where $G_Q$ is some quaternionic reflection group which is a subgroup of $G_L(\hat{D}_{k},\hat{D}_{k})$. As reviewed in appendix \ref{sec:QRG}, the natural candidates are the quaternionic reflection groups $G_L(\hat{D}_{k},H)$, for $H$ a normal subgroup of $\hat{D}_{k}$. Furthermore, for $L>2$, $H$ must be such that $\hat{D}_{k}/H$ is abelian. This restricts the choice of $H$ to only a few cases, depending on whether $k$ is even or odd.

\begin{itemize}
 \item The case of $k$ odd: here the maximal abelian quotient $\hat{D}_{k}/H$ is $\mathbb{Z}_4$, which is the center of the Lie group $D_k$ for odd $k$. This happens for $H=\mathbb{Z}_{k}$. The smallest quaternionic reflection group of the type we are considering here is then $G_L(\hat{D}_{k},\mathbb{Z}_{k})$. Besides this case, and the full group $G_L(\hat{D}_{k},\hat{D}_{k})$, there is one other quaternionic reflection group, $G_L(\hat{D}_{k},\mathbb{Z}_{2k})$, corresponding to the $\mathbb{Z}_2$ subgroup of $\mathbb{Z}_4$.
 \item The case of $k$ even: here the maximal abelian quotient $\hat{D}_{k}/H$ is $\mathbb{Z}_2\times \mathbb{Z}_2$, which is the center of the Lie group $D_k$ for even $k$. This again happens for $H=\mathbb{Z}_{k}$, and so again the smallest quaternionic reflection group of the type we are considering here is $G_L(\hat{D}_{k},\mathbb{Z}_{k})$. Besides this case, and the full group $G_L(\hat{D}_{k},\hat{D}_{k})$, there are two other quaternionic reflection groups, $G_L(\hat{D}_{k},\mathbb{Z}_{2k})$ and $G_L(\hat{D}_{k},\hat{D}_{\frac{k}{2}})$, corresponding to the $\mathbb{Z}_2$ subgroups of $\mathbb{Z}_2\times \mathbb{Z}_2$ \footnote{There are three distinct $\mathbb{Z}_2$ subgroups of $\mathbb{Z}_2\times \mathbb{Z}_2$, but only two of them lead to distinct quaternionic reflection groups.}.
\end{itemize}

We see then that there are $4$ possible options for $k$ even and three for $k$ odd. We next turn to the Lagrangian theories to check if the resulting structure can be matched.

As mentioned the Lagrangian comprises of an $O(M)_{2k}\times USp(2N)_{-k}$ gauge theory with a bifundamental hypermultiplet. The 1-form symmetries of this theory comes from two sources. One comes from the fact that a bifundamental hypermultiplet is invariant under the combined center of $O(M)$ and $USp(2N)$. This comes about as both act on it as a $-1$ and so the combination of both cancels. This provides a $\mathbb{Z}_2$ 1-form symmetry for all values of $N$, $M$ and $k$. Gauging this 1-form symmetry transforms the gauge group to $[O(M)_{2k}\times USp(2N)_{-k}]/\mathbb{Z}_2$.

The second source comes from the fact that we have a disconnected group $O(M)$. We can think of it as coming from an $SO(M)$ gauge group with a gauged parity transformation. In general, when gauging a discrete 0-form symmetry, we get a dual $d-2$ form symmetry \cite{Gaiotto:2014kfa}, which for $d=3$ is a 1-form symmetry. As such there should be a $\mathbb{Z}_2$ 1-form symmetry which is the dual of the gauged parity symmetry. Gauging this 1-form symmetry transforms the gauge group to $SO(M)_{2k}\times USp(2N)_k$. Note, that here there is a difference between the $M$ odd and $M$ even cases. When $M$ is odd, the 1-form symmetries we mentioned are the same. This comes about because the central element of $O(M)$ is not in $SO(M)$, and as such the resulting quotient is tantamount to making the group be $SO(M)$. However, for $M$ even, the central element is also in $SO(M)$ and $[O(M)_{2k}\times USp(2N)_{-k}]/\mathbb{Z}_2$ and $SO(M)_{2k}\times USp(2N)_{-k}$ are naively distinct theories. 

As such, we see that for $M$ odd the 1-form symmetry is $\mathbb{Z}_2$. For $M$ even, the 1-form symmetry is made from two $\mathbb{Z}_2$ groups, but this does not completely determines the group. As discussed in \cite{Cordova:2017vab}, the precise form of the group depends on the Chern-Simons level $k$. For $k$ odd the two $\mathbb{Z}_2$ groups form a non-trivial extension and lead to the group $\mathbb{Z}_4$, while for $k$ even we instead get the direct product $\mathbb{Z}_2 \times \mathbb{Z}_2$. The different global variants are then given by gauging different subgroups of the 1-form symmetry. Note that the resulting structure then precisely matches the possible reflection subgroups of $G_L(\hat{D}_{k},\hat{D}_{k})$, this then naturally suggests a correspondence between the two. To better understand this relation, we turn to consider the structure of the moduli space.

The fact that the moduli space of the $O(2N)_{2k}\times USp(2N)_{-k}$ theory is $Sym^N(\mathbb{C}^4/\hat{D}_k)$ was discussed from the Lagrangian perspective in \cite{Aharony:2008gk}. The idea is that the moduli space is spanned by a vev to the bifundamental hyper, which breaks the group to $N$ copies of $U(1)\times U(1)$, plus the discrete part of the $O$ group, which can act independently on each part ($O(2N)\rightarrow O(2)^N$). This results in a low energy theory consisting of $N$ copies of the $U(1)_{2k}\times U(1)_{-2k}$ SCFT, quotiented by the $O$ part. The moduli space of each $U(1)_{2k}\times U(1)_{-2k}$ copy is $\mathbb{C}^4/\mathbb{Z}_{2k}$. This comes about as one needs to quotient by residual $U(1)\times U(1)$ gauge transformations \cite{Aharony:2008ug}. Next we need to further quotient by the $O$ part, which turn the moduli space into $\mathbb{C}^4/\hat{D}_k$, as argued in \cite{Aharony:2008gk}. Combining the $N$ copies form the space $Sym^N(\mathbb{C}^4/\hat{D}_k)$.

Next we wish to repeat the analysis, but with different variants. We will be interested in two cases, one where the gauge group is $[O(2N) \times USp(2N)]/\mathbb{Z}_2$ and the other where the gauge group is $SO(2N) \times USp(2N)$. We shall begin with the former. In this case, we can again break the theory to $N$ copies of $U(1)_{2k}\times U(1)_{-2k}$, each also acted by the $O$ part, but now the $\mathbb{Z}_2$ quotient projects out the element associated with the center of the $O \times USp$ gauge group. Specifically, consider the $N$ copies of $\mathbb{C}^4/\mathbb{Z}_{2k}$, coming from the $N$ $U(1)_{2k}\times U(1)_{-2k}$ theories where we have not yet quotiented by the $O$ part. The $\mathbb{Z}_{2k}$ quotient originate from the $U(1)^N\times U(1)^N$ gauge redundancy. However, because of the $\mathbb{Z}_{2}$ quotient, the gauge transformation where the product of all the $N$ $\mathbb{Z}_{2k}$ elements is $-1$, is no longer in the gauge group, and so should no longer be quotiented by. As can be seen in appendix \ref{sec:QRG}, when combining the $N$ copies, the resulting group we must quotient by is now $G_N(\hat{D}_{k},\hat{D}_{\frac{k}{2}})$. Note that here $k$ must be even as for odd $k$ the $\mathbb{Z}_2$ 1-form symmetry is not a subgroup of the full 1-form symmetry group, but only a quotient, and as such cannot be gauged. 

Next we consider the case of $SO(2N) \times USp(2N)$. Again we first go on the moduli space where the gauge theory is broken to $N$ copies of $U(1)_{2k}\times U(1)_{-2k}$. However, note that as the group is now $SO$, we have $SO(2N)\rightarrow O(2)^N/\mathbb{Z}_2$, where the quotient is by the diagonal $O$ part. As such, while we can still act on each $U(1)_{2k}\times U(1)_{-2k}$ copy by the $O$ element, the total number of such actions on all the $N$ copies must be even. When combining the $N$ copies, we will now need to quotient by the group $G_N(\hat{D}_{k},\mathbb{Z}_{2k})$, as can be seen from appendix \ref{sec:QRG}. We summarize our results for the moduli space of the various cases in table \ref{N=5var}. In the next section, we shall test this against explicit computations using the superconformal index for selected values of $N$ and $k$. We shall see that this agrees with our expectation here.

This covers the moduli space in the case where the ranks of the two groups are equal, but still leaves the case with differing ranks. This case is quite subtle, where the moduli space can remain $\frac{\mathbb{H}^{2N}}{G_N(\hat{D}_k,\hat{D}_k)}$ for all variants in many cases, but can also change like the equal rank case in others. We note that in either case, the moduli space is still a quotient by a quaternionic reflection group, and so the question of the precise moduli space in each case is not important for supporting our conjecture. As such, and due to the technical nature of the discussion, we reserve the precise analysis to appendix \ref{app:MS}.

\begin{table}[htbp]
\begin{center}
	\begin{tabular}{|c|c|}
		\hline
		Variant & Moduli space  \\
		\hline
    $O(2N)_{2k}\times USp(2N)_{-k}$ & $\frac{\mathbb{H}^{2N}}{G_N(\hat{D}_k,\hat{D}_k)}$ \\
		\hline
		$SO(2N)_{2k}\times USp(2N)_{-k}$ & $\frac{\mathbb{H}^{2N}}{G_N(\hat{D}_k,\mathbb{Z}_{2k})}$ \\
		\hline
		$[O(2N)_{2k}\times USp(2N)_{-k}]/\mathbb{Z}_2$ & $\frac{\mathbb{H}^{2N}}{G_N(\hat{D}_k,\hat{D}_{\frac{k}{2}})}$ \\
		\hline
		$[SO(2N)_{2k}\times USp(2N)_{-k}]/\mathbb{Z}_2$ & $\frac{\mathbb{H}^{2N}}{G_N(\hat{D}_k,\mathbb{Z}_{k})}$ \\
		\hline
	\end{tabular}
\end{center}
\caption{The different variants of $OSp(2N|2N)$ based theories and their expected moduli spaces as quotients by quaternionic reflection groups.} 
\label{N=5var}
\end{table}

Finally, we want to comment on possible anomalies. So far we have analyzed the various cases with gauged 1-form symmetry, but such a gauging can only be done if the 1-form symmetry has no 't Hooft anomalies. It is actually known that the 1-form symmetries in this and related classes of theories can carry non-trivial 't Hooft anomalies that forbid their gauging \cite{Tachikawa:2019dvq,Comi:2023lfm}. As can be seen from the previous discussion, the moduli space seem to conform to our conjecture for all choices of global variants we considered here and so the presence or lack of an anomaly does not influence it. Nevertheless, we shall briefly review the condition for the presence of such an anomaly. Note that in the cases where it is present the global variants associated with its gauging do not exist.

To better understand the issue, consider the case of a simply-connected gauge group $G$ with a Chern-Simons term of level $k$. As is well known, the Chern-Simons term is not gauge invariant if $k$ is not an integer as under a large gauge transformation, the action can change by $e^{2\pi i k n_{inst}}$, where $n_{inst}$ is an integer. This comes about as we can consider representing the gauge transformation by adding a 4-th dimension and allowing the connection to vary continuously along the 4-th direction, with the boundary conditions being the original connection on one side and the one after the gauge transformation on the other. The change in the CS action under a gauge transformation is then given by $e^{2\pi i k n_{inst}}$, where $n_{inst}$ is the instanton number associated with the gauge bundle on the 4d interpolating space. For large gauge transformations, the connection in the bulk will not be flat and will in general have a non-trivial instanton number. Nevertheless, $n_{inst}$ is always an integer and so $e^{2\pi i k n_{inst}}$ would be one, as long as $k$ is an integer. The subtle issue is that $n_{inst}$ can be fractional for non simply-connected gauge groups. This in turn would put a restriction on $k$.

For the case at hand, we are mainly concerned with gauge groups $USp(2N)$ and $SO(2M)$, for which we can consider the diagonal quotient by their $\mathbb{Z}_2$ center. Here we shall quote the results for the quantization of the instanton number in these cases, taken from \cite{Aharony_2013rbl}. We refer the reader to the reference for further details and for the quantization in other cases. It is known that $n_{inst}$ is integer for $USp(4n)/\mathbb{Z}_2$ but can be half-integer for $USp(4n+2)/\mathbb{Z}_2$. Similarly, $n_{inst}$ can be half-integer for $SO(4m)/\mathbb{Z}_2$ and can be quarter integer ($\frac{i}{4}$) for $SO(4m+2)/\mathbb{Z}_2$. Now consider the case of $[SO(2M)_{2k}\times USp(2N)_{-k}]/\mathbb{Z}_2$, which is the case we are interested in. In this case, the change in the action under a gauge transformation would be $e^{2\pi i (k_{USp(2N)} n_{USp(2N)/\mathbb{Z}_2}-k_{SO(2M)} n_{SO(2M)/\mathbb{Z}_2})}$. Because of the diagonal quotient, a non-trivial $USp(2N)/\mathbb{Z}_2$ bundle would also be accompanied by a non-trivial $SO(2M)/\mathbb{Z}_2$ one, an vice versa. The phase we get under gauge transformation can be non-trivial if $k_{USp(2N)} n_{USp(2N)/\mathbb{Z}_2}-k_{SO(2M)} n_{SO(2M)/\mathbb{Z}_2} = k(n_{USp(2N)/\mathbb{Z}_2}-2 n_{SO(2M)/\mathbb{Z}_2})$ is fractional. For $USp(2N)/\mathbb{Z}_2$ this happens when $N$ is odd and $n_{inst}$ can be half-integer. For $SO(2M)/\mathbb{Z}_2$, due to the factor of $2$ from the Chern-Simons level, this also happens when $M$ is odd, in which case $2n_{inst}$ can be half-integer. Overall, we see that we can get a non-trivial phase if $k(N-M)=$ odd. When this happens, the $\mathbb{Z}_2$ part of the 1-form symmetry associated with the central element, has a 't Hooft anomaly forbidding the gauging, and the associated variant does not exist.  

\subsubsection*{$D(2|1;\alpha)$}
\label{MSD}

The case based on $D(2|1;\alpha)$ is special as it possesses two free parameters. This is related to the fact that the exceptional $D(2|1;\alpha)$ superalgebra has the free continuous parameter $\alpha$. The gauge content in this case is $SU(2)_{k_1}\times SU(2)_{k_2}\times SU(2)_{k_3}$, with the constraint $k^{-1}_1 + k^{-1}_2 + k^{-1}_3 = 0$, leaving a two integer parameter space free. The matter content consists of a trifundamental hypermultiplet. Next we shall discuss the structure of the moduli space in this case, following which we shall consider its 1-form symmetries and the moduli space of the quotients. Related discussions regarding some of the things explored here can also be found in \cite{Assel:2022row,Comi:2023lfm}.

We begin with the discussion on the CS levels. Without loss of generality, these can be taken to be $(k_1,k_2,k_3) = (- l p(p+s) , - l s(p+s) , l p s)$, with $p$, $s$ and $l$ integers. We can further assume that: $p\geq s>0$ and $gcd(p,s)=1$ (a non-trivial gcd can be absorbed into $l$). The reason for this is as follows. We seek three integers obeying the relation $k^{-1}_1 + k^{-1}_2 + k^{-1}_3 = 0$. Note, that two of the CS levels must have the same sign, and we shall assume these to be $k_1$ and $k_2$.  We can formally solve the constraint by setting $k_3 = -\frac{k_1 k_2}{k_1+k_2}$, but that is insufficient as $k_3$ must be integer. Next, assume that $gcd(k_1,k_2)=n$, for some integer $n$. This means that we can write $k_1 = p n$, $k_2 = s n$, with $gcd(p,s)=1$. We can further take $p,s>0$, as $k_1$ and $k_2$ have the same sign which we can absorb into $n$, and use the permutation symmetry of the CS levels to set $k_1\geq k_2 \rightarrow p\geq s$. Next, we see that for $k_3$ to be integer we must have that $m(p+s) = - n p s$, for some integer $m$. It then follows that: 

\be
m+m\frac{p}{s} = - n p \; , \; m+m\frac{s}{p} = - n s .
\ee

Here everything is integer save for $m\frac{p}{s}$ and $m\frac{s}{p}$. Therefore, these must be integer as well. However, as $gcd(p,s)=1$, we must have that $m=l p s$, for some integer $l$. We now see that:

\be
m(p+s) = - n p s \rightarrow l p s(p+s) = - n p s \rightarrow n = -l (p+s) .
\ee  

Plugging this into the expressions for the CS levels, we see that:

\be
(k_1,k_2,k_3) = (n p, n s, -\frac{n p s}{(p+s)})\rightarrow (k_1,k_2,k_3) = (- l p(p+s) , - l s(p+s) , l p s) ,
\ee 
where we can further require $p\geq s>0$ and $gcd(p,s)=1$ to avoid duplicities. 

Next, we consider the moduli space. As we have mentioned, analysis using the superconformal index, on which we shall explain more in section \ref{sec:excpC}, reveals that the moduli space is of the form $\mathbb{C}^4/\hat{D}_{\frac{I_0}{2}}$. Here $I_0$ is an integer determined from the following condition:

\be
2(m_1 + m_2) k_3 = - 2 m_1 k_1 = -2 m_2 k_2 = I .
\ee 

This condition arises for a monopole with magnetic charges $(m_1, m_2, m_1+m_2)$ to have a non-zero contribution to the index. The integer $I_0$ is then given by the minimal solution for this equation, that is for the case with the minimal allowed magnetic fluxes. For the global variant with gauge group $SU(2)^3$, so that all the $m_i$'s are integer, and if we use the previous presentation of the CS levels, we have that:  

\bea
2lps(m_1 + m_2) = 2 lp(p+s) k_1 & = & 2 ls(p+s) k_2 = I \rightarrow \\ \nonumber m^{min}_1 & = & s \; , \; m^{min}_2 = p \; , \; I_0 = 2lps(p+s) .
\eea
The moduli space is then $\mathbb{C}^4/\hat{D}_{lps(p+s)}$.

Next, we turn to analyze the 1-form symmetries and the possible global variants. Here the gauge group is $SU(2)^3$, whose center is $\mathbb{Z}^3_2$. The matter content is a trifundamental hyper so one $\mathbb{Z}_2$ combination of the center acts non-trivially on it, and only $\mathbb{Z}^2_2$ is realized as a 1-form symmetry. By gauging some subgroup of the 1-form symmetry, we can pass to the $SU(2)\times SO(4)$ and $[SU(2)\times SO(4)]/\mathbb{Z}_2$ variants. Note that for some of the variants it is possible to take $m^{min}_1 = \frac{s}{2}$, $m^{min}_2 = \frac{p}{2}$, for which we get a smaller value of $I_0 = lps(p+s)$. This depends on the CS levels and on the quotient taken. 

To illustrate this, consider the simplest case of $p=s=1$, $l=k$. In this case, the gauge group becomes $SU(2)_{k}\times SU(2)_{-2k}\times SU(2)_{-2k}$. Note that in this case, the theory becomes a variant of the $OSp(4|2)$ class considered previously, but with the $O$ group replaced by $Spin$. We can then compare our results with those argued for the $OSp$ cases. Specifically, for these cases we have that $m^{min}_1 = m^{min}_2 = 1$ and $I_0 = 4k$, so that the moduli space is $\mathbb{C}^4/\hat{D}_{2k}$.

We can next look at other variants. The most pertinent to our case would be the $SU(2)_{k}\times SO(4)_{-2k}$, where comparison against $OSp$ theories can be made. Note that we can now take $m^{min}_1 = m^{min}_2 = \frac{1}{2}$, $m_3 = m_1 + m_2 =1$, as this is a consistent magnetic flux for $SU(2)\times SO(4)$ ($m_3$ being the magnetic flux for the $SU(2)$ as $k_3=k$ for this choice). As such, the moduli space for this variant is $\mathbb{C}^4/\hat{D}_{k}$. Note that no smaller $I_0$ can be found and so this moduli space is also the one for the $[SU(2)\times SO(4)]/\mathbb{Z}_2$ variant. In particular note then that the $SU(2)\times O(4)$, $SU(2)\times SO(4)$ and $[SU(2)\times SO(4)]/\mathbb{Z}_2$ variants all have the moduli space, $\mathbb{C}^4/\hat{D}_{k}$. This is then an example of ABJ theories, with a rank difference, for which the moduli space remains the same for all variants.

Finally, we want to comment on possible anomalies. As we have seen in the discussion on the $OSp$ cases, some of the $\mathbb{Z}_2$ 1-form symmetries can be anomalous in certain cases. Specifically, whenever we gauge a $\mathbb{Z}_2$ subgroup of the 1-form symmetry, we pass from $SU(2)\times SU(2)$ to $[SU(2)\times SU(2)]/\mathbb{Z}_2$, with the three choices of choosing the two $SU(2)$ groups being mapped to the three different $\mathbb{Z}_2$ subgroups of the 1-form symmetry. Similarly to the analysis done for the $OSp$ cases, we see that the condition for the phase under large gauge transformations to be trivial in the $[SU(2)\times SU(2)]/\mathbb{Z}_2$ variant is $\tilde{k}_1 + \tilde{k}_2 =$ even, for $\tilde{k}_1$, $\tilde{k}_2$ the CS levels of the two $SU(2)$ groups. One can see from the general expression given for the CS levels, that either all are even, or one is even and the other two are odd. In the former case, all quotients can be taken and the $\mathbb{Z}^2_2$ 1-form symmetry has no 't Hooft anomalies. However, in the latter case, only one $\mathbb{Z}_2$ subgroup of the 1-form symmetry is free of 't Hooft anomalies. These results match those in \cite{Comi:2023lfm}, who looked at a similar gauge theory, but with slightly different matter content.

\subsubsection*{$F(4)$}

In this case, the gauge theory description is $SU(2)_{2k}\times Spin(7)_{-3k}$ with a hypermultiplet in the $({\bf 2}, {\bf 8})$. Note that this matter content is invariant under the diagonal $\mathbb{Z}_2$ center of $SU(2)\times Spin(7)$. As such, there should be a $\mathbb{Z}_2$ 1-form symmetry, whose gauging leads to the quotient gauge group $[SU(2)_{2k}\times Spin(7)_{-3k}]/\mathbb{Z}_2$. From the field theory analysis performed in appendix \ref{app:MS}, one can see that in this case the moduli space is $\mathbb{C}^4/\hat{D}_{3k}$ for $k>1$ and $\mathbb{C}^4/\hat{D}_1$ for $k=1$.

Finally, we can again ask whether or not there is a 't Hooft anomaly in the 1-form symmetry. We can answer this in the same way as before. Specifically, from \cite{Aharony_2013rbl} we see that the instanton number for $Spin(7)/\mathbb{Z}_2$ is always integer, so an anomaly can only arise from the $SU(2)$ part. However, its CS level is even, and we conclude that the 1-form symmetry in this case can always be gauged.
	\section{Exploring the moduli space using the Superconformal Index}
\label{sec:SCI}

Having explained our expectation for the moduli space in each case, we next proceed to check this using the superconformal index. For 3d $\mathcal{N}=4$ SCFTs, the moduli space can be conveniently studied using special limits of the superconformal index, see for instance \cite{Razamat_2014}. Specifically, the moduli space of $\mathcal{N}=4$ SCFTs consists of a Coulomb branch and a Higgs branch, both being Hyperkahler spaces. There exist limits of the supercoformal index, dubbed Higgs or Coulomb limit, that receive contributions only from operators spanning each branch, and should be equal to the Hilbert Series of each moduli space. These limits are then useful to study the form and structure of the moduli space.

For 3d $\mathcal{N}=5$ SCFTs, the moduli space is of the form $\mathbb{H}^{2r}/\Gamma$, with the Higgs and Coulomb branches being identical and equal to $\mathbb{H}^r/\Gamma$. This follows from the embedding of the $\mathcal{N}=4$ $SU(2)\times SU(2)$ R-symmetry in the $\mathcal{N}=5$ $USp(4)$ R-symmery, $USp(4)\rightarrow SU(2)\times SU(2)$, with one $SU(2)$ acting on the Coulomb branch and the other on the Higgs branch. Given that the $USp(4)$ R-symmetry rotates each $\mathbb{H}^2$, and that the two $SU(2)$ subgroups rotate each $\mathbb{H}$ individually, we see that the Higgs and Coulomb branch should both be $\mathbb{H}^r/\Gamma$. As such to probe $\Gamma$, it is sufficient to evaluate just one of them.

Next, we shall briefly set up the computation of the superconformal index and review the results extracted from it.

\subsection{The superconformal index}

The superconformal index of a $3d$ $\mathcal{N}=2$ theory is given as \cite{https://doi.org/10.48550/arxiv.1106.2484,Razamat_2014,Aharony_2013,Aharony_2013s,Kim_2009,Imamura_2011}
\begin{equation}
	\text{Tr}\left[(-1)^F e^{\beta H}x^{\Delta+j_3}\prod_{a}{}{t_a}^{F_a}\right] ,
\end{equation}
where $\Delta$ is the energy, $R$ is the $R$-charge, $j_3$ is the third component of the angular momentum rotating $S^2$ and $F_a$ are fugacities for global flavour symmetries. $H$ is given by
\begin{equation}
	H=\{Q,Q^\dagger\}=\Delta-R-j_3 ,
\end{equation}
for a given supercharge $Q$. For the $\mathcal{N}=5$ case, we compute the superconformal index with the following fugacities
\begin{equation}
	\text{Tr}\left[(-1)^F e^{\beta H}x^{\Delta+j_3}c^{F_c}\right]=\text{Tr}\left[(-1)^F e^{\beta H}x^{R}c^{F_c}\right] ,
\end{equation}
where $R$ is again the R-charge under the $\mathcal{N}=2$ $U(1)$ R-symmetry, and $c$ is the fugacity for the $SU(2)$ flavour symmetry (with associated flavour charge $F_c$) that is the commutant of the $U(1)$ in the $\mathcal{N}=5$ $USp(4)$ R-symmetry. This index can be written as an index for an $\mathcal{N}=4$ theory (\cite{Razamat_2014})
\begin{equation}
	\text{Tr}(-1)^F x^{2j_3+(R_H+R_C)}c^{2(R_H-R_C)}e^{-\beta(\Delta-R_H-R_C-j_3)} .
\end{equation}

Here $R_H$ and $R_C$ label the $SO(4)_R\sim SU(2)_H\times SU(2)_C$ R-symmetry of the theory. The subscripts $H$ and $C$ denote Higgs and Coulomb respectively, where the reason for this will be clear in a moment. Let us redefine the fugacities in the following way
\begin{equation}
	\frac{x}{c^2}=\tau^2\;\;\; x c^2=\tilde{\tau}^2 
\end{equation}
The superconformal Index can be written as
\begin{equation}
	\text{Tr}\left[(-1)^F e^{\beta H}\tau^{2\Delta-R-\frac{F}{2}}\tilde{\tau}^{2\Delta-R+\frac{F}{2}}\right]=\text{Tr}\left[(-1)^F e^{\beta H}\tau^{2\Delta-2R_C}\tilde{\tau}^{2\Delta-2R_H}\right]
\end{equation}
We will be concerning ourselves with the following limits of this index
\begin{equation}
	x\to0,c\to0\;\;\  \text{such that}\;\; \frac{x}{c^2}=\tau^2 \ \ \text{fixed}\;\;\; (\tilde{\tau}\to 0)\;\;\; (\text{Coulomb limit})
\end{equation}
\begin{equation}
	x\to0,c\to \infty\;\;\  \text{such that}\;\; x c^2=\tilde{\tau}^2 \ \ \text{fixed}\;\;\; (\tau\to 0)\;\;\; (\text{Higgs limit})
\end{equation}
The Coulomb limit counts states with $\Delta=R_C$, while the Higgs limit counts states with $\Delta=R_H$. For an $\mathcal{N}=5$ theory, the Higgs branch limit and Coulomb branch limit are identical due to the enchanced supersymmetry. This follows as the two are related by taking $c\rightarrow c^{-1}$, which is just the action of the $SU(2)$ Weyl group. 

For Lagrangian theories which we consider, the superconformal index in $\mathcal{N}=2$ notation takes a simple form (more details regarding the index and the derivation of the formula below are given in \ref{app:SCIdetails})\footnote{Note that we divide by the Weyl group $|W|$ of the gauge algebra and not the Weyl group of residual gauge algebra of the monopole configuration which is usually written in literature (see for example \cite{https://doi.org/10.48550/arxiv.1106.2484}). The division by residual gauge group is done when one summing over Weyl inequivalent choices of monopoles. However we sum over all the monopoles. }:
\begin{equation}
	\label{eq:N5indexsimped}
	\mathcal{I}=\frac{1}{|W|}\sum_{s}\int\prod_j \frac{dz_j}{2\pi i z_j}x^{\epsilon_0} e^{-S_{CS}(h,s)}PE\left[\text{ind}_{\text{vec}}+\text{ind}_{\text{chir}} \right] ,
\end{equation}
where $\text{ind}_{chir}$ and $\text{ind}_{vec}$ denote the single particle index for chiral and vector multiplet respectively. These are given by
\begin{equation}
	\text{ind}_{\text{chir}}=\left(c+\frac{1}{c}\right)\left(\frac{x^{1/2}-x^{3/2}}{1-x^2}\right)\left(\sum_{\rho\in R_\Phi} e^{i\rho(h)}x^{2|\rho(s)|}\right) ,
\end{equation}	
\begin{equation}
	\text{ind}_{\text{vec}}=-\sum_{\alpha\in ad(G)}e^{i\alpha(h)}x^{2|\alpha(s)|} ,
\end{equation}
with $S_{CS}$ denoting the contribution from the Chern-Simons term
\begin{equation}
	S_{CS}(h,s)=2k\text{Tr}_{CS}(h\cdot s) ,
\end{equation}
and 
\begin{equation}
	\epsilon_0=\sum_\Phi(1-\Delta_{\Phi})\sum_{\rho\in R_\Phi}|\rho(s)|-\sum_{\alpha\in ad(G)}|\alpha(s)| .
\end{equation}
The notation is explained in detail in \ref{app:SCIdetails} but for convenience we mention the details needed to understand this section. $\alpha$ denotes the roots and $\rho$ denotes the weights of the representation of the Lie algebra. $h$ takes values in the maximal torus of the group and $s$ takes values in the Cartan subalgebra of the gauge group. $h$ parameterizes the $S^1$ Wilson lines of the gauge field and $s$ parameterize the GNO charge of the monopole configuration of the gauge field. We follow the normalization in which the components of $s$ take half integer values. $\Delta_\Phi$ is the scaling dimension for the chiral multiplet $\Phi$ and for us it is $\frac{1}{2}$. $c$ denotes the fugacity for the $SU(2)$ flavour symmetry. The single particle index for the chiral multiplet can be written in terms of $\tau$ and $\tilde{\tau}$
\begin{equation}
	\label{eq:chirind}
	\left(\frac{\tau^{1/2}+\tilde{\tau}^{1/2}}{1+\tau^{1/2}\tilde{\tau}^{1/2}}\right)\left(\sum_{\tau\in R_\Phi} e^{i\rho(h)}x^{2|\rho(s)|}\right).
\end{equation}	
This makes the invariance under exchange of $\tau$ and $\tilde{\tau}$ manifest. Therefore taking either Higgs or Coulomb limit will give the same result.


Now let us explain how we perform the computation of the moduli space limit of the Superconformal Index. We first perform the substitution $c=\frac{\sqrt{x}}{\tau}$ (Coulomb limit) or $c=\frac{\tau}{\sqrt{x}}$ (Higgs limit). Equation \ref{eq:chirind} becomes 
\begin{equation}
	\frac{x+\tau^2}{\tau(1+x)}\left(\sum_{\tau\in R_\Phi} e^{i\rho(h)}x^{2|\rho(s)|}\right).
\end{equation}	
Next step is to take the $x\to 0$ limit. We divide theories into two categories based on the sign of $\epsilon_0$
\begin{enumerate}
	\item $\epsilon_0$ is either zero or postive
	\item $\epsilon_0$ can be either negative, zero or positive.
\end{enumerate}
In theories where $\epsilon_0\geq 0$, one only needs to identify the monopoles with $\epsilon_0=0$ and sum over their contributions to the integral, as monopoles with $\epsilon_0> 0$ have no contributions in this limit due to the factor of $x^{\epsilon_0}$. Note that not every solution to the equation $\epsilon_0=0$ will contribute to the index. Some of those solutions contribute identically $0$ to the index. While considering different global variants of the theory one needs to be careful about which non-trivial solutions will contribute in that global variant. The case of $\epsilon_0\geq 0$ is particulary simple because the limit $x\to 0$ can be taken directly at the level of single particle indices before evaluating the integral. Which essentially means that equation \ref{eq:N5indexsimped} simplifies to
\begin{equation}
	\begin{split}
		\mathcal{I}&=\frac{1}{|W|}\sum_{s}\int\prod_j \frac{dz_j}{2\pi i z_j} e^{-S_{CS}(h,s)}\mathcal{I}_{PE}(z_i,s_i)\\
		\mathcal{I}_{PE}(z_i,s_i)&=PE\left[\lim_{x\to 0}\left(-\sum_{\alpha\in ad(G)}e^{i\alpha(h)}x^{2|\alpha(s)|}+\tau\left(\sum_{\tau\in R_\Phi} e^{i\rho(h)}x^{2|\rho(s)|}\right) \right)\right]
	\end{split}
\end{equation}
where the sum $\sum_{s}$ , is restricted to the monopoles such that $\epsilon_0=0$.

The cases when $\epsilon_0$ can take negative values, requires some care. The contribution to $x\to 0$ comes not only from $\epsilon_0=0$ monopoles. As a series expansion in $x$,   $PE\left[\text{ind}_{\text{vec}}+\text{ind}_{\text{chir}} \right]$ may contain a term with $x^{-\epsilon_0}$. This will cancel $x^{\epsilon_0}$ and contribute to the index in the $x\to 0$ limit. The strategy we use is to first restrict to a subspace of monopoles and also restrict to certain lowest value of $\epsilon_0$, say $\epsilon_0^*$. Enumerate the monopoles which have a given value $\epsilon_0$.   To make the computation faster, we expand $\text{ind}_{\text{chir}}$ to order $x^{\epsilon_0}$ and finally expand $PE\left[\text{ind}_{\text{vec}}+\text{ind}_{\text{chir}} \right]$ to $x^{\epsilon_0}$ for a given monopole. We simply evaluate the integral on coefficient $x^{\epsilon_0}$ for each such monopole and finally sum their contributions. We repeat this step for all values of $\epsilon\geq \epsilon_0^*$. Restricting to just the coefficient of $x^{\epsilon_0}$ is sufficient because to have a nice $x\to 0$ limit, as a series expansion in $x$, the negative powers of $x$ should not contribute and the positive powers of $x$ go to zero in the $x\to 0$ limit.

The cases where $\epsilon_0$ can be negative is important in dualities between ABJ theories of the form $U(N+l)_k\times U(N)_{-k}=U(N)_k\times U(N+k-l)_{-k}$ \cite{Aharony:2008gk}. Such an effect is explained in appendix \ref{sec:UUuneqrank}. A similar effect exists for theories of $OSp$ type (see appendix \ref{sec:OSpuneqrank}). For simplicity, we have restricted ourselves to the $OSp$ theories where $\epsilon_0\geq 0$. In the next subsection \ref{sec:excpC} we will consider $F(4)$ theory where monopoles with $\epsilon_0<0$ play an important role for different values of $k$.


\begin{table}[ht]
	\centering
	\resizebox{\textwidth}{!}{
		\begin{tabular}{|c|c|c|}
			\hline
			Theory & Moduli Space & Cases checked\\
			\hline
			$SO(2)_{-2k}\times USp(2)_k$    & $\mathbb{H}/\mathbb{Z}_{2k}=\mathbb{H}/G_{1}(\hat{D}_k,\mathbb{Z}_{2k})$ &\\
			$\left[SO(2)_{-2k}\times USp(2)_k\right]/\mathbb{Z}_2$    & $\mathbb{H}/\mathbb{Z}_{k}=\mathbb{H}/G_{1}(\hat{D}_k,\mathbb{Z}_{k})$ &  \\
			$O(2)_{-2k}\times USp(2)_k$    & $\mathbb{H}/\hat{D}_k=\mathbb{H}/G_{1}(\hat{D}_k,\hat{D}_k)$ & \\
			$\left[O(2)_{-2k}\times USp(2)_k\right]/\mathbb{Z}_2$    & $\mathbb{H}/\hat{D}_{\frac{k}{2}}=\mathbb{H}/G_{1}(\hat{D}_k,\hat{D}_{\frac{k}{2}})$ & \\
			\hline
			\hline
			$SO(N)_{-2k}\times USp(2)_k$    & $\mathbb{H}/\hat{D}_{k}=\mathbb{H}/G_{1}(\hat{D}_k,\hat{D}_k)$ & \makecell{$3\leq N\leq 4$}\\
			
			$\left[SO(N)_{-2k}\times USp(2)_k\right]/\mathbb{Z}_2$    & $\mathbb{H}/\hat{D}_{k}=\mathbb{H}/G_{1}(\hat{D}_k,\hat{D}_k)$ & \\
			\hline	
	\end{tabular}}
	\caption{List of theories for which the superconformal index was matched with the expected Hilbert series for rank $1$ quaternionic reflection groups. The last column displays the values of $N$ for which the check was made. The first block shows the results for $SO(2)_{-2k}\times USp(2)_{k}$ and its various variants. For the rank $1$ cases we have closed form expressions.
	}
	\label{tab:resultsconfirmedrank1}
\end{table}

\begin{table}[ht]
	\centering
	\resizebox{\textwidth}{!}{
		\begin{tabular}{|c|c|c|}
			\hline
			Theory & Moduli Space & Order matched\\
			\hline
			$SO(4)_{-2k}\times USp(4)_k$    & $\mathbb{H}^2/G_{2}(\hat{D}_k,\mathbb{Z}_{2k})$ & \makecell{$k=2,3,4$\;\; $O(\tau^{10})$} \\
			\hline
			$\left[SO(4)_{-2k}\times USp(4)_k\right]/\mathbb{Z}_2$    & $\mathbb{H}^2/G_{2}(\hat{D}_k,\mathbb{Z}_k)$ & \makecell{$k=2,3,4$\;\; $O(\tau^{10})$}\\
			\hline
			$O(4)_{-2k}\times USp(4)_k$    & $\mathbb{H}^2/G_{2}(\hat{D}_k,\hat{D}_k)$ & \makecell{$k=2,3,4$\;\; $O(\tau^{10})$}\\
			\hline
			$\left[O(4)_{-2k}\times USp(4)_k\right]/\mathbb{Z}_2$    & $\mathbb{H}^2/G_{2}(\hat{D}_k,\hat{D}_{\frac{k}{2}})$ & \makecell{$k=2,4$\;\; $O(\tau^{10})$}\\
			\hline
			\hline
			\hline
			$SO(5)_{-2k}\times USp(4)_k$    & $\mathbb{H}^2/G_{2}(\hat{D}_k,\hat{D}_k)$ & $k=2,3,4$ \;\; $O(\tau^{10})$\\
			\hline
			\hline
			$SO(6)_{-2k}\times USp(4)_k$    & $\mathbb{H}^2/G_{2}(\hat{D}_k,\hat{D}_k)$ & $k=2,3,4$ \;\; $O(\tau^{10})$\\
			\hline
			$\left[SO(6)_{-2k}\times USp(4)_k\right]/\mathbb{Z}_2$    & $\mathbb{H}^2/G_{2}(\hat{D}_k,\hat{D}_k)$ & $k=2,4$ \;\; $O(\tau^{10})$\\
			\hline
			$O(6)_{-2k}\times USp(4)_k$    & $\mathbb{H}^2/G_{2}(\hat{D}_k,\hat{D}_k)$& $k=2,3,4$ \;\; $O(\tau^{10})$\\
			\hline
			$\left[O(6)_{-2k}\times USp(4)_k\right]/\mathbb{Z}_2$    & $\mathbb{H}^2/G_{2}(\hat{D}_k,\hat{D}_k)$ & $k=2,4$ \;\; $O(\tau^{10})$\\
			\hline
	\end{tabular}}
	\caption{List of theories for which the superconformal index was matched with the expected Hilbert series for rank $2$ quaternionic reflection groups. The last column displays the values of $k$ and the order upto which the matching was done.
	}
	\label{tab:resultsconfirmedrank2}
\end{table}
\begin{table}[ht]
	\centering
	\resizebox{\textwidth}{!}{
		\begin{tabular}{|c|c|c|}
			\hline
			Theory & Moduli Space & Order matched\\
			\hline
			$SO(6)_{-2k}\times USp(6)_k$    & $\mathbb{H}^3/G_{3}(\hat{D}_k,\mathbb{Z}_{2k})$ & \makecell{$k=1,2,3$\;\; $O(\tau^{6})$} \\
			\hline
			$\left[SO(6)_{-2k}\times USp(6)_k\right]/\mathbb{Z}_2$    & $\mathbb{H}^3/G_{3}(\hat{D}_k,\mathbb{Z}_k)$ & \makecell{$k=1,2,3$\;\; $O(\tau^{6})$}\\
			\hline
			$O(6)_{-2k}\times USp(6)_k$    & $\mathbb{H}^3/G_{2}(\hat{D}_k,\hat{D}_k)$ & \makecell{$k=1,2,3$\;\; $O(\tau^{6})$}\\
			\hline
			$\left[O(6)_{-2k}\times USp(6)_k\right]/\mathbb{Z}_2$    & $\mathbb{H}^3/G_{3}(\hat{D}_k,\hat{D}_{\frac{k}{2}})$ & \makecell{$k=2$\;\; $O(\tau^{6})$}\\
			\hline
			
	\end{tabular}}
	\caption{List of theories for which the superconformal index was matched with the expected Hilbert series for rank $3$ quaternionic reflection groups. The last column displays the values of $k$ and the order upto which the matching was done.
	}
	\label{tab:resultsconfirmedrank3}
\end{table}

Using the aforementioned strategy, we have calculated the Higgs branch limit (or Coulomb branch limit as the two are identical) for the various known Lagrangian $\mathcal{N}=5$ SCFTs, and compare them against the expectations laid out in the previous section. Specifically, we have looked at cases based on the $OSp$ superalgebras for which $\epsilon_0\geq 0$, where the index computation is simplified significantly. These cases include the equal rank case, $O(2N)_{-2k}\times USp(2N)_k$, which are our main intrest here, as well as the cases $O(2N+1)_{-2k}\times USp(2N)_k$ and $O(2N+2)_{-2k}\times USp(2N)_k$. The full list of cases, and the results for them are summarized in table \ref{tab:resultsconfirmedrank1}, \ref{tab:resultsconfirmedrank2} and \ref{tab:resultsconfirmedrank3}. Note that here, outside of the relatively simple rank $1$ cases, technical limitations led to the computation being only carried out for selected number of cases with small rank and Chern-Simons levels. In all cases we find agreement with our expectations.

An illustrative review of the computation for the $OSp(4|4)$ case can be found in appendix \ref{sec:exdet}. To match the results in these cases with our expectations, one would require the Hilbert series of $\mathbb{C}^{2r}$ quotiented by a quaternionic reflection group. As this is just a quotient of a flat space, the Hilbert series can be computed by counting invariants under the group made from the $\mathbb{C}^{2r}$ coordinates. A review of this procedure, as well as a list of the invariants for selected cases, can be found in appendix \ref{app:HSQRG}.

Finally, we have also looked at the cases based on the exceptional Lie Superalgebras. This includes the ones based on $D(2|1,\alpha)$, for which $\epsilon_0 \geq 0$ and the index computation is significantly simplified. Here, we used the index computation to determine the moduli space reported previously. This brings us to the cases based on the $G(3)$ and $F(4)$ Lie Superalgebras, which are the only cases we studied which allow $\epsilon_0 < 0$, and so require greater care. The results for cases based on the exceptional Lie Superalgebras are summarized in tables \ref{tab:summarizecompute} and \ref{tab:keq1caseF(4)}. In these cases as well, we find agreement with our expectations. As the computation here is either more involved due to the presence of monopoles for which $\epsilon_0 < 0$, or was used to determine the moduli space, we shall expand upon the index computation for these cases in the next subsection. This would also allow us to illustrate the computation strategy for cases with both $\epsilon_0 \geq 0$ and without.

\subsection{Exceptional Lie-Superalgebras}
\label{sec:excpC}

Here we would like to discuss the Index computation for theories based on exceptional Lie-superalagebras. Let us discuss them one by one. 
\begin{table}
	\centering
	\resizebox{\textwidth}{!}{
		\begin{tabular}{|c|c|c|c|c|}
			\hline
			\small \makecell{Lie \\ superalgebra} &\makecell{\small Perturbative\\ \small contribution} & \small \makecell{Monopoles and\\ their contribution} & Total & \makecell{\small Moduli\\ \small Space} \\
			
			\hline
			\rule{0pt}{3ex}
			$D(2|1,\alpha)$& $\frac{1}{1-\tau^4}$ & $	\left(s,p,s+p\right):\;\;\; \frac{1}{8}\frac{\tau^{I_0}}{(1-\tau^2)}$ & 	$ \frac{1-\tau^{2I_0+4}}{(1-\tau^4)(1-\tau^{I_0})(1-\tau^{I_0+2})}$ & $\frac{\mathbb{H}^2}{\hat{D}_{I_0/2}}$\\
			\hline
			
			\rule{0pt}{4.5ex}
			$G(3)$& $\frac{1}{1-\tau^4}$ & \makecell{	$\left(\frac{3}{2},\frac{1}{2},2\right):\;\;\; \frac{1}{12}\frac{\tau^{12k}}{(1-\tau^2)}$\\$\left(0,1,2\right):\;\;\; \frac{1}{12}\frac{\tau^{12k}}{(1-\tau^2)}$} & 	$\frac{1-\tau^{24k+4}}{(1-\tau^4)(1-\tau^{12k})(1-\tau^{12k+2})}$ & $\frac{\mathbb{H}^2}{\hat{D}_{6k}}$\\
			\hline
			\rule{0pt}{3ex}
			$F(4)_{k>1}$& $\frac{1}{1-\tau^4}$ & $\left(2,0,1,3\right):\;\;\; \frac{1}{16}\frac{\tau^{12k}}{(1-\tau^2)}$ & 	$\frac{1-\tau^{24k+4}}{(1-\tau^4)(1-\tau^{12k})(1-\tau^{12k+2})}$ & $\frac{\mathbb{H}^2}{\hat{D}_{6k}}$\\
			
			\hline
			\rule{0pt}{3ex}
			$F(4)_{k>1}/\mathbb{Z}_2$& $\frac{1}{1-\tau^4}$ & $\left(1,0,\frac{1}{2},\frac{3}{2}\right):\;\;\; \frac{1}{16}\frac{\tau^{6k}}{(1-\tau^2)}$ & 	$\frac{1-\tau^{12k+4}}{(1-\tau^4)(1-\tau^{6k})(1-\tau^{6k+2})}$ & $\frac{\mathbb{H}^2}{\hat{D}_{3k}}$\\		
			\hline
	\end{tabular}}
	\caption{Summary of Hilbert series computation for exceptional Lie superalgebras. Here, the second column of the table shows the contribution from the perturbative sector, while the third column shows the monopoles that contribute to the superconformal index. The full Hilbert series and corresponding moduli space is displayed in the fourth and fifth column respectively, where for $D(2|1;\alpha)$ $s,p\in \mathbb{Z}$ and $I_0=2l p s(p+s)$. The monopoles mentioned in the table show the minimal solution to the equation $\epsilon_0=0$, and all the Weyl reflections and multiples need to be added to get the full contribution to the index (see text for more details). In these cases no monopoles with $\epsilon_0<0$ were found to contribute to the index so the total result is just the one for monopoles with $\epsilon_0=0$, and a closed form expression can be written. We stress that for $G(3)$ and $F(4)$ we have only checked monopoles with $\epsilon_0<0$ up to a given order so despite the closed form expression, the index was only checked to these orders (see tables \ref{tab:G(3)numcheck}, \ref{tab:F(4)numcheck} and \ref{tab:F(4)z2numcheck}). The only exception is the $F(4)$ theory with $k=1$, for which contributing monopoles with $\epsilon_0<0$ were found, and so this case is treated separately in table \ref{tab:keq1caseF(4)}.}
	\label{tab:summarizecompute}
\end{table}

\begin{table}
	\centering
	\resizebox{\textwidth}{!}{
		\begin{tabular}{|c|c|c|c|c|}
			\hline
			\small \makecell{Lie \\ superalgebra} &\makecell{\small Perturbative\\ \small contribution} & \small \makecell{
			Monopole \\ contribution}
			 & Total & \makecell{\small Moduli\\ \small Space} \\
			\hline
			\rule{0pt}{3ex}
			$F(4)_{k=1}$& $\frac{1}{1-\tau^4}$ &
			\begin{tabular}{c|c|c} 
					$\epsilon_0$\;\; & monopoles & contribution\\
				\hline
				$-2$&$\left(1,0,0,1\right)$&$ \frac{1}{24}\frac{\tau^{4}}{(1-\tau^2)}$\\ $-2$&$\left(1,0,1,2\right)$&$ \frac{1}{48}\frac{\tau^{8}}{(1-\tau^2)}$\\$0$ &$\left(2,0,1,3\right)$&$\frac{1}{16}\frac{\tau^{12}}{(1-\tau^2)}$\\ $-2$&$\left(\frac{1}{2}, \frac{5}{2}, \frac{3}{2}, 4\right)$&$ \frac{1}{48}\frac{\tau^{16}}{(1-\tau^2)}$\\ $-2$&$\left(3,0,2,5\right)$&$ \frac{1}{48}\frac{\tau^{20}}{(1-\tau^2)}$\\ $0$&$\left(4,0,2,6\right)$&$ \frac{1}{16}\frac{\tau^{24}}{(1-\tau^2)}$
			\end{tabular}
			 & 	\makecell{$1+2 \tau ^4+\tau ^6+3 \tau ^8+2 \tau ^{10}+4 \tau ^{12}+3 \tau ^{14}$\\$+5 \tau ^{16}+4 \tau ^{18}+6 \tau ^{20}+5 \tau ^{22}+7 \tau ^{24}+O\left(\tau ^{25}\right)$} & $\frac{\mathbb{H}^2}{\hat{D}_{2}}$\\		
			\hline
			\rule{0pt}{3ex}
			$F(4)_{k=1}/\mathbb{Z}_2$& $\frac{1}{1-\tau^4}$ & \begin{tabular}{c|c|c}
					$\epsilon_0$\;\; & monopoles & contribution\\
				\hline $-2$&$\left(\frac{1}{2},\frac{1}{2},0,\frac{1}{2}\right)$&$ \frac{1}{12}\frac{\tau^{2}}{(1-\tau^2)}$\\ $-2$&$\left(1,0,0,1\right)$&$ \frac{1}{24}\frac{\tau^{4}}{(1-\tau^2)}$\\$0$&$\left(1,0,\frac{1}{2},\frac{3}{2}\right)$&$\frac{1}{16}\frac{\tau^{6}}{(1-\tau^2)}$\\ $-2$&$\left(1,0,1,2\right)$&$ \frac{1}{48}\frac{\tau^{8}}{(1-\tau^2)}$\\ $-2$&$\left(\frac{3}{2},\frac{1}{2},1,\frac{5}{2}\right)$&$ \frac{1}{48}\frac{\tau^{10}}{(1-\tau^2)}$\\$0$&$\left(2,0,1,3\right)$&$ \frac{1}{16}\frac{\tau^{12}}{(1-\tau^2)}$
			\end{tabular} & 	$1+\tau ^2+3 \tau ^4+3 \tau ^6+5 \tau ^8+5 \tau ^{10}+7 \tau ^{12}+O\left(\tau ^{13}\right)$ & $\frac{\mathbb{H}^2}{\hat{D}_{1}}$\\		
			\hline
	\end{tabular}}
	\caption{Hilbert series for the $F(4)$ and $F(4)/\mathbb{Z}_2$ theory for $k=1$. The third column displays the monopole, its $\epsilon_0$ value and its contribution. One needs to take all possible Weyl reflections of the monopole and sum to get the total contribution. In table \ref{tab:summarizecompute}, a single monopole was a representative for its Weyl reflections and also its multiples, while in this table we do display multiples, but not Weyl reflection.}
	\label{tab:keq1caseF(4)}
\end{table}
\subsubsection*{\underline{$D(2|1,\alpha)$}}
The SCFT based on $D(2|1,\alpha)$ consists of the gauge group $SU(2)_{k_1} \times SU(2)_{k_2}\times SU(2)_{k_3}$ with a hypermultiplet in the $(\mathbf{2},\mathbf{2},\mathbf{2})$. We label the monopoles by $(m_1,m_2,m_3)$ using the Cartans of $SU(2)^3$. The SUSY enhancement condition is
\begin{equation}
	\frac{1}{k_1}+\frac{1}{k_2}+\frac{1}{k_3}=0
\end{equation}
As discussed in section \ref{sec:reviewN5}, the solution to the SUSY enchancement condition can be written as
\begin{equation}
	\label{eq:susyenhanced21a}
	(k_1,k_2,k_3)=(-lp(p+s),-ls(p+s),lps)
\end{equation}

This theory has $\epsilon_0\geq 0$ and therefore we can simply restrict to monopoles with $\epsilon_0=0$. Computing the perturbative sector one gets
\begin{equation}
	\frac{1}{1-\tau^4}
\end{equation}
In the $x\to 0$ limit of the index, the contributing monopoles satisfy
\begin{equation}
	\label{eq:minmonD21a}
	\begin{split}
		\epsilon_0=&-2 \left| m_1\right| -2 \left| m_2\right| -2 \left| m_3\right| +  \left| m_1-m_2-m_3\right| +\left| m_1+m_2-m_3\right|
		+\left| m_1-m_2+m_3\right|\\ & +\left| m_1+m_2+m_3\right|=0
	\end{split}
\end{equation}
The above equation is satisfied whenever $\pm m_1\pm m_2\pm m_3=0$. The solutions can be written in terms of the $s$ and $p$ variables used to satisfy the SUSY enhancement condition \ref{eq:susyenhanced21a}. The choice of the minimal solution to \ref{eq:minmonD21a}  is made such that the Dirac quantization condition is obeyed. For fixed values of $s$ and $p$, the monpoles that contribute are $(\pm s,\pm p,\pm(s+p))$ and all their multiples. The monopole $(s,p,s+p)$ has a contribution
\begin{equation}
	\frac{1}{8}\frac{\tau^{I_0}}{(1-\tau^2)}\;,
\end{equation}
where $I_0=2lps(p+s)$. Summing over all the multiples gives 
\begin{equation}
	\sum_{n\in\mathbb{Z}_{\geq 0}}	\frac{1}{8}\frac{\tau^{I_0}}{(1-\tau^2)}=	\frac{1}{8}\frac{\tau^{I_0}}{(1-\tau^2)(1-\tau^{I_0})}
\end{equation}
Taking into account all the signs of $(\pm s,\pm p,\pm(s+p))$, there are in total $2^3=8$ terms with the same contribution to the index. Therefore the total is
\begin{equation}
	\frac{1}{1-\tau^4}+\frac{2^3}{8}\frac{\tau^{I_0}}{(1-\tau^2)(1-\tau^{I_0})}=	\frac{1-\tau^{2I_0+4}}{(1-\tau^4)(1-\tau^{I_0})(1-\tau^{I_0+2})}
\end{equation} 
which is the Hilbert series for  $\mathbb{H}^2/\hat{D}_{lps(p+s)}$.
The $\alpha$ of $D(2|1,\alpha)$ is related to the Chern-Simons levels as follows (see \cite{frappat1996dictionary})\footnote{Commutators for $D(2|1;\alpha)$ are given on page 29 of \cite{frappat1996dictionary}. There $\alpha=s_2/s_1$. The $s_i$ are inversely proportional to the $k_i$ here ie. $k_i\propto \frac{1}{s_i}$ }
\begin{equation}
	\alpha=\frac{k_1}{k_2}=\frac{p}{s} .
\end{equation}

\subsubsection*{\underline{$G(3)$}}

The SCFT based on $G(3)$ consists of the gauge groups $\left(G_2\right)_{k_1} \times USp(2)_{k_2}$ with a hypermultiplet in the $(\mathbf{7},\mathbf{2})$. We choose an $SU(2)^2$ subgroup of $G_2$ and label the monopoles by $(m_1,m_2,n_1)$ using the Cartans of $SU(2)^3\subset G_2\times USp(2)$. The $m_i$'s are used for $G_2$ and $n_1$ is used for $USp(2)$. The SUSY enchancement condition is 
\begin{equation}
	3k_1+4k_2=0
\end{equation}
We choose $k_1=-4k$ and $k_2=3k$. Computing the perturbative sector one gets
\begin{equation}
	\frac{1}{1-\tau^4}
\end{equation}
This theory contains monopoles which have negative $\epsilon_0$. But first we restrict to the case when $\epsilon_0=0$. The contributing monopoles satisfy
\begin{equation}
	\label{eq:minmonG3}
	\begin{split}
		\epsilon_0=&-2 \left| m_1\right| -\left| m_1-3 m_2\right| -\left| m_1-m_2\right| -2 \left| m_2\right| -\left| m_1+m_2\right| -\left|
		m_1+3 m_2\right|\\& -\left| n_1\right| +\left| -2 m_2+n_1\right| +\left| -m_1-m_2+n_1\right| +\left| m_1-m_2+n_1\right|
		+\left| -m_1+m_2+n_1\right| \\&+\left| m_1+m_2+n_1\right| +\left| 2 m_2+n_1\right|=0
	\end{split}
\end{equation}
We find that this equation is satisfied by $(\pm \frac{3}{2},\pm \frac{1}{2},\pm 2)$ and $(0,\pm 1,\pm 2)$ and all their multiples. The choice of the minimal solution  to \ref{eq:minmonG3} is made such that the Dirac quantization condition is obeyed.
The monopole $(\frac{3}{2},\frac{1}{2}, 2)$ and $(0,1, 2)$ both individually have the contribution
\begin{equation}
	\frac{1}{12}\frac{\tau^{12k}}{(1-\tau^2)}
\end{equation}
and summing over all the multiples gives 
\begin{equation}
	\sum_{n\in\mathbb{Z}_{> 0}}	\frac{1}{12}\frac{\tau^{12k}}{(1-\tau^2)}=	\frac{1}{12}\frac{\tau^{12k}}{(1-\tau^2)(1-\tau^{12k})}
\end{equation}
Since we have $(\pm \frac{3}{2},\pm \frac{1}{2},\pm 2)$ and $(0,\pm 1,\pm 2)$, there are in total $2^3\times 2+2^2=12$ terms with the same contribution to the index. Therefore the total is
\begin{equation} \label{G3index}
	\frac{1}{1-\tau^4}+\frac{2^3\times 2+2^2 }{12}\frac{\tau^{12k}}{(1-\tau^2)(1-\tau^{12k})}=	\frac{1-\tau^{24k+4}}{(1-\tau^4)(1-\tau^{12k})(1-\tau^{12k+2})}
\end{equation}
which is the Hilbert series for $\frac{\mathbb{H}^2}{\hat{D}_{6k}}$.

Now, let us include the effect of monopoles which have $\epsilon_0<0$. We have summarized the numerical checks in table \ref{tab:G(3)numcheck}. As mentioned above, this theory also contains monopoles with $\epsilon_0<0$, although for low values of $k$ we haven't found any such non-trivial contributions and the result seems to hold for all $k$. As such, the closed form expression, \eqref{G3index}, appears to fully capture the moduli space.  

\begin{table}
	\begin{tabular}{|c|c|c|c|c|}
		\hline
		$k$ & Moduli Space &  Order matched & range of $m_{i=1,2}$ & range of $n_1$\\
		\hline
		$k=1$ & $\mathbb{H}^2/\hat{D}_{6}$ & $O(\tau^{24})$ & $-3,-5/2,...,5/2,3$ & $-6,...,6$\\
		$k=2$ & $\mathbb{H}^2/\hat{D}_{12}$ & $O(\tau^{24})$ & $-3,-5/2,...,5/2,3$ & $-6,...,6$ \\
		$k=3$ & $\mathbb{H}^2/\hat{D}_{18}$ & $O(\tau^{24})$ & $-3,-5/2,...,5/2,3$ & $-6,...,6$\\
		\hline
	\end{tabular}
	\caption{$G(3)$ moduli space checks. For the cases mentioned in this table, the minimum value of $\epsilon_0$ to which the check was performed was $\epsilon_0^*=-4$}
	\label{tab:G(3)numcheck}
\end{table}

\subsubsection*{\underline{$F(4)$}}
The SCFT based on $F(4)$ consists of the gauge groups $Spin(7)_{k_1} \times USp(2)_{k_2}$ with a hypermultiplet in the $(\mathbf{8},\mathbf{2})$. We choose an $SU(2)^3$ subgroup of $Spin(7)$ and label the monopoles by $(m_1,m_2,m_3,n_1)$ using the Cartans of $SU(2)^4\subset Spin(7)\times USp(2)$. The $m_i$'s are used for $Spin(7)$ and $n_1$ is used for $USp(2)$. The SUSY enchancement condition is 
\begin{equation}
	2k_1+3k_2=0
\end{equation}
We choose $k_1=-3k$ and $k_2=2k$. Computing the perturbative sector one gets
\begin{equation}
	\frac{1}{1-\tau^4}
\end{equation}
This theory contains monopoles which have negative $\epsilon_0$. But first we restrict to the case when $\epsilon_0=0$. The contributing monopoles satisfy
\begin{equation}
	\label{eq:minmonF4}
	\begin{split}
		\epsilon_0=&-2 \left| m_1\right| -\left| m_1-m_2\right| -2 \left| m_2\right| -\left| m_1+m_2\right | -2 \left| m_3\right| -\left|
		-m_1-m_2+2 m_3\right|\\ & -\left| m_1-m_2+2 m_3\right| -\left| -m_1+m_2+2 m_3\right|-\left| m_1+m_2+2 m_3\right| +\left|
		-m_1+m_3-n_1\right|\\ & +\left| m_1+m_3-n_1\right|  +\left| -m_2+m_3-n_1\right| +\left| m_2+m_3-n_1\right| -2 \left|
		n_1\right| +\left| -m_1+m_3+n_1\right|\\ & +\left| m_1+m_3+n_1\right| +\left| -m_2+m_3+n_1\right| +\left|
		m_2+m_3+n_1\right|=0
	\end{split}
\end{equation}
We find that this equation is satisfied by  $(\pm 2,0,\pm 1,\pm 3)$ and $(0,\pm 2,\pm 1,\pm 3)$ and all their multiples. The choice of the minimal monopole solution to \ref{eq:minmonF4} is made such that the Dirac quantization condition is obeyed.
The monopole $(2,0,1,3)$ has the contribution
\begin{equation}
	\frac{1}{16}\frac{\tau^{12k}}{(1-\tau^2)}
\end{equation}
and summing over all the multiples gives 
\begin{equation}
	\sum_{n\in\mathbb{Z}_{> 0}}\frac{1}{16}\frac{\tau^{12k n}}{(1-\tau^2)}=\frac{1}{16}\frac{\tau^{12k}}{(1-\tau^2)(1-\tau^{12k})}
\end{equation}
Since we have  $(\pm 2,0,\pm 1,\pm 3)$ and $(0,\pm 2,\pm 1,\pm 3)$, there are in total $2^3\times 2=16$ terms with the same contribution to the index.. Therefore the total is
\begin{equation}
	\frac{1}{1-\tau^4}+\frac{2^3\times 2}{16}\frac{\tau^{12k}}{(1-\tau^2)(1-\tau^{12k})}=\frac{1-\tau^{24k+4}}{(1-\tau^4)(1-\tau^{12k})(1-\tau^{12k+2})}
\end{equation}
which is the Hilbert series for $\frac{\mathbb{H}^2}{\hat{D}_{6k}}$. We can also consider the variant $\frac{Spin(7)_{k_1} \times USp(2)_{k_2}}{\mathbb{Z}_2}$ in which case, the Hilbert series is $\frac{\mathbb{H}^2}{\hat{D}_{3k}}$.

\begin{table}
	\begin{tabular}{|c|c|c|c|c|}
		\hline
		$k$ & Moduli Space &   Order matched & range of $m_{i=1,2,3}$ & range of $n_1$\\
		\hline
		$k=1$ & $\mathbb{H}^2/\hat{D}_{2}$ & $O(\tau^{24})$ & $-3,-5/2,...,5/2,3$ & $-6,...,6$\\
		$k=2$ & $\mathbb{H}^2/\hat{D}_{12}$ & $O(\tau^{24})$ & $-3,-5/2,...,5/2,3$ & $-6,...,6$ \\
		$k=3$ &  $\mathbb{H}^2/\hat{D}_{18}$ & $O(\tau^{24})$ & $-3,-5/2,...,5/2,3$ & $-6,...,6$\\
		\hline
	\end{tabular}
	\caption{$F(4)$ moduli space checks. For the cases mentioned in this table, the minimum value of $\epsilon_0$ to which the check was performed was $\epsilon_0^*=-3$}
	\label{tab:F(4)numcheck}
\end{table}

\begin{table}
	\begin{tabular}{|c|c|c|c|c|}
		\hline
		$k$ & Moduli Space &   Order matched & range of $m_{i=1,2,3}$ & range of $n_1$\\
		\hline
		$k=1$ & $\mathbb{H}^2/\hat{D}_{1}$ & $O(\tau^{12})$ & $-3,-5/2,...,5/2,3$ & $-3,-5/2,...,5/2,3$\\
		$k=2$ & $\mathbb{H}^2/\hat{D}_{6}$ & $O(\tau^{12})$ & $-3,-5/2,...,5/2,3$ & $-3,-5/2,...,5/2,3$ \\
		$k=3$ &  $\mathbb{H}^2/\hat{D}_{9}$ & $O(\tau^{12})$ & $-3,-5/2,...,5/2,3$ & $-3,-5/2,...,5/2,3$\\
		\hline
	\end{tabular}
	\caption{$F(4)$ quotiented by $\mathbb{Z}_2$ moduli space checks. For the cases mentioned in this table, the minimum value of $\epsilon_0$ to which the check was performed was $\epsilon_0^*=-3$}
	\label{tab:F(4)z2numcheck}
\end{table}

Now, let us include the effect of monopoles which have $\epsilon_0<0$. $k=1$ case is a little special since here we find contributions from monopoles with $\epsilon_0<0$. In the checks that we performed, we didn't find such contribution for $k>1$ and the moduli space we expect there is $\frac{\mathbb{H}^2}{\hat{D}_{6k}}$ as mentioned in table \ref{tab:summarizecompute}. The monopoles for the $k=1$ case are mentioned in \ref{tab:keq1caseF(4)}.  We have summarized the numerical checks in table \ref{tab:F(4)numcheck} and \ref{tab:F(4)z2numcheck}. Summing the monopoles mentioned in the table, we get the following Hilbert series for $F(4)$ theory and its $\mathbb{Z}_2$ quotient version.
\begin{equation}
	\mathcal{I}_{F(4)}=1+2 \tau ^4+\tau ^6+3 \tau ^8+2 \tau ^{10}+4 \tau ^{12}+3 \tau ^{14}+5 \tau ^{16}+4 \tau ^{18}+6 \tau ^{20}+5 \tau ^{22}+7 \tau ^{24}+O\left(\tau ^{25}\right),
\end{equation}

\begin{equation}
	\mathcal{I}_{F(4)/\mathbb{Z}_2}=1+\tau ^2+3 \tau ^4+3 \tau ^6+5 \tau ^8+5 \tau ^{10}+7 \tau ^{12}+O\left(\tau ^{13}\right).
\end{equation}
These match the the Hilbert Series of $\mathbb{H}^2/\hat{D}_{2}$ and $\mathbb{H}^2/\hat{D}_{1}$ to order $\tau^{24}$ and $\tau^{12}$ respectively. This matches our expectations for the moduli space for $k=1$ mentioned in section \ref{sec:expmodspace}, which is discussed in detail in appendix \ref{app:f4}.

Given the monopoles in table \ref{tab:keq1caseF(4)}, we observe that there is a conjectural set of monopoles that reproduce the Hilbert Series in closed form. We observe that the monopoles $(1,0,0,1)$, $(1,0,1,2)$,  $(2,0,1,3)$ and their products and powers can reproduce the result. For the monopole $(2,0,1,3)$ we can consider any multiple of this monopole, and summing all of these, we get the contribution $\frac{\tau^{12}}{(1-\tau^2)(1-\tau^{12})}$. The monopoles $(1,0,0,1)$, $(1,0,1,2)$ and their Weyl reflections require the gauginos, which are fermions, to be gauge invariant and we cannot take arbitrary powers of these, though these can be multiplied with the $(2,0,1,3)$ monopole. Summing all the multiples of $(2,0,1,3)$ multiplied with $(1,0,0,1)$ and $(1,0,1,2)$ gives the contribution $\frac{\tau^4}{(1-\tau^2)(1-\tau^{12})}$ and $\frac{\tau^8}{(1-\tau^2)(1-\tau^{12})}$ respectively. Summing all these contributions we get 
\begin{equation}
	\begin{split}
		&\frac{1}{1-\tau^4}+\frac{\tau^{12}}{(1-\tau^2)(1-\tau^{12})}+\frac{\tau^4}{(1-\tau^2)(1-\tau^{12})}+\frac{\tau^8}{(1-\tau^2)(1-\tau^{12})}\\=&\frac{1-\tau^{12}}{(1-\tau^4)(1-\tau^{4})(1-\tau^{6})}.
	\end{split}
\end{equation}
This is the Hilbert Series for $\mathbb{H}^2/\hat{D}_{2}$.

The $F(4)/\mathbb{Z}_2$ theory also admits a similar closed form expression. For this case, we find the $(2,0,1,3)$ monopole is replaced with the more basic $(1,0,\frac{1}{2},\frac{3}{2})$, which now obeys Dirac quantization. In addition we also have the monopoles $(\frac{1}{2},\frac{1}{2},0,\frac{1}{2})$, $(1,0,0,1)$ and their Weyl reflections, which require the gauginos to be gauge invariant. We can again sum over all the multiples of the $(1,0,\frac{1}{2},\frac{3}{2})$ monopole, and its products with the $(\frac{1}{2},\frac{1}{2},0,\frac{1}{2})$ and $(1,0,0,1)$ monopoles, resulting in the Hilbert Series:
\begin{equation}
	\begin{split}
		&\frac{1}{1-\tau^4}+\frac{\tau^6}{(1-\tau^{2})(1-\tau^{6})}+\frac{\tau^2}{(1-\tau^2)(1-\tau^{6})}+\frac{\tau^4}{(1-\tau^2)(1-\tau^{6})}\\=&\frac{1-\tau^{8}}{(1-\tau^4)(1-\tau^{2})(1-\tau^{4})}.
	\end{split}
\end{equation}
This is the Hilbert Series for $\mathbb{H}^2/\hat{D}_{2}$.

\section{Conclusions}
\label{sec:concl}

We have proposed a correspondence between 3d $\mathcal{N}=5$ SCFTs and quaternionic reflection groups, where the moduli space of  $\mathcal{N}=5$ SCFTs is of the form of $\mathbb{H}^{2r}/\Gamma$ with $\Gamma$ a quaternionic reflection group. More specifically, we  proposed that any 3d $\mathcal{N}=5$ SCFT can be realized through the gauging of a discrete 0-form symmetry of an $\mathcal{N}=5$ SCFT whose moduli space is of the form listed above. This suggests that we can uniquely associate a quaternionic reflection group with every family of 3d $\mathcal{N}=5$ SCFTs, related to one another through the gauging of discrete symmetries. The associated quaternionic reflection group is then the one quotienting the moduli space of the mother SCFT, from which all other members of the family can be realized by gauging discrete symmetries (this can be defined as the mother SCFT having no non-anomalous 1-form symmetries). This extends a similar conjecture in \cite{Tachikawa:2019dvq}, regarding 3d $\mathcal{N}\geq 6$ SCFTs and 4d $\mathcal{N}\geq 3$ SCFTs, so as to cover all SCFTs with ten or more supercharges in three and four dimensions.

We have tested the above proposal by examining the moduli space of all known 3d $\mathcal{N}=5$ SCFTs, and observe that they seem to conform to the above proposal. Here we have mostly examined Lagrangian theories, where the moduli space can be studied from the Lagrangian description. A powerful tool to study the moduli space in these cases is the superconformal index, and we have employed this tool to test our expectations regarding the moduli space of theories of interest, and even used it to determine it in certain cases. Our results then elucidate the moduli spaces of 3d $\mathcal{N}=5$ SCFTs, and their variants, in cases where it was not known before.

There are several directions where our results can be extended. First, it would be interesting if the classification of Lagrangian 3d $\mathcal{N}=6$ SCFTs of \cite{Schnabl:2008wj} can be extended also to 3d $\mathcal{N}=5$ SCFTs. This might lead to new 3d $\mathcal{N}=5$ SCFTs on which we could check our proposal. It would also be interesting to check for the presence of non-invertible higher form symmetries, which if present would require gauging to get to a variant devoid of 1-form symmetries.

Another interesting direction is to consider non-Lagrangian $\mathcal{N}=5$ SCFTs. Specifically, there is a known family of 3d $\mathcal{N}=5$ SCFTs, corresponding to M$2$-branes probing a $\mathbb{C}^{4}/\hat{E}_k$ singularity, for which no Lagrangian description is known. It would be interesting to understand its 1-form symmetries and what variants exist whose moduli space is a quotient by a smaller group. Interestingly, the spectrum of possible quaternionic reflection groups seems to suggest that the 1-form symmetry be the center of the associated ADE group\footnote{This means $\mathbb{Z}_3$ for $\hat{E}_6$, $\mathbb{Z}_2$ for $\hat{E}_7$ and trivial for $\hat{E}_8$.}. This indeed works for the A and D-type cases, and it would be interesting to check this for the E-type cases. Indeed, 1-form symmetries of non-Lagrangian SCFTs, like 4d $\mathcal{N}=3$ SCFTs, have been determined from a string theory construction, see \cite{Aharony:2016kai,Etheredge:2023ler,Amariti:2023hev}, so this might be possible also in this case. It is also interesting to determine if there exists additional SCFTs in this class, corresponding to the unequal rank case of the $SU(N|N)$ and $OSp(2N|2N)$ cases. 

Finally, the quaternionic reflection groups that seem to appear in the known $\mathcal{N}=5$ SCFTs all belong to a certain infinite family. While this is the only possible family of quaternionic reflection groups that exists for every rank, there exist many quaternionic reflection groups at rank 2, and a handful of cases at higher rank, for which there is no known associated $\mathcal{N}=5$ SCFT. This might hint at the existence of many yet-to-be discovered $\mathcal{N}=5$ SCFTs, and it is interesting if these can be found. Such a goal might motivate novel QFT (both perturbative and non-perturbative) and string theory constructions that would be interesting in their own right.

\section*{Acknowledgments}
We would like to thank Oren Bergman, Yichul Choi, Mykola Dedushenko, Leonardo Rastelli, Matteo Sacchi, Yaman Sanghavi 
and Yuji Tachikawa for helpful comments and discussions. GZ is partially supported by the Simons Foundation grant 815892 and by the Israel Science Foundation under grant no. 759/23. AD is supported in part by NSF grant PHY-2210533, and the Simons Foundation grants 397411 (Simons Collaboration on the Nonperturbative Bootstrap) and 681267 (Simons Investigator Award).

\appendix

\section{Quaternionic reflection groups}
\label{sec:QRG}

Reflection groups are, as their name suggests, groups spanned by reflections. For instance, given the space $\mathbb{R}^r$, a reflection is a transformation acting on one of the coordinates as $x_i \rightarrow - x_i$. Note that reflections can also act on one combination of coordinates. For instance, the permutation of two coordinates, say $x_1$ and $x_2$, is a reflection as: $x_1 - x_2 \rightarrow - (x_1 - x_2)$. A reflection group is then a group whose basic generators can be chosen to be reflections. For instance, the permutation group acting on $\mathbb{R}^r$ by permuting its coordinates, is a reflection group, since any permutation can be build from ones exchanging only two coordinates. Note that a reflection group then refers not just to its definition as an abstract group, but also to its action on a specific real space. This action then defines a specific representation of the group, such that the group action can be taken as $r\times r$ real matrices acting on the vector of coordinates. Note that the condition that the group is generated by reflections implies that the aforementioned representation can be generated by matrices whose eigenvalues are all $1$, save for one that is $-1$.  

One common example of reflection groups are Weyl groups, which are generated by reflections around the roots of a Lie algebra. However, these are not the only possible reflection groups. For instance, the Dihedral group $D_k$, which is the symmetry group of the regular $k$-gon is a reflection group, while it is a Weyl group only if $k=1,2,3,4$ and $6$. Weyl groups play an important role in physics, where they often appear as symmetry groups quotienting a flat moduli space due to residual gauge invariance. Reflection groups provide a generalization of Weyl groups, and as such one might expect for these to also appear as quotients of certain moduli spaces. Indeed, it was noted in \cite{Tachikawa:2019dvq} that the moduli space of $3d$ $\mathcal{N}=8$ SCFTs is often of the form of a flat moduli space quotiented by a reflection group, which was then used to put forward a proposal for the classification of $3d$ $\mathcal{N}=8$ SCFTs.

So far, we assumed the space in question was a real space. However, we can also consider generalizations to other division algebras, most notably complex numbers. So what is a reflection of a complex number? For real numbers, we called a coordinate transformation under which $x\rightarrow \xi x$ a reflection if $|\xi|=1$. It is then natural to generalize this to complex numbers by defining a reflection as the coordinate transformation $z\rightarrow \xi z$, $|\xi|=1$. We can then consider groups acting on $\mathbb{C}^r$ that can be generated by reflections. These are referred to as complex reflection groups, and the previously mentioned groups acting on a real space through real reflections are called real reflection groups. Complex reflection groups then provide another generalization of Weyl groups, and it has been suggested that these may play a role in the classification of $4d$ $\mathcal{N}=3$ and $3d$ $\mathcal{N}=6$ SCFTs \cite{Tachikawa:2019dvq}. The appearance of these groups in relation to SCFTs in $4d$ and $3d$ has lead to their introduction to physics, with several overviews on these groups now available in the physics literature \cite{Tachikawa:2019dvq,Kaidi:2022lyo}. As such, we shall not elaborate on these here and refer the reader to the cited references for further details. Instead, here we shall be concerned with the additional generalization from real and complex numbers to quaternions, that lead to the definition of quaternionic reflection groups.

Similarly to complex reflection groups, quaternionic reflection groups are groups generated by quaternionic reflections, that is a unimodular transformation of a quternionic coordinate. As before, the group then naturally acts on $\mathbb{H}^r$ such that the generating reflections act non-trivially only on one combination of coordinates, where the action is a quaternionic reflection. As such, they define not only an abstract group, but a specific representation of it, such that it is given by $r\times r$ quaternionic unitary matrices, determining its action on the $r$ coordinates of $\mathbb{H}^r$, which have only one non-unit eigenvalue of the form of a unimodular quaternion. 

Recall first that the quaternionic numbers are defined to have the general form $a + i b + j c + k d$, for $a$, $b$, $c$ and $d$ real numbers. Here the objects $i$, $j$ and $k$ obey the following relations: $i^2 = j^2 = k^2 = -1$, $i j = - j i = k$. These can be conveniently represented through matrices:

\be \nonumber
i = \begin{pmatrix}
  i & 0  \\
  0 & -i
\end{pmatrix} \; , \;
j = \begin{pmatrix}
  0 & 1  \\
  -1 & 0
\end{pmatrix} \; , \;
k = \begin{pmatrix}
  0 & i  \\
  i & 0
\end{pmatrix},
\ee
which form the quaternion group. This implies that a general quaternionic number can be represented by the matrix: 

\be \nonumber
\begin{pmatrix}
  a+bi & c+id  \\
  -c+id & a-bi
\end{pmatrix}.
\ee

A unimodular quaternion further obeys that $a^2 + b^2 + c^2 + d^2 = 1$. This means that unimodular quaternions are given by elements of $SU(2)$. The basic quaternionic reflections are then given by discrete subgroups of $SU(2)$. These are given by the lift to $SU(2)$ of the discrete subgroups of $SO(3)$, which in turn are given by symmetries of three dimensional solids. These are given by the cyclic groups $\mathbb{Z}_k$, the binary dihedral group $\hat{D}_k$, and the binary polyhedral groups $\hat{T}$, $\hat{O}$ and $\hat{I}$. These should not be confused with their non-binary friends. Specifically, given a $3d$ solid we can consider its symmetry under only rotations or with rotations and reflections, and also differentiate by whether the symmetry acts only on bosons or on bosons and fermions. The four possibilities distinguish whether the discrete group we get is a subgroup of $SO(3)$, $O(3)$, $Spin(3)=SU(2)$, or one of the $Pin(3)$ groups, and we get different groups in each case. Especially confusing are the $O(3)$ and $Spin(3)$ cases as these are different $\mathbb{Z}_2$ extensions of $SO(3)$, and as such the resulting discrete groups are different but have the same order.

\begin{table}[!h]
\centering
\resizebox{\textwidth}{!}{
\begin{tabular}{|c|c|c|}
\cline{2-3}
 \multicolumn{1}{c|}{} & Bosons only & Bosons and fermions \\
\hline 
 \multirow{2}{*}{Rotations only} & Discrete subgroups of $SO(3)$ & Discrete subgroups of $Spin(3)=SU(2)$ \\
  & Rotational/chiral groups & Binary groups \\
\hline
 \multirow{2}{*}{Rotations and reflections} & Discrete subgroups of $O(3)$ & Discrete subgroups of $Pin(3)$ \\
  & Reflection/full groups & Pinor groups \\
\hline
\end{tabular}}
\caption{Different possible symmetry groups of solids depending on whether we allow reflections and/or fermions.}
\label{RRgroups}
\end{table}

For example, the standard dihedral group $D_k$, is the rotational symmetry of a prism whose base is the regular $k$-gon, that is it is the discrete subgroup of $SO(3)$ that maps this prism to itself. If we consider instead its lift to $SU(2)$, then we get the binary dihedral group $\hat{D}_k$. 

Similarly, we can consider the rotational symmetries of the regular platonic solids, the tetrahedral, $T$, octahedral, $O$, and icosahedral, $I$, groups. As abstract groups these are equal to: $T=A_4$, the alternating group for four elements, $O=S_4$, the permutation group on four elements, $I=A_5$, the alternating group for five elements. This gives the symmetries as subgroups of $SO(3)$. If we consider their lifts to $SU(2)$ then we get the binary versions, $\hat{T}$, $\hat{O}$ and $\hat{I}$. As abstract groups we have that $\hat{T}=SL(2,3)$, the group of all $2 \times 2$ unit determinant matrices over the field of integers modulo $3$, and $\hat{I}=SL(2,5)$, the group of all $2 \times 2$ unit determinant matrices over the field of integers modulo $5$. If we allow reflections, then we get a real reflection group, of the type previously discussed. Specifically, the real reflection group associated with the tetrahedron is $S_4$, the permutation group on four elements, with the cube/octahedron is $S_4 \times \mathbb{Z}_2$, and with the dodecahedron/icosahedron is $H_3=A_5 \times \mathbb{Z}_2$. The former two are the Weyl groups of $SU(4)$ and $USp(6)/SO(7)$ respectively.

Returning to quaternionic reflection groups, the binary groups mentioned above give the full list of quaternionic reflection groups acting on the one dimensional quaternionic vector space $\mathbb{H}$. For the case of $\mathbb{Z}_k$, the group is actually a complex reflection group, and if further $k=2$, then it is a real reflection group. The problem then is to find how these can be combined to form quaternionic reflection groups acting on the higher dimensional quaternionic vector space $\mathbb{H}^n$. In fact, there is a complete classification of quaternionic reflection groups by Cohen \cite{Cohen:1980qrg}, similarly to the classification of real reflection groups by Coxeter and complex reflection groups by Shephard and Todd. Besides the groups appearing as real or complex reflection groups, the quaternionic reflection groups consist of an infinite family $G_n (K,H)$, with a special case $G (K,H,\alpha)$ for $n=2$, a family of rank $r=2$ cases $E(H)$, and $13$ new exceptional cases (including an extension of the Hall-Janko simple group by $\mathbb{Z}_2$). Here we shall be mainly concerned with the infinite family $G_n (K,H)$, which is the one appearing in the moduli space of known $3d$ $\mathcal{N}=5$ SCFT. We next summarize the structure of the groups in this family, for both the $n>2$ and $n=2$ cases. 

\subsection{The groups $G_n (K,H)$ and $G (K,H,\alpha)$}

Here we consider the groups $G_n (K,H)$ and $G (K,H,\alpha)$, which can be thought of as the quaternionic versions of Weyl groups of the classical groups. Specifically, these are defined by a choice of groups $K$ and $H$, both discrete subgroups of $SU(2)$ of the type we discussed previously. These always contain the permtation group, as well as all elements of the form $diag(h,1,1,...,1)$ for $h$ an element of $H$. The structure of these groups differ between $n=2$ and $n>2$, and we next discuss each in turn.  

\subsubsection{$G_n (K,H)$ for $n>2$}

We first consider the case of $n>2$. Specifically, we consider the groups $G_n (K,H)$, for $n>2$. These are defined as the quaternionic reflection groups acting on the space $\mathbb{H}^n$, which are generated by permutations of the $n$ coordinates as well as the transformations given by the matrices $M=diag(k_1,k_2,k_3,...,k_n)$, where the $k_i$'s are all elements of the group $K$ and we demand that $Det(M) = k_1 k_2 ... k_n$ is an element of $H$. For this to be consistent we must demand that $H$ is a normal subgroup of $K$ such that $K/H$ is an abelian group\footnote{This can also be phrased as the condition that $H$ contains the commutator subgroup of $K$.}. This follows as the $k_i$'s are quaternions and so are not commutative. As such, the action of permutations on $M$ will change the determinant in such a way that it may no longer be in $H$. However, all elements of $K$ commute up to an element in the commutator subgroup, and so if this group is contained in $H$, we are guaranteed that $Det(M)$ will still be in $H$ for all permutations.

It is instructive to consider another argument for why $K/H$ should necessarily be abelian. Specifically, assume $n=3$ and consider the product:

\be
\begin{pmatrix}
  k_1 & 0 & 0  \\
  0 & k^{-1}_1 & 0 \\
	0 & 0 & 1
\end{pmatrix} \begin{pmatrix}
  k_2 & 0 & 0  \\
  0 & 1 & 0 \\
	0 & 0 & k^{-1}_2
\end{pmatrix} \begin{pmatrix}
  k^{-1}_1 & 0 & 0  \\
  0 & k_1 & 0 \\
	0 & 0 & 1
\end{pmatrix} \begin{pmatrix}
  k^{-1}_2 & 0 & 0  \\
  0 & 1 & 0 \\
	0 & 0 & k_2
	\end{pmatrix} = \begin{pmatrix}
  k_1 k_2 k^{-1}_1 k^{-1}_2 & 0 & 0  \\
  0 & 1 & 0 \\
	0 & 0 & 1
	\end{pmatrix} .
\ee

Here all four matrices have unit determinant so are in $H$ and so part of the group for any $k_1$ and $k_2$ in $K$. Consistency then necessitates that $k_1 k_2 k^{-1}_1 k^{-1}_2$ be in $H$ for any $k_1$ and $k_2$ in $K$, implying that $H$ contains the commutator subgroup. It is straightforward to generalize this to $n>3$, where the matrices act as the identity operator on the additional direction. However, this argument fails if $n<3$. Indeed, as we shall see in the next section, for $n=2$ there are quaternionic reflection groups associated with $K$ and $H$ for which $K/H$ is non-abelian.  

Next we want to discuss the possible groups in this family. First we note that if $K=Z_k$ the resulting groups are complex reflection groups. If we then take $H=Z_r$, for $k=rp$, then we have that $G_n (Z_{rp},Z_r)$ becomes the family of complex reflection groups usually denoted as $G(k,p,n)$. Here, since the cyclic groups are abelian, any subgroup is normal and gives an abelian quotient. Note that for $k=1,2$ these reduce to real reflection groups, with $G_n (\mathbb{I},\mathbb{I})$ being the Weyl group of $SU(n+1)$, $G_n (\mathbb{Z}_2,\mathbb{Z}_2)$ being the Weyl group of $USp(2n)/SO(2n+1)$, and $G_n (\mathbb{Z}_2,\mathbb{I})$ being the Weyl group of $SO(2n)$. As such we see that this family of quaternionic reflection groups can be thought of as a generalization of the Weyl groups of classical groups.

We are then left to consider the purely quaternionic cases. For this we need to consider the possible normal subgroups of the binary dihedral and polyhedral groups. We have listed the necessary data in table \ref{subgroups}, and we shall next discuss each in turn. Note, that the largest abelian quotient one can get for each case is identical to the center of the corresponding ADE group.  
 
\renewcommand{\arraystretch}{1.5}

\begin{table}[!h]
\centering
\begin{tabular}{|c|c|c|}
\hline
 Group & Normal subgroup & Quotient group \\
\hline 
 \multirow{6}{*}{$\hat{D}_k$} & $\hat{D}_k$ & $\mathbb{I}$ \\
\cline{2-3}
  & $\hat{D}_{\frac{k}{2}}$ ($k$ even) & $\mathbb{Z}_{2}$ \\
\cline{2-3}
  & $\mathbb{Z}_{2k}$ & $\mathbb{Z}_{2}$ \\
\cline{2-3}
  & \multirow{2}{*}{$\mathbb{Z}_{\frac{k}{l}}$ ($l$ divisor of $k$)} & $D_{2l}$ ($\frac{k}{l}$ even) \\
	\cline{3-3}
  &  & $\hat{D}_{l}$ ($\frac{k}{l}$ odd) \\
\cline{2-3}
  & $\mathbb{I}$ & $\hat{D}_k$ \\
\hline
\hline
 \multirow{4}{*}{$\hat{T}$} & $\hat{T}$ & $\mathbb{I}$ \\
\cline{2-3}
 & $\hat{D}_2$ & $\mathbb{Z}_3$ \\
\cline{2-3}
 & $\mathbb{Z}_2$ & $A_4$ \\
\cline{2-3}
 & $\mathbb{I}$ & $\hat{T}$ \\
\hline
\hline
\multirow{5}{*}{$\hat{O}$} & $\hat{O}$ & $\mathbb{I}$ \\
\cline{2-3}
 & $\hat{T}$ & $\mathbb{Z}_2$ \\
\cline{2-3}
 & $\hat{D}_2$ & $S_3$ \\
\cline{2-3}
 & $\mathbb{Z}_2$ & $S_4$ \\
\cline{2-3}
 & $\mathbb{I}$ & $\hat{O}$ \\
\hline
\hline
\multirow{3}{*}{$\hat{I}$} & $\hat{I}$ & $\mathbb{I}$ \\
\cline{2-3}
 & $\mathbb{Z}_2$ & $A_4$ \\
\cline{2-3}
 & $\mathbb{I}$ & $\hat{I}$ \\
\hline
\end{tabular}
\caption{Summary of the normal subgroups of discrete subgroups of $SU(2)$.}
\label{subgroups}
\end{table}

We begin by considering the binary dihedral group, $\hat{D}_k$. This group is defined using two generators $a$ and $x$ obeying the following relations: $x^2 = a^k$, $a^{2k} = 1$, $x^{-1} a x = a^{-1}$. Given that $x^{-1} a^{-1} x a = a^2$, we see that the commutator subgroup here is $\mathbb{Z}_k$ spanned by $a^2$. The quotient group is $\mathbb{Z}_2 \times \mathbb{Z}_2$ for $k$ even, and $\mathbb{Z}_4$ for $k$ odd. The other normal subgroups containing this group are the full $\hat{D}_k$ group, $\mathbb{Z}_{2k}$ and $\hat{D}_{\frac{k}{2}}$ for $k$ even. 

We can describe the associated quaternionic reflection groups as follows. First we use a quaternionic presentation of $\hat{D}_k$ as: $x=j$, $a=e^{\frac{\pi i}{k}}$. We can then write a generic element of $\hat{D}_k$ as $e^{\frac{\theta\pi i}{k}}j^{\rho}$, for $\theta=0,1,2,...,2k-1$ and $\rho=0,1$. The group $G_n (\hat{D}_k,H)$ is then generated by permutations of the $n$ coordinates of $\mathbb{H}^n$ together with all transformations of the form $diag(e^{\frac{\theta_1\pi i}{k}}j^{\rho_1}, e^{\frac{\theta_2\pi i}{k}}j^{\rho_2}, ... , e^{\frac{\theta_n\pi i}{k}}j^{\rho_n})$ such that the determinant is in $H$. We then have that:

\begin{itemize}
\item 
$H=\hat{D}_k$: all values of $\theta_i$, $\rho_i$ are admissible.
\item
$H=\mathbb{Z}_{2k}$: all values of $\theta_i$ are admissible, but the values of $\rho_i$ are restricted to obey $\sum^n_{i=1} \rho_i = $ even number.
\item
$H=\hat{D}_{\frac{k}{2}}$: all values of $\rho_i$ are admissible, but the values of $\theta_i$ are restricted to obey $\sum^n_{i=1} \theta_i = $ even number.
\item
$H=\mathbb{Z}_{k}$: The values of both $\theta_i$ and $\rho_i$ are restricted. First these must obey that $\sum^n_{i=1} \rho_i = $ even number. For $k$ even, we also need to enforce that $\sum^n_{i=1} \theta_i = $ even number. For $k$ odd, however, we instead need to enforce that $\sum^n_{i=1} \theta_i = $ even number if $\sum^n_{i=1} \rho_i = 0$ mod 4, while $\sum^n_{i=1} \theta_i = $ odd number if $\sum^n_{i=1} \rho_i = 2$ mod 4.
\end{itemize}

We will not deal much with the binary polyhedral groups here, but we shall briefly comment on the options in this case for completeness. For $\hat{T}$, there are two possible groups, corresponding to the cases of $H=\hat{T}$, where there is no restriction, and $H=\hat{D}_2$. A similar story exist for $\hat{O}$, where we again have two options: the unrestricted case $H=\hat{O}$, and $H=\hat{T}$. Finally, for $\hat{I}$, we have only one case, $H=\hat{I}$. This corresponds to the fact that $E_8$ has trivial center.   

\subsubsection{The case of $n=2$ and $G (K,H,\alpha)$}

As we mentioned, in the $n=2$ case we can also support normal subgroups which do not contain the commutator subgroup. In fact there can even be more than one group corresponding to each choice of $K$ and $H$. This extra choice is given by an automorphism of the quotient group $\frac{K}{H}$ denoted by $\alpha$. The resulting group, refereed to as $G (K,H,\alpha)$ acts on $\mathbb{H}^2$, spanned by the coordinates $z_1, z_2$. It is generated by the permutation of the two coordinate and the transformations $(z_1, z_2) \rightarrow (k z_1, \alpha(k) h z_2)$, where $k$ is an element of $K$, $h$ an element of $H$, and $\alpha$ is an automorphism obeying some restrictions. The reflections in the group are given by:

\be
\begin{pmatrix}
  h & 0 \\
  0 & 1
	\end{pmatrix}, \begin{pmatrix}
  1 & 0 \\
  0 & h
	\end{pmatrix}, \begin{pmatrix}
  0 & l \\
  l^{-1} & 0
	\end{pmatrix} ,
\ee
where $l$ are all elements of $K$ such that $\alpha(l) = l^{-1} h'$, that is they are mapped by the automorphism $\alpha$ to their inverses up to elements of $H$. We note that for this to be a quaternionic reflection group all elements of type $l$ must generate the group $K$. This in turn gives a restriction on $\alpha$.  

We begin by considering the case where $K$ contains the commutator subgroup. In this case there is a unique choice of $\alpha$ and the group just reduces to the previous case. This follows as:

\be
\begin{pmatrix}
  l_1 & 0 \\
  0 & l^{-1}_1
	\end{pmatrix} \begin{pmatrix}
  l_2 & 0 \\
  0 & l^{-1}_2
	\end{pmatrix} = \begin{pmatrix}
  l_1 l_2 & 0 \\
  0 & l^{-1}_1 l^{-1}_2
	\end{pmatrix} = \begin{pmatrix}
  l_1 l_2 & 0 \\
  0 & l^{-1}_2 l^{-1}_1 l_1 l_2 l^{-1}_1 l^{-1}_2
	\end{pmatrix} .
\ee    
Here the two matrices we are multiplying are part of the group and obey that their determinant is the identity. However, the determinant of their product may not be the identity\footnote{Note that as the elements are quaternions, they do not commute, and so the determinant of a product of matrices is not necessarily equal to the product of determinants.}. Nevertheless, the determinant is always in the commutator subgroup, and so if it is contained in $H$, then the determinant will also be in $H$. As such these cases are just $G_2 (K,H)$, and we can ignore the automorphism $\alpha$. 

\begin{table}[!h]
\centering
\resizebox{\textwidth}{!}{
\begin{tabular}{|c|c|c|c|c|}
\hline
 Group & $\alpha$ & Order of & Number of & Number of \\
 & & group & reflections & possible $\alpha$ \\
\hline 
 & $\alpha_r$, for $r$ odd, & & $\frac{2m}{l} gcd(2l,r+1)$ & Number of integers \\
$G(\hat{D}_{2m},\mathbb{Z}_{\frac{2m}{l}},\alpha)$ & $0<r\leq l$, $l=$ & $\frac{32m^2}{l}$ & $+$ & $q$, $p$ for $l = q p$, \\
 & $gcd(l,\frac{r+1}{2})gcd(l,\frac{r-1}{2})$ & & $\frac{2m}{l} gcd(2l,r-1)$ & $gcd(q,p)=1$ \\
\hline
 & $\beta_r$, for $r$ odd, & & $\frac{2m+1}{l} gcd(2l,r+1)$ & Number of integers \\
$G(\hat{D}_{2m+1},\mathbb{Z}_{\frac{2m+1}{l}},\alpha)$ & $0<r\leq l$, $l=$ & $\frac{8(2m+1)^2}{l}$ & $+$ & $q$, $p$ for $l = q p$, \\
 & $gcd(l,\frac{r+1}{2})gcd(l,\frac{r-1}{2})$ & & $\frac{2m+1}{l} gcd(2l,r-1)$ & $gcd(q,p)=1$ \\
\hline
 & $\alpha_r$, for $0<r\leq m$, & & $2gcd(k,r+1)$ & Number of integers \\
$G(\hat{D}_{k=2m+1},\mathbb{Z}_2,\alpha)$ & $k=gcd(k,r+1)$ & $16k$ & $+$ & $q$, $p$ for $k = q p$, \\
 & $\cdot gcd(k,r-1)$ & & $2gcd(k,r-1)$ & $gcd(q,p)=1$ \\
\hline
 & $\beta_r$, for $r$ odd, &  & $gcd(2k,r+1)$ & Number of integers \\
$G(\hat{D}_k,\mathbb{I},\alpha)$ & $0<r\leq k$, $k=$ & $8k$ & $+$ & $q$, $p$ for $k = q p$,    \\
& $gcd(k,\frac{r+1}{2})gcd(k,\frac{r-1}{2})$ & & $gcd(2k,r-1)$ & $gcd(q,p)=1$ \\
\hline
\hline
$G(\hat{T},\mathbb{Z}_2,\alpha)$ & conj. by $(12)$ & $96$ & $12$ & $1$ \\
\hline
$G(\hat{T},\mathbb{I},\alpha)$ & conj. by $(i-j)$ & $48$ & $12$ & $1$ \\
\hline
\hline
$G(\hat{O},\hat{D}_{2},\alpha)$ & $1$ & $768$ & $32$ & $1$ \\
\hline
$G(\hat{O},\mathbb{Z}_2,\alpha)$ & $1$ & $192$ & $14$ & $1$ \\
\hline
$G(\hat{O},\mathbb{I},\alpha)$ & conj. by $k$ & $96$ & $18$ & $2$ \\
$G(\hat{O},\mathbb{I},\alpha)$ & outer automorp. of $\hat{O}$ & & $14$ & \\
\hline
\hline
$G(\hat{I},\mathbb{Z}_2,\alpha)$ & $1$ & $480$ & $32$ & $2$ \\
 & conj. by $(12)$ & & $20$ & \\
\hline
$G(\hat{I},\mathbb{I},\alpha)$ & conj. by $j$ & $240$ & $30$ & $2$ \\
 & order $2$ outer automorp. & & $20$ & \\
\hline
\end{tabular}}
\caption{Possible cases of $G(K,H,\alpha)$. Here $\alpha_r$ is the auomorphism of $D_k$ given by $\alpha_r (x) = x$ and $\alpha_r (y)= y^r$, where we take $D_k$ to be the group given by $x^2 = y^k = (x y)^2 = 1$. Similarly, $\beta_r$ is the automorphism of $\hat{D}_k$ given by $\beta_r (j) = -j$ and $\beta_r (e^{\frac{\pi i}{k}})= e^{\frac{r\pi i}{k}}$.}
\label{QRgroupr2}
\end{table}

The cases when $H$ does not contain the commutator subgroup, though, leads to new reflections groups. The possible choices of $\alpha$ were studied in \cite{Cohen:1980qrg}, and we summarize them in table \ref{QRgroupr2}. We next consider some examples.    

\subsubsection*{Example: $G(\hat{D}_k,\mathbb{I},\alpha)$}

As an example, we consider the case of $G(\hat{D}_k,\mathbb{I},\alpha)$. Here $H$ is trivial so all basic reflections involve a permutation. We have that $\alpha=\beta_r$, where $\beta_r (j) = -j$, $\beta_r (e^{\frac{\pi i}{k}}) = e^{\frac{r \pi i}{k}}$, for $r$ an odd integer obeying: $0<r\leq k$ and $k = gcd(k,\frac{r+1}{2})gcd(k,\frac{r-1}{2})$. We note that $r=1$ is always possible, implying that for every $k$ we have the group $G(\hat{D}_k,I,\beta_1)$ spanned by the elements:

\be
\begin{pmatrix}
  0 & 1 \\
  1 & 0
	\end{pmatrix}, \begin{pmatrix}
  j e^{\frac{m \pi i}{k}}  & 0 \\
  0 & -j e^{\frac{m \pi i}{k}}
	\end{pmatrix}, \begin{pmatrix}
  e^{\frac{m \pi i}{k}}  & 0 \\
  0 & e^{\frac{m \pi i}{k}}
	\end{pmatrix},
\ee
for $m=0,1,2,...,2k-1$, plus the product of the first element with the other two. 

For many values of $k$ this is the only option. However, there are cases for which there are additional possibilities. Specifically, the number of possible values of $r$ is equal to the number of different ways one can write $k$ as the product of two integers, $k = q p$ such that $gcd(q,p)=1$. Obviously, we can always write $k=1 \cdot k$, which corresponds to the $r=1$ case. We can prove this statement as follows. We must have that $k = gcd(k,\frac{r+1}{2})gcd(k,\frac{r-1}{2})$, implying that $k$ must have the form of a product of two integers. We then have that $r+1 = 2 q s$, $r-1 = 2 p t$, for some integers $t$ and $s$. Taking the difference of the two equations we see that $q s - p t = 1$. It follows from Bezout’s theorem that this can only be solved if $gcd(p,q)=1$. This also implies that $gcd(t,q)=1$ and $gcd(p,s)=1$, which is necessary so that $gcd(k,q s)=q$ and $gcd(k,p t)=t$. 

Taking the sum of the equations we have that $r = q s + p t$. Say we found a solution for $s$ and $t$ such that $q s - p t = 1$. We can generate a new solution by taking $s\rightarrow s + x p$ and $t\rightarrow t + x q$, for $x$ an integer. This changes $r$ by $r\rightarrow r + 2 k x$. We can use this to force $-k \leq r \leq k$. Furthermore we can consider exchanging $q$ and $p$, which affects the solution by $s \rightarrow -t$, $t \rightarrow -s$. This changes $r\rightarrow -r$, which we can now use to further force $0<r \leq k$. As a result, we see that for every decomposition of $k$ as the product of two integers with no common divisor, we have precisely one solution for $r$ obeying the restriction. Furthermore, this $r$ is unique as the two integers are given by $gcd(k,\frac{r+1}{2})$ and $gcd(k,\frac{r-1}{2})$ and so the pair are in one-to-one correspondence with $r$.

Next we want to consider an example. The first $k$ for which there is more than one option is $k=6$ as $k = 1 \cdot 6 = 2 \cdot 3$. For $k = 1 \cdot 6$ we have the solution $r=1$, while for $k= 2 \cdot 3$ we have that $r=5$. We discussed the case $r=1$ previously, where we have that the reflections are given by:     

\be \nonumber
\begin{pmatrix}
  0 & 1 \\
  1 & 0
\end{pmatrix},
\begin{pmatrix}
  0 & -1 \\
  -1 & 0
\end{pmatrix},
\begin{pmatrix}
  0 & j e^{\frac{m \pi i}{6}} \\
  -j e^{\frac{m \pi i}{6}} & 0
\end{pmatrix},
\ee
for $m=0,1,2,...,11$.

However for $r=5$, the reflections are given by:

\be \nonumber
\begin{pmatrix}
  0 & e^{\frac{m \pi i}{3}} \\
  e^{\frac{5m \pi i}{3}} & 0
\end{pmatrix},
\ee
for $m=0,1,2,...,5$, and
\be \nonumber
\begin{pmatrix}
  0 & j e^{\frac{m \pi i}{2}} \\
  -j e^{\frac{5m \pi i}{2}} & 0
\end{pmatrix},
\ee
for $m=0,1,2,3$.

It is straightforward to see that these define a reflection group as the elements appearing span the group $\hat{D}_6$. However, we note that the number of reflections in this group is $10$, while the number for $r=1$ is $14$. Thus, these are indeed different reflection groups associated with $K=\hat{D}_6$ and $H=\mathbb{I}$. 

\section{Calculations}
	In this section we explicitly discuss an example and match it with the Hilbert series computed though an explicit counting of the invariants for the respective quaternionic reflection groups.
	\label{sec:calc}
\subsection{Some details about the superconformal index}
	\label{app:SCIdetails}
	Here we collect some details about the $3d$ superconformal index and show how it reduces to equation \ref{eq:N5indexsimped} for the $\mathcal{N}=5$ theories that we consider in this paper. When we have a theory with a Lagrangian description, the superconformal index is given by 
	\begin{equation}
		\label{index3d}
		\mathcal{I}=\frac{1}{|W|}\sum_{s_j}\int\prod_j \frac{dz_j}{2\pi i z_j} e^{-S_{CS}(h,s)}e^{ib_0(h)}x^{\epsilon_0}\prod_a t_a^{q_{0a}}PE\left[\text{ind}_{\text{vec}}+\text{ind}_{\text{chir}} \right]
	\end{equation}
	where $\text{ind}_{\text{chir}}$ and $\text{ind}_{\text{vec}}$ denote the single particle index for the chiral and vector multiplets respectively. They are given as follows:
	\begin{equation}
		\text{ind}_{\text{chir}}(e^{ih_j},s_j;t_a,x)=\sum_{\Phi}\sum_{\rho\in R_\Phi}\left(e^{i\rho(h)}\prod_a t_a^{f_a(\Phi)}\frac{x^{2|\rho(s)|+\Delta_\Phi}}{1-x^2}-e^{-i\rho(h)}\prod_a t_a^{-f_a(\Phi)}\frac{x^{2|\rho(s)|+2-\Delta_\Phi}}{1-x^2}\right)
	\end{equation}	
	\begin{equation}
		\label{eq:indvec}
		\text{ind}_{\text{vec}}(e^{ih_j},s_j;t_a,x)=-\sum_{\alpha\in ad(G)}e^{i\alpha(h)}x^{2|\alpha(s)|}
	\end{equation}
	Furthermore
	\begin{equation}
		\begin{split}    \epsilon_0&=\sum_\Phi(1-\Delta_{\Phi})\sum_{\rho\in R_\Phi}|\rho(s)|-\sum_{\alpha\in ad(G)}|\alpha(s)|\\
			q_{0a}&=-\sum_\Phi\sum_{\rho\in\Phi}|\rho(s)|f_a(\Phi)\\
			b_0(h)&=-\sum_\Phi\sum_{\rho\in R_\Phi}|\rho(s)|\rho(h)
		\end{split}
	\end{equation}
	and $S_{CS}$ denotes the contribution from the Chern-Simons term:
	\begin{equation}
		S_{CS}(h,s)=2k\text{Tr}_{CS}(h\cdot s) .
	\end{equation}
	The notation used in the above expressions is as follows.
		\paragraph{Notation}
	\begin{itemize}
		\item $G$: gauge algebra.
		\item $\alpha$: roots of $G$.
		\item $\rho$:  weights.
		\item $h$ takes values in the maximal torus of the group.  Parametrizes the $S^1$ Wilson lines of the gauge field.
		\item $s$ takes values in the Cartan of the gauge group. Parametrize the GNO charge of the monopole configuration of the gauge field. We follow the normalization in which the components of $s$ take half integer values.
		\item $\Delta_{\Phi}$: scaling dimension of the field.
		\item $t_a$ parametrize the maximal torus of the global symmetry group.
		\item $f_a(\Phi)$: charge of $\Phi$ under the $U(1)$ subgroup of the corresponding $t_a$.
		\item $|W|$ : order of Weyl group of $G$.
	\end{itemize}	
	Now, let us specialize to the class of theories considered in this paper. These are made up of bifundamental chiral multiplets of the gauge group $G_1\times G_2$ where $G_1$ and $G_2$ are simple Lie groups (except the case with $SU(2)\times SU(2)\times SU(2)$ gauge group). The chirals are in the bifundamental of the gauge group and are either pseudoreal or in the direct sum of a representation and its complex conjugate. Since these theories have $\mathcal{N}=5$ SUSY, they have an $SU(2)$ flavour symmetry from the point of view of $\mathcal{N}=2$ SUSY. We use $c$ to denote the fugacity of this $SU(2)$ flavour symmetry. For these cases, the single particle index for the chiral multiplet is given by
	\begin{equation}
		\label{eq:indchirsimped}
		\text{ind}_{\text{chir}}=\left(c+\frac{1}{c}\right)\left(\frac{x^{1/2}-x^{3/2}}{1-x^2}\right)\left(\sum_{\rho\in R_\Phi} e^{i\rho(h)}x^{2|\rho(s)|}\right) .
	\end{equation}	
	Also, for these theories
	\begin{equation}
		e^{ib_0(h)}=1,\;\;\;\;\prod_{a} t^{{q_0}_a}=1 .
	\end{equation}
	Here, $e^{ib_0(h)}=1$ because for every positive value of $\rho(h)$ there exists a negative value for $\rho(h)$. Since we only refine by an $SU(2)$ flavour symmetry and the hypermultiplets are in the doublet of this $SU(2)$, it implies $\prod_{a} t^{{q_0}_a}=1$.
	Putting all these together we obtain a simpler expression for the Index 
	\begin{equation}
		\label{eq:N5indexsimpedapp}
		\mathcal{I}=\frac{1}{|W|}\sum_{s}\int\prod_j \frac{dz_j}{2\pi i z_j}x^{\epsilon_0} e^{-S_{CS}(h,s)}PE\left[\text{ind}_{\text{vec}}+\text{ind}_{\text{chir}} \right] ,
	\end{equation}
	where $\text{ind}_{\text{vec}}$ and $\text{ind}_{\text{chir}}$ are given in equations \ref{eq:indvec} and \ref{eq:indchirsimped} respectively.

			\subsection{\texorpdfstring{$SO(4)\times USp(4)$}{} and its global variants}
	\label{sec:exdet}

	We will consider the different global variants of the $SO(4)_{-2k}\times USp(4)_k$ theory. We display the results of the index calculation upto $\tau^{10}$ for $k=2,3,4$ for the cases  $SO(4)_{-2k}\times USp(4)_k$, $O(4)_{-2k}\times USp(4)_k$ and $\left[SO(4)_{-2k}\times USp(4)_k\right]/\mathbb{Z}_2$. For the case $\left[SO(4)_{-2k}\times USp(4)_k\right]/\mathbb{Z}_2$,	we compute it only for $k=2,4$ since this variant exists only for even $k$. For computing the variants with orthogonal groups we follow \cite{Hwang_2011,Hwang_2011C} (see also \cite{Aharony_2013}). 
	
	\paragraph{\underline{$SO(4)_{-2k}\times USp(4)_k$}}
	For $SO(4)_{-2k}\times USp(4)_k$ theory, the contributions of monopoles are given in table \ref{tab:so4usp4halfint}. The result of adding the monopoles listed in the table, gives the following 
	\begin{equation}
		\begin{split}
			\mathcal{I}_{SO(4)_{-2k}\times USp(4)_k} &=\frac{1}{\left(-1+\tau ^4\right)^2}+\frac{\tau ^{2k}}{\left(-1+\tau ^2\right)^2 \left(1-\tau ^{2k}\right)}+\frac{2 \tau
				^{4k}}{\left(-1+\tau ^2\right)^2 \left(1+\tau ^2\right) \left(1-\tau ^{4k}\right)} \\ &+\frac{2 \tau ^{6k}}{\left(-1+\tau
				^2\right)^2 \left(1-\tau ^{6k}\right)}+...
		\end{split}
	\end{equation}
	The series expansion of the above series for $k=2,3,4$ is given in table \ref{tab:so4usp4ktab}. 
	This matches the expected Hilbert series of $\mathbb{H}^2/G_2(\hat{D}_k,\mathbb{Z}_{2k})$ for $k=2,3,4$.

	\begin{table}[htpb]
		\centering
		\begin{adjustbox}{center}
			\begin{tabular}{|c|c|c|}
				\hline
				Contribution & Monopole & Total Contribution  \\
				\hline
				$\frac{1}{\left(1-\tau ^4\right)^2}$   & $(0,0,0,0)$ &$\frac{1}{\left(1-\tau ^4\right)^2}$\\ $\frac{\tau ^{2k}}{16\left(-1+\tau ^2\right)^2 \left(1-\tau ^{2k}\right)}$ 
				& $(\frac{1}{2},\frac{1}{2},0,1)$ & $\frac{\tau ^{2k}}{\left(-1+\tau ^2\right)^2 \left(1-\tau ^{2k}\right)}$\\
				$\frac{\tau ^{4k}}{8\left(-1+\tau ^2\right)^2 \left(1+\tau ^2\right) \left(1-\tau ^{4k}\right)}$&$(0,1,1,1)$ & $\frac{2\tau ^{4k}}{\left(-1+\tau ^2\right)^2 \left(1+\tau ^2\right) \left(1-\tau ^{4k}\right)}$\\
				$\frac{ \tau ^{8k}}{32\left(-1+\tau ^2\right)^2 \left(1-\tau ^{8k}\right)}$ & $(1,2,1,3)$ & $\frac{ 2\tau ^{8k}}{\left(-1+\tau ^2\right)^2 \left(1-\tau ^{8k}\right)}$
				\\
				$\frac{ \tau ^{6k}}{32\left(-1+\tau ^2\right)^2 \left(1-\tau ^{6k}\right)}$ & $(\frac{1}{2},\frac{3}{2},1,2)$ & $\frac{ 2\tau ^{6k}}{\left(-1+\tau ^2\right)^2 \left(1-\tau ^{6k}\right)}$
				\\
				$\frac{ \tau ^{12k}}{32\left(-1+\tau ^2\right)^2 \left(1-\tau ^{12k}\right)}$ & $(2,3,1,5)$ & $\frac{ 2\tau ^{12k}}{\left(-1+\tau ^2\right)^2 \left(1-\tau ^{12k}\right)}$\\
				\hline
			\end{tabular}
		\end{adjustbox}
		\caption{$SO(4)_{-2k}\times USp(4)_k$, the contributions are given for all positive integer multiples of the monopole. Once all possible signs and multiples are taken into account we get the total contribution. This table shows the half integer contributions}
		\label{tab:so4usp4halfint}
	\end{table}
	
	\begin{table}[htpb]
		\centering
		\begin{adjustbox}{center}
			\begin{tabular}{|c|c|c|c|}
				\hline
				$k$ & Hilbert Series & Table \\
				\hline
				$2$ & $1+3 \tau ^4+2 \tau ^6+9 \tau ^8+8 \tau ^{10}+...$ & \ref{tab:invsDkZ2k}\\
				$3$ &$1+2 \tau ^4+\tau ^6+5 \tau ^8+3 \tau ^{10}+...$ & \ref{tab:invsDkZ2kk3}\\
				$4$ & $1+2 \tau ^4+4 \tau ^8+2 \tau ^{10}+...$  &\ref{tab:invsDkZ2kk4}\\
				\hline
			\end{tabular}
		\end{adjustbox}
		\caption{Table for $SO(4)_{-2k}\times USp(4)_k$. The moduli space equals $\mathbb{H}^2/G_2(\hat{D}_k,\mathbb{Z}_{2k})$. The last column shows the table number which contains the list of invariants corresponding to the Hilbert Series.}
		\label{tab:so4usp4ktab}
	\end{table}

	\paragraph{\underline{$O(4)_{-2k}\times USp(4)_k$}}
	For computing the case of $O(4)_{-2k}\times USp(4)_k$, we need to add the $\mathbb{Z}_2$ projected part which we denote by $\mathcal{I}^{\mathbb{Z}_2}_{SO(4)_{-2k}\times USp(4)_k}$. The $O(4)_{-2k}\times USp(4)_k$ index is given by
	\begin{equation}
		\mathcal{I}_{O(4)_{-2k}\times USp(4)_k}= \frac{\mathcal{I}_{SO(4)_{-2k}\times USp(4)_k}+\mathcal{I}^{\mathbb{Z}_2}_{SO(4)_{-2k}\times USp(4)_k}}{2}
	\end{equation}
	where
	\begin{equation}
		\label{eq:projecto4usp4}
		\mathcal{I}^{\mathbb{Z}_2}_{O(4)_{-2k}\times USp(4)_k}= \frac{1}{1-\tau^8}+
		\frac{\tau ^{2k}}{\left(1-\tau ^4\right) \left(1-\tau ^{2k}\right)}+...
	\end{equation}
	The first term in equation \ref{eq:projecto4usp4} denotes the contribution from the perturbative part and the second term from the monopole $\left(\frac{1}{2},\frac{1}{2},0,1\right)$.
	This matches with the Hilbert Series of $\mathbb{H}^2/G_2(\hat{D}_k,\hat{D}_k)$ and the series expansions for $k=2,3,4$ are given in table \ref{tab:o4usp4ktab}.
	\begin{table}[htpb]
		\centering
		\begin{adjustbox}{center}
			\begin{tabular}{|c|c|c|c|}
				\hline
				$k$ & Hilbert Series & Table  \\
				\hline
				$2$ & $1+2 \tau ^4+\tau ^6+6 \tau ^8+4 \tau ^{10}+...$ & \ref{tab:invsDkDk}\\
				$3$ &$1+\tau ^4+\tau ^6+3 \tau ^8+2 \tau ^{10}+...$& \ref{tab:invsDkDk3}\\
				$4$ & $1+\tau ^4+3 \tau ^8+\tau ^{10}+...$ & \ref{tab:invsDkDk4}\\
				\hline
			\end{tabular}
		\end{adjustbox}
		\caption{Table for $O(4)_{-2k}\times USp(4)_k$. The moduli space equals $\mathbb{H}^2/G_2(\hat{D}_k,\hat{D}_{k})$. The last column shows the table number which contains the list of invariants corresponding to the Hilbert Series.}
		\label{tab:o4usp4ktab}
	\end{table}
	\paragraph{\underline{$\left(SO(4)_{-2k}\times USp(4)_k\right)/\mathbb{Z}_2$}}
	Quotienting $SO(4)_{-2k}\times USp(4)_k$ by $\mathbb{Z}_2$ changes the allowed monopoles. Before the $\mathbb{Z}_2$ quotient, the allowed monopoles were given in table \ref{tab:so4usp4ktab}. After the $\mathbb{Z}_2$ quotient, the monopoles $(0,1,1,1)$, $(1,2,1,3)$ and $(2,3,1,5)$ can be made half integer (ie $(0,\frac{1}{2},\frac{1}{2},\frac{1}{2})$, $(\frac{1}{2},1,\frac{1}{2},\frac{3}{2})$ and $(1,\frac{3}{2},\frac{1}{2},\frac{5}{2})$ ) and will still obey the Dirac quantization condition. The Hilbert series can be written as follows
	
	\begin{equation}
		\begin{split}
			\mathcal{I}_{\frac{SO(4)_{-2k}\times USp(4)_k}{\mathbb{Z}_2}}&= \frac{1}{\left(-1+\tau ^4\right)^2}+\frac{\tau ^{2k}}{\left(-1+\tau ^2\right)^2 \left(1-\tau ^{2k}\right)}+\frac{2 \tau
				^{2k}}{\left(-1+\tau ^2\right)^2 \left(1+\tau ^2\right) \left(1-\tau ^{2k}\right)}\\&+\frac{2 \tau ^{6k}}{\left(-1+\tau
				^2\right)^2 \left(1-\tau ^{6k}\right)}+
			\frac{2 \tau ^{4k}}{\left(-1+\tau
				^2\right)^2 \left(1-\tau ^{4k}\right)}+\frac{2 \tau ^{6k}}{\left(-1+\tau
				^2\right)^2 \left(1-\tau ^{6k}\right)}
			+...
		\end{split}
	\end{equation}
	
	The series expansion for $k=2,3,4$ are given in table \ref{tab:so4usp4modz2kseries}. This matches the Hilbert series of $\mathbb{H}^2/G_2(\hat{D}_k,\mathbb{Z}_k)$ for respective values of $k$.
	
	\begin{table}[htpb]
		\centering
		\begin{adjustbox}{center}
			\begin{tabular}{|c|c|c|c|}
				\hline
				$k$ & Hilbert Series & Table \\
				\hline
				$2$ & $1+5 \tau ^4+4 \tau ^6+15 \tau ^8+16 \tau ^{10}+...$ &\ref{tab:invsDkZk}\\
				$3$ &$1+2 \tau ^4+3 \tau ^6+7 \tau ^8+7 \tau ^{10}+...$ &\ref{tab:invsDkZk3}\\
				$4$ & $1+2 \tau ^4+6 \tau ^8+4 \tau ^{10}+...$ &\ref{tab:invsDkZk4}\\
				\hline
			\end{tabular}
		\end{adjustbox}
		\caption{Table for $\left[SO(4)\times USp(4)\right]/\mathbb{Z}_2$. The moduli space equals $\mathbb{H}^2/G_2(\hat{D}_k,\mathbb{Z}_{k})$. The last column shows the table number which contains the list of invariants corresponding to the Hilbert Series.}
		\label{tab:so4usp4modz2kseries}
	\end{table}
	
	\paragraph{\underline{$\left(O(4)_{-2k}\times USp(4)_k\right)/\mathbb{Z}_2$}}
	Quotienting $O(4)_{-2k}\times USp(4)_k$ by $\mathbb{Z}_2$ works similarly to the $\left(SO(4)_{-2k}\times USp(4)_k\right)/\mathbb{Z}_2$ case. We denote the $\mathbb{Z}_2$ projected part by 	$\mathcal{I}^{\mathbb{Z}_2}_{SO(4)_{-2k}\times USp(4)_k}$. The Hilbert series is given as
	\begin{equation}
		\mathcal{I}_{\frac{SO(4)_{-2k}\times USp(4)_k}{\mathbb{Z}_2}}= \frac{\mathcal{I}_{\frac{SO(4)_{-2k}\times USp(4)_k}{\mathbb{Z}_2}}+\mathcal{I}^{\mathbb{Z}_2}_{\frac{SO(4)_{-2k}\times USp(4)_k}{\mathbb{Z}_2}}}{2}
	\end{equation}
	where
	\begin{equation}
		\label{eq:projecto4usp4modz2}
		\mathcal{I}^{\mathbb{Z}_2}_{O(4)_{-2k}\times USp(4)_k}= \frac{1}{1-\tau^8}+
		\frac{\tau ^{2k}}{\left(1-\tau ^4\right) \left(1-\tau ^{2k}\right)}+...
	\end{equation}
	The first and second terms in equation \ref{eq:projecto4usp4modz2} are the same as for equation \ref{eq:projecto4usp4} because the $(0,0,0,0)$ and $\left(\frac{1}{2},\frac{1}{2},0,1\right)$ can't be made further half integer and contribute the same. The series for $k=2$ and $4$ is given table \ref{tab:o4usp4modz2kseries}. These match the Hilber series for  $\mathbb{H}^2/G_2(\hat{D}_k,\hat{D}_{\frac{k}{2}})$.
	\begin{table}[htpb]
		\centering
		\begin{adjustbox}{center}
			\begin{tabular}{|c|c|c|c|}
				\hline
				$k$ & Hilbert Series & Table \\
				\hline
				$2$ & $1+3 \tau ^4+2\tau ^6+9 \tau ^8+8 \tau ^{10}+...$ & \ref{tab:invsDkDkb2}\\
				$4$ & $1+\tau ^4+4 \tau ^8+2 \tau ^{10}+...$& \ref{tab:invsDkDkb2k4} \\
				\hline
			\end{tabular}
		\end{adjustbox}
		\caption{Table for $\left[O(4)\times USp(4)\right]/\mathbb{Z}_2$. The moduli space equals $\mathbb{H}^2/G_2(\hat{D}_k,\hat{D}_{\frac{k}{2}})$. The last column shows the table number which contains the list of invariants corresponding to the Hilbert Series.}
		\label{tab:o4usp4modz2kseries}
	\end{table}

	Note that the numerators of the contribution from monopoles go like $\tau^{nk}$. If $\tau^{Nk}$ is the numerator of the highest monopole upto which we compute the series, we could only be certain of the series upto $\tau^{Nk}$. Therefore, the higher the value of $k$ is, the more accurate the series gets to higher order. If $k$ is taken to be large, the Hilbert series gets contribution from the perturbative sector until the smallest monopoles starts contributing. Therefore the check of lower orders is not a good one since essentially one is only comparing the perturbative sector. The issue with comparing the superconformal index result to the Hilbert series obtained by computing the invariants is that as we increase $k$, it gets harder to compute higher order terms by computing invariants using the quaternioinc reflection group. Choosing $k=2,3,4$ and evaluating series upto $\tau^{10}$, captures both the effect of perturbative and monopole sectors.
	\subsection{Hilbert series of quaternionic reflection groups}
	\label{app:HSQRG}
	
	In this section we collect the invariants for some of the quaternionic reflection groups. We briefly discuss the matrix presentation for $\hat{D}_k$. This construction is easy to generalize to the case when we have more than one quaternion. We parametrize a quaternionic variable $Q$ by two complex variables $Z_1$ and $Z_2$.
	\begin{equation}
		Q=Z_1+j Z_2
	\end{equation}
	The group $\hat{D}_k$ has the generators $j$ and $e^{\frac{\theta\pi i}{k}}$. The action of $j$ is 
	\begin{equation}
		j Q=-Z_2+jZ_1
	\end{equation}
	and the action of $e^{\frac{\theta\pi i}{k}}$ as
	\begin{equation}
		e^{\frac{\theta\pi i}{k}} Q= e^{\frac{\theta\pi i}{k}} Z_1+j e^{-\frac{\theta\pi i}{k}}Z_2 
	\end{equation}
	All these actions can be represented in matrix form as
	\begin{equation}
		Q=\begin{pmatrix}
			Z_1 \\ Z_2
		\end{pmatrix}
	\end{equation}
	\begin{equation}
		j=\begin{pmatrix}
			0 && -1\\
			1 && 0
		\end{pmatrix}\;\;\;e^{\frac{\theta\pi i}{k}}=\begin{pmatrix}
			e^{\frac{\theta\pi i}{k}} && 0\\
			0 && e^{-\frac{\theta\pi i}{k}}
		\end{pmatrix}
	\end{equation}
	The invariats under this action are
	\begin{equation}
	\label{eq:xyzdhat}
		X=Z_1^2Z_2^2,\;\;\;Y=Z_1^{2k}+Z_2^{2k},\;\;\; Z=Z_1Z_2(Z_1^{2k}-Z_2^{2k})
	\end{equation}
	These obey the relation
	\begin{equation}
		XY^2-4X^{k+1}=Z^2
	\end{equation}
	This is the relation for the $\hat{D}_k$ simple surface singularity. It is straightforward to write the Hilbert series for this. 
	\begin{equation}
		\frac{(1-\tau^{4k+4})}{(1-\tau^4)(1-\tau^{2k})(1-\tau^{2k+2})}
	\end{equation}
	For other cases, obtaining a closed form for the Hilbert series is not straightforward. We compute the invariants order by order in $\tau$ and match it with the Index calculation. 
	
	At a given order $\tau^n$, first we need to write down all possible words that can be formed by the letters at that order. Then we act with the generators of the quaternionic reflection group on these words and solve for words that are invariant under that action. Let us do this exercise for $\mathbb{H}/\hat{D}_2$ upto $\tau^8$.	For $\tau^2$, $\tau^4$, $\tau^6$ and $\tau^8$, the possible letters are
	\begin{equation}
		\begin{split}
			\tau^2:\; & Z_1^2,\;Z_1 Z_2,Z_2^2 \\
			\tau^4:\; & Z_1^4,\;Z_1^3 Z_2,\;Z_1^2 Z_2^2,\;Z_1 Z_2^3,\;Z_2^4 \\
			\tau^6:\; & Z_1^6,\;Z_1^5 Z_2,\;Z_1^4 Z_2^2,\;Z_1^3 Z_2^3,\;Z_1^2 Z_2^4,\;Z_1 Z_2^5,Z_2^6\\
			\tau^8:\; & Z_1^8,\;Z_1^7 Z_2,\;Z_1^6 Z_2^2,\;Z_1^5 Z_2^3,\;Z_1^4 Z_2^4,\;Z_1^3 Z_2^5,\;Z_1^2 Z_2^6,\;Z_1 Z_2^7\;,Z_2^8
		\end{split}
	\end{equation}
	Acting with the generators described above, one finds the following invariants.
	\begin{equation}
		\begin{split}
		\tau^2:\; & 0 \\
		\tau^4:\; & Z_1^2 Z_2^2,\;Z_1^4+Z_2^4 \\
		\tau^6:\; & Z_1^5 Z_2-Z_1 Z_2^5\\
		\tau^8:\; & Z_1^4 Z_2^4,\; Z_1^6 Z_2^2+Z_1^2 Z_2^6,\; Z_1^8+Z_2^8
	    \end{split}
	\end{equation}
	which can be written in terms of $X$, $Y$ and $Z$ defined in \ref{eq:xyzdhat} as follows
	\begin{equation}
	\begin{split}
		\tau^2:\;& 0 \\
		\tau^4:\; &X,\; Y \\
		\tau^6:\; &Z \\
		\tau^8:\; &X^2,\; XY,\; Y^2
	\end{split}
	\end{equation} 
	Note that we have written $Y^2$ instead of $Z_1^8+Z_2^8$. The reason being $Z_1^8+Z_2^8= Y^2-2X^2$ and therefore $X^2,XY$ and $Y^2$ are also linearly independent.
	The number of these invariants equals the coefficient of $\tau^2$, $\tau^4$, $\tau^6$ and $\tau^8$ in the Hilbert Series
	\begin{equation}
		\frac{1-\tau ^{12}}{\left(1-\tau ^4\right) \left(1-\tau ^4\right) \left(1-\tau ^6\right)}=1+2 \tau ^4+\tau ^6+3 \tau ^8+O\left(\tau ^9\right)
	\end{equation}
	This way of computing invariants is systematic and is easier to implement on a computer. Using the generators of various quaternionic reflection groups, their Hilbert series can be computed order by order as described above. The tables in this section collect the invariants for some of the moduli spaces of rank $2$. We label the two quaternions by $\left(Z_1,Z_2\right)$ and $\left(\tilde{Z_1},\tilde{Z_2}\right)$ where $Z_i,\tilde{Z_i}\in \mathbb{C}$. The transformations for the group $G_2(\hat{D}_k,H)$ are generated by the permutations of the coordinates of $\mathbb{H}^2$ and all the transformations of the form  $diag(e^{\frac{\theta_1\pi i}{k}}j^{\rho_1}, e^{\frac{\theta_2\pi i}{k}}j^{\rho_2})$ such that the determinant is in $H$, as described in appendix \ref{sec:QRG}. The tables display the invariants at a given order $\tau^{n}$ corresponding to different moduli spaces. The number of invariants at a given order equal the coefficient of $\tau^n$ for the corresponding Hilbert Series which are given in tables \ref{tab:so4usp4ktab},\ref{tab:o4usp4ktab},\ref{tab:so4usp4modz2kseries} and \ref{tab:o4usp4modz2kseries}.


	\begin{table}[htpb]
		\begin{adjustbox}{center}
			\centering
			\resizebox{\textwidth}{!}{
				\begin{tabular}{|c|c|c|}
					\hline
					Level & $\#$ of invariants & Invariants\\
					\hline
					$\tau^2$ & $0$ &\\
					\hline
					\rule{0pt}{3ex}
					$\tau^4$ & $3$ &$\begin{array}{c}
						Z_1 Z_2 \tilde{Z_1} \tilde{Z_2} \\
						\tilde{Z_1}{}^2 \tilde{Z_2}{}^2+Z_1^2 Z_2^2 \\
						\tilde{Z_1}{}^4+\tilde{Z_2}{}^4+Z_1^4+Z_2^4 \\
					\end{array}$\\
					\hline
					\rule{0pt}{3ex}
					$\tau^6$ & $2$ &$ \begin{array}{c}
						Z_1 Z_2 \tilde{Z_1}{}^4+Z_1^4 \tilde{Z_2} \tilde{Z_1}-Z_2^4 \tilde{Z_2} \tilde{Z_1}-Z_1 Z_2 \tilde{Z_2}{}^4 \\
						\tilde{Z_2} \tilde{Z_1}{}^5-\tilde{Z_2}{}^5 \tilde{Z_1}-Z_1 Z_2^5+Z_1^5 Z_2 \\
					\end{array}$\\
					\hline
					\rule{0pt}{3ex}
					$\tau^8$ & $9$ &$\begin{array}{c}
						Z_1^2 Z_2^2 \tilde{Z_1}{}^2 \tilde{Z_2}{}^2 \\
						Z_1 Z_2 \tilde{Z_1}{}^3 \tilde{Z_2}{}^3+Z_1^3 Z_2^3 \tilde{Z_1} \tilde{Z_2} \\
						Z_1^4 \tilde{Z_2}{}^4+Z_2^4 \tilde{Z_1}{}^4 \\
						\tilde{Z_1}{}^4 \tilde{Z_2}{}^4+Z_1^4 Z_2^4 \\
						Z_1^2 Z_2^2 \tilde{Z_1}{}^4+Z_1^4 \tilde{Z_2}{}^2 \tilde{Z_1}{}^2+Z_2^4 \tilde{Z_2}{}^2 \tilde{Z_1}{}^2+Z_1^2 Z_2^2 \tilde{Z_2}{}^4 \\
						Z_1^4 \tilde{Z_1}{}^4+Z_2^4 \tilde{Z_2}{}^4 \\
						Z_1 Z_2 \tilde{Z_2} \tilde{Z_1}{}^5+Z_1 Z_2^5 \tilde{Z_2} \tilde{Z_1}+Z_1^5 Z_2 \tilde{Z_2} \tilde{Z_1}+Z_1 Z_2 \tilde{Z_2}{}^5 \tilde{Z_1} \\
						\tilde{Z_2}{}^2 \tilde{Z_1}{}^6+\tilde{Z_2}{}^6 \tilde{Z_1}{}^2+Z_1^2 Z_2^6+Z_1^6 Z_2^2 \\
						\tilde{Z_1}{}^8+\tilde{Z_2}{}^8+Z_1^8+Z_2^8 \\
					\end{array}$\\
					\hline
					\rule{0pt}{3ex}
					$\tau^{10}$ & $8$ &$\begin{array}{c}
						Z_1^3 Z_2^3 \tilde{Z_1}{}^4+Z_1^4 \tilde{Z_2}{}^3 \tilde{Z_1}{}^3-Z_2^4 \tilde{Z_2}{}^3 \tilde{Z_1}{}^3-Z_1^3 Z_2^3 \tilde{Z_2}{}^4 \\
						Z_2^4 \tilde{Z_2} \tilde{Z_1}{}^5-Z_1 Z_2^5 \tilde{Z_1}{}^4-Z_1^4 \tilde{Z_2}{}^5 \tilde{Z_1}+Z_1^5 Z_2 \tilde{Z_2}{}^4 \\
						Z_1^2 Z_2^2 \tilde{Z_2} \tilde{Z_1}{}^5-Z_1 Z_2^5 \tilde{Z_2}{}^2 \tilde{Z_1}{}^2+Z_1^5 Z_2 \tilde{Z_2}{}^2 \tilde{Z_1}{}^2-Z_1^2 Z_2^2 \tilde{Z_2}{}^5 \tilde{Z_1} \\
						Z_1^4 \tilde{Z_2} \tilde{Z_1}{}^5+Z_1^5 Z_2 \tilde{Z_1}{}^4-Z_2^4 \tilde{Z_2}{}^5 \tilde{Z_1}-Z_1 Z_2^5 \tilde{Z_2}{}^4 \\
						Z_1 Z_2 \tilde{Z_2}{}^2 \tilde{Z_1}{}^6-Z_1 Z_2 \tilde{Z_2}{}^6 \tilde{Z_1}{}^2-Z_1^2 Z_2^6 \tilde{Z_2} \tilde{Z_1}+Z_1^6 Z_2^2 \tilde{Z_2} \tilde{Z_1} \\
						\tilde{Z_2}{}^3 \tilde{Z_1}{}^7-\tilde{Z_2}{}^7 \tilde{Z_1}{}^3-Z_1^3 Z_2^7+Z_1^7 Z_2^3 \\
						Z_1 Z_2 \tilde{Z_1}{}^8+Z_1^8 \tilde{Z_2} \tilde{Z_1}-Z_2^8 \tilde{Z_2} \tilde{Z_1}-Z_1 Z_2 \tilde{Z_2}{}^8 \\
						\tilde{Z_2} \tilde{Z_1}{}^9-\tilde{Z_2}{}^9 \tilde{Z_1}-Z_1 Z_2^9+Z_1^9 Z_2 \\
					\end{array}$\\
					\hline
				\end{tabular}
			}
		\end{adjustbox}
		\caption{Invariants for $\mathbb{H}^2/G_2(\hat{D}_k,\mathbb{Z}_{2k})$ at $k=2$}
		\label{tab:invsDkZ2k}
		
	\end{table}

		\begin{table}[htpb]
		\begin{adjustbox}{center}
			\centering
			\resizebox{\textwidth}{!}{
				\begin{tabular}{|c|c|c|}
					\hline
					Level & $\#$ of invariants & Invariants\\
					\hline
					$\tau^2$ & $0$ &\\
					\hline
					\rule{0pt}{3ex}
					$\tau^4$ & $2$ &$ \begin{array}{c}
						Z_1 Z_2 \tilde{Z_1} \tilde{Z_2} \\
						\tilde{Z_1}{}^2 \tilde{Z_2}{}^2+Z_1^2 Z_2^2 \\
					\end{array}$\\
					\hline
					$\tau^6$ & $1$ & $\begin{array}{c}
						\tilde{Z_1}{}^6+\tilde{Z_2}{}^6+Z_1^6+Z_2^6 \\
					\end{array}$\\
					\hline
					\rule{0pt}{3ex}
					$\tau^8$ & $5$ &$\begin{array}{c}
						Z_1^2 Z_2^2 \tilde{Z_1}{}^2 \tilde{Z_2}{}^2 \\
						Z_1 Z_2 \tilde{Z_1}{}^3 \tilde{Z_2}{}^3+Z_1^3 Z_2^3 \tilde{Z_1} \tilde{Z_2} \\
						\tilde{Z_1}{}^4 \tilde{Z_2}{}^4+Z_1^4 Z_2^4 \\
						Z_1 Z_2 \tilde{Z_1}{}^6+Z_1^6 \tilde{Z_2} \tilde{Z_1}-Z_2^6 \tilde{Z_2} \tilde{Z_1}-Z_1 Z_2 \tilde{Z_2}{}^6 \\
						\tilde{Z_2} \tilde{Z_1}{}^7-\tilde{Z_2}{}^7 \tilde{Z_1}-Z_1 Z_2^7+Z_1^7 Z_2 \\
					\end{array}$\\
					\hline
					\rule{0pt}{3ex}
					$\tau^{10}$ & $3$ &$\begin{array}{c}
						Z_1^2 Z_2^2 \tilde{Z_1}{}^6+Z_1^6 \tilde{Z_2}{}^2 \tilde{Z_1}{}^2+Z_2^6 \tilde{Z_2}{}^2 \tilde{Z_1}{}^2+Z_1^2 Z_2^2
						\tilde{Z_2}{}^6 \\
						Z_1 Z_2 \tilde{Z_2} \tilde{Z_1}{}^7+Z_1 Z_2^7 \tilde{Z_2} \tilde{Z_1}+Z_1^7 Z_2 \tilde{Z_2} \tilde{Z_1}+Z_1 Z_2
						\tilde{Z_2}{}^7 \tilde{Z_1} \\
						\tilde{Z_2}{}^2 \tilde{Z_1}{}^8+\tilde{Z_2}{}^8 \tilde{Z_1}{}^2+Z_1^2 Z_2^8+Z_1^8 Z_2^2 \\
					\end{array}$\\
					\hline
				\end{tabular}
			}
		\end{adjustbox}
		\caption{Invariants for $\mathbb{H}^2/G_2(\hat{D}_k,\mathbb{Z}_{2k})$ at $k=3$}
		\label{tab:invsDkZ2kk3}
		
	\end{table}

	\begin{table}[htpb]
		\begin{adjustbox}{center}
			\centering
			\resizebox{!}{!}{
				\begin{tabular}{|c|c|c|}
					\hline
					Level & $\#$ of invariants & Invariants\\
					\hline
					$\tau^2$ & $0$ &\\
					\hline
					\rule{0pt}{3ex}
					$\tau^4$ & $2$ &$ \begin{array}{c}
						Z_1 Z_2 \tilde{Z_1} \tilde{Z_2} \\
						\tilde{Z_1}{}^2 \tilde{Z_2}{}^2+Z_1^2 Z_2^2 \\
					\end{array}$\\
					\hline
					$\tau^6$ & $0$ & \\
					\hline
					\rule{0pt}{3ex}
					$\tau^8$ & $4$ &$\begin{array}{c}
						Z_1^2 Z_2^2 \tilde{Z_1}{}^2 \tilde{Z_2}{}^2 \\
						Z_1 Z_2 \tilde{Z_1}{}^3 \tilde{Z_2}{}^3+Z_1^3 Z_2^3 \tilde{Z_1} \tilde{Z_2} \\
						\tilde{Z_1}{}^4 \tilde{Z_2}{}^4+Z_1^4 Z_2^4 \\
						\tilde{Z_1}{}^8+\tilde{Z_2}{}^8+Z_1^8+Z_2^8 \\
					\end{array}$\\
					\hline
					\rule{0pt}{3ex}
					$\tau^{10}$ & $2$ &$\begin{array}{c}
						Z_1 Z_2 \tilde{Z_1}{}^8+Z_1^8 \tilde{Z_2} \tilde{Z_1}-Z_2^8 \tilde{Z_2} \tilde{Z_1}-Z_1 Z_2 \tilde{Z_2}{}^8 \\
						\tilde{Z_2} \tilde{Z_1}{}^9-\tilde{Z_2}{}^9 \tilde{Z_1}-Z_1 Z_2^9+Z_1^9 Z_2 \\
					\end{array}$\\
					\hline
				\end{tabular}
			}
		\end{adjustbox}
		\caption{Invariants for $\mathbb{H}^2/G_2(\hat{D}_k,\mathbb{Z}_{2k})$ at $k=4$}
		\label{tab:invsDkZ2kk4}
		
	\end{table}

\begin{table}[htpb]
	\begin{adjustbox}{center}
		\centering
		\resizebox{\textwidth}{!}{
			\begin{tabular}{|c|c|c|}
				\hline
				Level & $\#$ of invariants & Invariants\\
				\hline
				$\tau^2$ & $0$ &\\
				\hline
				\rule{0pt}{3ex}
				$\tau^4$ & $2$ &$\begin{array}{c}
					\tilde{Z_1}{}^2 \tilde{Z_2}{}^2+Z_1^2 Z_2^2 \\
					\tilde{Z_1}{}^4+\tilde{Z_2}{}^4+Z_1^4+Z_2^4 \\
				\end{array}$\\
				\hline
				\rule{0pt}{3ex}
				$\tau^6$ & $1$ &$ \begin{array}{c}
					\tilde{Z_2} \tilde{Z_1}{}^5-\tilde{Z_2}{}^5 \tilde{Z_1}-Z_1 Z_2^5+Z_1^5 Z_2 \\
				\end{array}$\\
				\hline
				\rule{0pt}{3ex}
				$\tau^8$ & $6$ &$\begin{array}{c}
					Z_1^2 Z_2^2 \tilde{Z_1}{}^2 \tilde{Z_2}{}^2 \\
					\tilde{Z_1}{}^4 \tilde{Z_2}{}^4+Z_1^4 Z_2^4 \\
					Z_1^2 Z_2^2 \tilde{Z_1}{}^4+Z_1^4 \tilde{Z_2}{}^2 \tilde{Z_1}{}^2+Z_2^4 \tilde{Z_2}{}^2 \tilde{Z_1}{}^2+Z_1^2 Z_2^2 \tilde{Z_2}{}^4 \\
					Z_1^4 \tilde{Z_1}{}^4+Z_1^4 \tilde{Z_2}{}^4+Z_2^4 \tilde{Z_1}{}^4+Z_2^4 \tilde{Z_2}{}^4 \\
					\tilde{Z_2}{}^2 \tilde{Z_1}{}^6+\tilde{Z_2}{}^6 \tilde{Z_1}{}^2+Z_1^2 Z_2^6+Z_1^6 Z_2^2 \\
					\tilde{Z_1}{}^8+\tilde{Z_2}{}^8+Z_1^8+Z_2^8 \\
				\end{array}$\\
				\hline
				\rule{0pt}{3ex}
				$\tau^{10}$ & $4$ &$\begin{array}{c}
					Z_1^2 Z_2^2 \tilde{Z_2} \tilde{Z_1}{}^5-Z_1 Z_2^5 \tilde{Z_2}{}^2 \tilde{Z_1}{}^2+Z_1^5 Z_2 \tilde{Z_2}{}^2
					\tilde{Z_1}{}^2-Z_1^2 Z_2^2 \tilde{Z_2}{}^5 \tilde{Z_1} \\
					Z_1^4 \tilde{Z_2} \tilde{Z_1}{}^5+Z_2^4 \tilde{Z_2} \tilde{Z_1}{}^5-Z_1 Z_2^5 \tilde{Z_1}{}^4+Z_1^5 Z_2
					\tilde{Z_1}{}^4-Z_1^4 \tilde{Z_2}{}^5 \tilde{Z_1}-Z_2^4 \tilde{Z_2}{}^5 \tilde{Z_1}-Z_1 Z_2^5
					\tilde{Z_2}{}^4+Z_1^5 Z_2 \tilde{Z_2}{}^4 \\
					\tilde{Z_2}{}^3 \tilde{Z_1}{}^7-\tilde{Z_2}{}^7 \tilde{Z_1}{}^3-Z_1^3 Z_2^7+Z_1^7 Z_2^3 \\
					\tilde{Z_2} \tilde{Z_1}{}^9-\tilde{Z_2}{}^9 \tilde{Z_1}-Z_1 Z_2^9+Z_1^9 Z_2 \\
				\end{array}$\\
				\hline
			\end{tabular}
		}
	\end{adjustbox}
	\caption{Invariants for $\mathbb{H}^2/G_2(\hat{D}_k,\hat{D}_k)$ at $k=2$}
	\label{tab:invsDkDk}
	
\end{table}

\begin{table}[htpb]
	\begin{adjustbox}{center}
		\centering
		\resizebox{!}{!}{
			\begin{tabular}{|c|c|c|}
				\hline
				Level & $\#$ of invariants & Invariants\\
				\hline
				$\tau^2$ & $0$ &\\
				\hline
				\rule{0pt}{3ex}
				$\tau^4$ & $1$ &$\begin{array}{c}
					\tilde{Z_1}{}^2 \tilde{Z_2}{}^2+Z_1^2 Z_2^2 \\
				\end{array}$\\
				\hline
				\rule{0pt}{3ex}
				$\tau^6$ & $1$ &$ \begin{array}{c}
					\tilde{Z_1}{}^6+\tilde{Z_2}{}^6+Z_1^6+Z_2^6 \\
				\end{array}$\\
				\hline
				\rule{0pt}{3ex}
				$\tau^8$ & $3$ &$\begin{array}{c}
					Z_1^2 Z_2^2 \tilde{Z_1}{}^2 \tilde{Z_2}{}^2 \\
					\tilde{Z_1}{}^4 \tilde{Z_2}{}^4+Z_1^4 Z_2^4 \\
					\tilde{Z_2} \tilde{Z_1}{}^7-\tilde{Z_2}{}^7 \tilde{Z_1}-Z_1 Z_2^7+Z_1^7 Z_2 \\
				\end{array}$\\
				\hline
				\rule{0pt}{3ex}
				$\tau^{10}$ & $2$ &$\begin{array}{c}
					Z_1^2 Z_2^2 \tilde{Z_1}{}^6+Z_1^6 \tilde{Z_2}{}^2 \tilde{Z_1}{}^2+Z_2^6 \tilde{Z_2}{}^2 \tilde{Z_1}{}^2+Z_1^2 Z_2^2
					\tilde{Z_2}{}^6 \\
					\tilde{Z_2}{}^2 \tilde{Z_1}{}^8+\tilde{Z_2}{}^8 \tilde{Z_1}{}^2+Z_1^2 Z_2^8+Z_1^8 Z_2^2 \\
				\end{array}$\\
				\hline
			\end{tabular}
		}
	\end{adjustbox}
	\caption{Invariants for $\mathbb{H}^2/G_2(\hat{D}_k,\hat{D}_k)$ at $k=3$}
	\label{tab:invsDkDk3}
	
\end{table}

\begin{table}[htpb]
	\begin{adjustbox}{center}
		\centering
		\resizebox{!}{!}{
			\begin{tabular}{|c|c|c|}
				\hline
				Level & $\#$ of invariants & Invariants\\
				\hline
				$\tau^2$ & $0$ &\\
				\hline
				\rule{0pt}{3ex}
				$\tau^4$ & $1$ &$\begin{array}{c}
					\tilde{Z_1}{}^2 \tilde{Z_2}{}^2+Z_1^2 Z_2^2 \\
				\end{array}$\\
				\hline
				\rule{0pt}{3ex}
				$\tau^6$ & $0$ &\\
				\hline
				\rule{0pt}{3ex}
				$\tau^8$ & $3$ &$\begin{array}{c}
					Z_1^2 Z_2^2 \tilde{Z_1}{}^2 \tilde{Z_2}{}^2 \\
					\tilde{Z_1}{}^4 \tilde{Z_2}{}^4+Z_1^4 Z_2^4 \\
					\tilde{Z_1}{}^8+\tilde{Z_2}{}^8+Z_1^8+Z_2^8 \\
				\end{array}$\\
				\hline
				\rule{0pt}{3ex}
				$\tau^{10}$ & $1$ &$\begin{array}{c}
					\tilde{Z_2} \tilde{Z_1}{}^9-\tilde{Z_2}{}^9 \tilde{Z_1}-Z_1 Z_2^9+Z_1^9 Z_2 \\
				\end{array}$\\
				\hline
			\end{tabular}
		}
	\end{adjustbox}
	\caption{Invariants for $\mathbb{H}^2/G_2(\hat{D}_k,\hat{D}_k)$ at $k=4$}
	\label{tab:invsDkDk4}
	
\end{table}

	\begin{table}[htpb]
		\begin{adjustbox}{center}
			\centering
			\resizebox{\textwidth}{!}{
				\begin{tabular}{|c|c|c|}
					\hline
					Level & $\#$ of invariants & Invariants\\
					\hline
					$\tau^2$ & $0$ &\\
					\hline
					\rule{0pt}{3ex}
					$\tau^4$ & $5$ &$\begin{array}{c}
						Z_1 Z_2 \tilde{Z_1} \tilde{Z_2},\;
						Z_1^2 \tilde{Z_2}{}^2+Z_2^2 \tilde{Z_1}{}^2\;
						\tilde{Z_1}{}^2 \tilde{Z_2}{}^2+Z_1^2 Z_2^2,\;
						Z_1^2 \tilde{Z_1}{}^2+Z_2^2 \tilde{Z_2}{}^2 \\
						\tilde{Z_1}{}^4+\tilde{Z_2}{}^4+Z_1^4+Z_2^4 \\
					\end{array}$\\
					\hline
					\rule{0pt}{3ex}
					$\tau^6$ & $4$ &$ \begin{array}{c}
						Z_2^2 \tilde{Z_2} \tilde{Z_1}{}^3-Z_1 Z_2^3 \tilde{Z_1}{}^2-Z_1^2 \tilde{Z_2}{}^3 \tilde{Z_1}+Z_1^3 Z_2 \tilde{Z_2}{}^2 \\
						Z_1^2 \tilde{Z_2} \tilde{Z_1}{}^3+Z_1^3 Z_2 \tilde{Z_1}{}^2-Z_2^2 \tilde{Z_2}{}^3 \tilde{Z_1}-Z_1 Z_2^3 \tilde{Z_2}{}^2 \\
						Z_1 Z_2 \tilde{Z_1}{}^4+Z_1^4 \tilde{Z_2} \tilde{Z_1}-Z_2^4 \tilde{Z_2} \tilde{Z_1}-Z_1 Z_2 \tilde{Z_2}{}^4 \\
						\tilde{Z_2} \tilde{Z_1}{}^5-\tilde{Z_2}{}^5 \tilde{Z_1}-Z_1 Z_2^5+Z_1^5 Z_2 \\
					\end{array}$\\
					\hline
					\rule{0pt}{3ex}
					$\tau^8$ & $15$ &$\begin{array}{c}
						Z_1^2 Z_2^2 \tilde{Z_1}{}^2 \tilde{Z_2}{}^2,\;
						Z_2 Z_1^3 \tilde{Z_1} \tilde{Z_2}{}^3+Z_2^3 Z_1 \tilde{Z_1}{}^3 \tilde{Z_2} \\
						Z_1 Z_2 \tilde{Z_1}{}^3 \tilde{Z_2}{}^3+Z_1^3 Z_2^3 \tilde{Z_1} \tilde{Z_2},\;
						Z_2 Z_1^3 \tilde{Z_1}{}^3 \tilde{Z_2}+Z_2^3 Z_1 \tilde{Z_1} \tilde{Z_2}{}^3 \\
						Z_1^4 \tilde{Z_2}{}^4+Z_2^4 \tilde{Z_1}{}^4, \;
						Z_2^2 \tilde{Z_2}{}^2 \tilde{Z_1}{}^4+Z_1^2 Z_2^4 \tilde{Z_1}{}^2+Z_1^2 \tilde{Z_2}{}^4 \tilde{Z_1}{}^2+Z_1^4 Z_2^2 \tilde{Z_2}{}^2 \\
						\tilde{Z_1}{}^4 \tilde{Z_2}{}^4+Z_1^4 Z_2^4,\;
						Z_1^2 Z_2^2 \tilde{Z_1}{}^4+Z_1^4 \tilde{Z_2}{}^2 \tilde{Z_1}{}^2+Z_2^4 \tilde{Z_2}{}^2 \tilde{Z_1}{}^2+Z_1^2 Z_2^2 \tilde{Z_2}{}^4\\
						Z_1^2 \tilde{Z_2}{}^2 \tilde{Z_1}{}^4+Z_2^2 \tilde{Z_2}{}^4 \tilde{Z_1}{}^2+Z_1^4 Z_2^2 \tilde{Z_1}{}^2+Z_1^2 Z_2^4 \tilde{Z_2}{}^2 \\
						Z_1^4 \tilde{Z_1}{}^4+Z_2^4 \tilde{Z_2}{}^4 ,\;
						Z_1 Z_2 \tilde{Z_2} \tilde{Z_1}{}^5+Z_1 Z_2^5 \tilde{Z_2} \tilde{Z_1}+Z_1^5 Z_2 \tilde{Z_2} \tilde{Z_1}+Z_1 Z_2 \tilde{Z_2}{}^5 \tilde{Z_1} \\
						Z_2^2 \tilde{Z_1}{}^6+Z_2^6 \tilde{Z_1}{}^2+Z_1^6 \tilde{Z_2}{}^2+Z_1^2 \tilde{Z_2}{}^6 ,\;
						\tilde{Z_2}{}^2 \tilde{Z_1}{}^6+\tilde{Z_2}{}^6 \tilde{Z_1}{}^2+Z_1^2 Z_2^6+Z_1^6 Z_2^2 \\
						Z_1^2 \tilde{Z_1}{}^6+Z_1^6 \tilde{Z_1}{}^2+Z_2^6 \tilde{Z_2}{}^2+Z_2^2 \tilde{Z_2}{}^6 ,\;
						\tilde{Z_1}{}^8+\tilde{Z_2}{}^8+Z_1^8+Z_2^8 \\
					\end{array}$\\
					\hline
					\rule{0pt}{3ex}
					$\tau^{10}$ & $16$ &$\begin{array}{c}
						Z_1 Z_2^3 \tilde{Z_2}{}^2 \tilde{Z_1}{}^4-Z_1^2 Z_2^4 \tilde{Z_2} \tilde{Z_1}{}^3-Z_1^3 Z_2 \tilde{Z_2}{}^4 \tilde{Z_1}{}^2+Z_1^4 Z_2^2 \tilde{Z_2}{}^3 \tilde{Z_1} \\
						Z_1^3 Z_2^3 \tilde{Z_1}{}^4+Z_1^4 \tilde{Z_2}{}^3 \tilde{Z_1}{}^3-Z_2^4 \tilde{Z_2}{}^3 \tilde{Z_1}{}^3-Z_1^3 Z_2^3 \tilde{Z_2}{}^4 \\
						Z_1^3 Z_2 \tilde{Z_2}{}^2 \tilde{Z_1}{}^4+Z_1^4 Z_2^2 \tilde{Z_2} \tilde{Z_1}{}^3-Z_1 Z_2^3 \tilde{Z_2}{}^4 \tilde{Z_1}{}^2-Z_1^2 Z_2^4 \tilde{Z_2}{}^3 \tilde{Z_1} \\
						Z_2^4 \tilde{Z_2} \tilde{Z_1}{}^5-Z_1 Z_2^5 \tilde{Z_1}{}^4-Z_1^4 \tilde{Z_2}{}^5 \tilde{Z_1}+Z_1^5 Z_2 \tilde{Z_2}{}^4 \\
						Z_2^2 \tilde{Z_2}{}^3 \tilde{Z_1}{}^5-Z_1^2 \tilde{Z_2}{}^5 \tilde{Z_1}{}^3-Z_1^3 Z_2^5 \tilde{Z_1}{}^2+Z_1^5 Z_2^3 \tilde{Z_2}{}^2 \\
						Z_1^2 Z_2^2 \tilde{Z_2} \tilde{Z_1}{}^5-Z_1 Z_2^5 \tilde{Z_2}{}^2 \tilde{Z_1}{}^2+Z_1^5 Z_2 \tilde{Z_2}{}^2 \tilde{Z_1}{}^2-Z_1^2 Z_2^2 \tilde{Z_2}{}^5 \tilde{Z_1} \\
						Z_1^2 \tilde{Z_2}{}^3 \tilde{Z_1}{}^5-Z_2^2 \tilde{Z_2}{}^5 \tilde{Z_1}{}^3+Z_1^5 Z_2^3 \tilde{Z_1}{}^2-Z_1^3 Z_2^5 \tilde{Z_2}{}^2 \\
						Z_1^4 \tilde{Z_2} \tilde{Z_1}{}^5+Z_1^5 Z_2 \tilde{Z_1}{}^4-Z_2^4 \tilde{Z_2}{}^5 \tilde{Z_1}-Z_1 Z_2^5 \tilde{Z_2}{}^4 \\
						Z_1 Z_2^3 \tilde{Z_1}{}^6-Z_2^6 \tilde{Z_2} \tilde{Z_1}{}^3+Z_1^6 \tilde{Z_2}{}^3 \tilde{Z_1}-Z_1^3 Z_2 \tilde{Z_2}{}^6 \\
						Z_1 Z_2 \tilde{Z_2}{}^2 \tilde{Z_1}{}^6-Z_1 Z_2 \tilde{Z_2}{}^6 \tilde{Z_1}{}^2-Z_1^2 Z_2^6 \tilde{Z_2} \tilde{Z_1}+Z_1^6 Z_2^2 \tilde{Z_2} \tilde{Z_1} \\
						Z_1^3 Z_2 \tilde{Z_1}{}^6+Z_1^6 \tilde{Z_2} \tilde{Z_1}{}^3-Z_2^6 \tilde{Z_2}{}^3 \tilde{Z_1}-Z_1 Z_2^3 \tilde{Z_2}{}^6 \\
						Z_2^2 \tilde{Z_2} \tilde{Z_1}{}^7-Z_1 Z_2^7 \tilde{Z_1}{}^2-Z_1^2 \tilde{Z_2}{}^7 \tilde{Z_1}+Z_1^7 Z_2 \tilde{Z_2}{}^2 \\
						\tilde{Z_2}{}^3 \tilde{Z_1}{}^7-\tilde{Z_2}{}^7 \tilde{Z_1}{}^3-Z_1^3 Z_2^7+Z_1^7 Z_2^3 ,\;
						Z_1^2 \tilde{Z_2} \tilde{Z_1}{}^7+Z_1^7 Z_2 \tilde{Z_1}{}^2-Z_2^2 \tilde{Z_2}{}^7 \tilde{Z_1}-Z_1 Z_2^7 \tilde{Z_2}{}^2 \\
						Z_1 Z_2 \tilde{Z_1}{}^8+Z_1^8 \tilde{Z_2} \tilde{Z_1}-Z_2^8 \tilde{Z_2} \tilde{Z_1}-Z_1 Z_2 \tilde{Z_2}{}^8 ,\;
						\tilde{Z_2} \tilde{Z_1}{}^9-\tilde{Z_2}{}^9 \tilde{Z_1}-Z_1 Z_2^9+Z_1^9 Z_2 \\
					\end{array}$\\
					\hline
				\end{tabular}
			}
		\end{adjustbox}
		\caption{Invariants for $\mathbb{H}^2/G_2(\hat{D}_k,\mathbb{Z}_k)$ at $k=2$}
		\label{tab:invsDkZk}
		
	\end{table}
	
	\begin{table}[htpb]
		\begin{adjustbox}{center}
			\centering
			\resizebox{\textwidth}{!}{
				\begin{tabular}{|c|c|c|}
					\hline
					Level & $\#$ of invariants & Invariants\\
					\hline
					$\tau^2$ & $0$ &\\
					\hline
					\rule{0pt}{3ex}
					$\tau^4$ & $2$ &$\begin{array}{c}
						Z_1 Z_2 \tilde{Z_1} \tilde{Z_2} \\
						\tilde{Z_1}{}^2 \tilde{Z_2}{}^2+Z_1^2 Z_2^2 \\
					\end{array}$\\
					\hline
					\rule{0pt}{3ex}
					$\tau^6$ & $3$ &$ \begin{array}{c}
						Z_1^3 \tilde{Z_2}{}^3+Z_2^3 \tilde{Z_1}{}^3 \\
						Z_1^3 \tilde{Z_1}{}^3-Z_2^3 \tilde{Z_2}{}^3 \\
						\tilde{Z_1}{}^6+\tilde{Z_2}{}^6+Z_1^6+Z_2^6 \\
					\end{array}$\\
					\hline
					\rule{0pt}{3ex}
					$\tau^8$ & $7$ &$\begin{array}{c}
						Z_1^2 Z_2^2 \tilde{Z_1}{}^2 \tilde{Z_2}{}^2 \\
						Z_1 Z_2 \tilde{Z_1}{}^3 \tilde{Z_2}{}^3+Z_1^3 Z_2^3 \tilde{Z_1} \tilde{Z_2} \\
						Z_2^3 \tilde{Z_2} \tilde{Z_1}{}^4-Z_1 Z_2^4 \tilde{Z_1}{}^3-Z_1^3 \tilde{Z_2}{}^4 \tilde{Z_1}+Z_1^4 Z_2 \tilde{Z_2}{}^3 \\
						\tilde{Z_1}{}^4 \tilde{Z_2}{}^4+Z_1^4 Z_2^4 \\
						Z_1^3 \tilde{Z_2} \tilde{Z_1}{}^4+Z_1^4 Z_2 \tilde{Z_1}{}^3+Z_2^3 \tilde{Z_2}{}^4 \tilde{Z_1}+Z_1 Z_2^4 \tilde{Z_2}{}^3 \\
						Z_1 Z_2 \tilde{Z_1}{}^6+Z_1^6 \tilde{Z_2} \tilde{Z_1}-Z_2^6 \tilde{Z_2} \tilde{Z_1}-Z_1 Z_2 \tilde{Z_2}{}^6 \\
						\tilde{Z_2} \tilde{Z_1}{}^7-\tilde{Z_2}{}^7 \tilde{Z_1}-Z_1 Z_2^7+Z_1^7 Z_2 \\
					\end{array}$\\
					\hline
					\rule{0pt}{3ex}
					$\tau^{10}$ & $7$ &$\begin{array}{c}
						Z_2 Z_1^4 \tilde{Z_1} \tilde{Z_2}{}^4+Z_2^4 Z_1 \tilde{Z_1}{}^4 \tilde{Z_2} \\
						Z_1^4 Z_2 \tilde{Z_1}{}^4 \tilde{Z_2}-Z_1 Z_2^4 \tilde{Z_1} \tilde{Z_2}{}^4 \\
						Z_2^3 \tilde{Z_2}{}^2 \tilde{Z_1}{}^5+Z_1^2 Z_2^5 \tilde{Z_1}{}^3+Z_1^3 \tilde{Z_2}{}^5 \tilde{Z_1}{}^2+Z_1^5 Z_2^2 \tilde{Z_2}{}^3 \\
						Z_1^3 \tilde{Z_2}{}^2 \tilde{Z_1}{}^5+Z_1^5 Z_2^2 \tilde{Z_1}{}^3-Z_2^3 \tilde{Z_2}{}^5 \tilde{Z_1}{}^2-Z_1^2 Z_2^5 \tilde{Z_2}{}^3 \\
						Z_1^2 Z_2^2 \tilde{Z_1}{}^6+Z_1^6 \tilde{Z_2}{}^2 \tilde{Z_1}{}^2+Z_2^6 \tilde{Z_2}{}^2 \tilde{Z_1}{}^2+Z_1^2 Z_2^2 \tilde{Z_2}{}^6 \\
						Z_1 Z_2 \tilde{Z_2} \tilde{Z_1}{}^7+Z_1 Z_2^7 \tilde{Z_2} \tilde{Z_1}+Z_1^7 Z_2 \tilde{Z_2} \tilde{Z_1}+Z_1 Z_2 \tilde{Z_2}{}^7 \tilde{Z_1} \\
						\tilde{Z_2}{}^2 \tilde{Z_1}{}^8+\tilde{Z_2}{}^8 \tilde{Z_1}{}^2+Z_1^2 Z_2^8+Z_1^8 Z_2^2 \\
					\end{array}$\\
					\hline
				\end{tabular}
			}
		\end{adjustbox}
		\caption{Invariants for $\mathbb{H}^2/G_2(\hat{D}_k,\mathbb{Z}_k)$ at $k=3$}
		\label{tab:invsDkZk3}
		
	\end{table}

		\begin{table}[htpb]
		\begin{adjustbox}{center}
			\centering
			\resizebox{!}{!}{
				\begin{tabular}{|c|c|c|}
					\hline
					Level & $\#$ of invariants & Invariants\\
					\hline
					$\tau^2$ & $0$ &\\
					\hline
					\rule{0pt}{3ex}
					$\tau^4$ & $2$ &$\begin{array}{c}
						Z_1 Z_2 \tilde{Z_1} \tilde{Z_2} \\
						\tilde{Z_1}{}^2 \tilde{Z_2}{}^2+Z_1^2 Z_2^2 \\
					\end{array}$\\
					\hline
					\rule{0pt}{3ex}
					$\tau^6$ & $0$ &\\
					\hline
					\rule{0pt}{3ex}
					$\tau^8$ & $6$ &$\begin{array}{c}
						Z_1^2 Z_2^2 \tilde{Z_1}{}^2 \tilde{Z_2}{}^2 \\
						Z_1 Z_2 \tilde{Z_1}{}^3 \tilde{Z_2}{}^3+Z_1^3 Z_2^3 \tilde{Z_1} \tilde{Z_2} \\
						Z_1^4 \tilde{Z_2}{}^4+Z_2^4 \tilde{Z_1}{}^4 \\
						\tilde{Z_1}{}^4 \tilde{Z_2}{}^4+Z_1^4 Z_2^4 \\
						Z_1^4 \tilde{Z_1}{}^4+Z_2^4 \tilde{Z_2}{}^4 \\
						\tilde{Z_1}{}^8+\tilde{Z_2}{}^8+Z_1^8+Z_2^8 \\
					\end{array}$\\
					\hline
					\rule{0pt}{3ex}
					$\tau^{10}$ & $4$ &$\begin{array}{c}
						Z_2^4 \tilde{Z_2} \tilde{Z_1}{}^5-Z_1 Z_2^5 \tilde{Z_1}{}^4-Z_1^4 \tilde{Z_2}{}^5 \tilde{Z_1}+Z_1^5 Z_2
						\tilde{Z_2}{}^4 \\
						Z_1^4 \tilde{Z_2} \tilde{Z_1}{}^5+Z_1^5 Z_2 \tilde{Z_1}{}^4-Z_2^4 \tilde{Z_2}{}^5 \tilde{Z_1}-Z_1 Z_2^5
						\tilde{Z_2}{}^4 \\
						Z_1 Z_2 \tilde{Z_1}{}^8+Z_1^8 \tilde{Z_2} \tilde{Z_1}-Z_2^8 \tilde{Z_2} \tilde{Z_1}-Z_1 Z_2 \tilde{Z_2}{}^8 \\
						\tilde{Z_2} \tilde{Z_1}{}^9-\tilde{Z_2}{}^9 \tilde{Z_1}-Z_1 Z_2^9+Z_1^9 Z_2 \\
					\end{array}$\\
					\hline
				\end{tabular}
			}
		\end{adjustbox}
		\caption{Invariants for $\mathbb{H}^2/G_2(\hat{D}_k,\mathbb{Z}_k)$ at $k=4$}
		\label{tab:invsDkZk4}
		
	\end{table}
	

	
	\begin{table}[htpb]
		\begin{adjustbox}{center}
			\centering
			\resizebox{\textwidth}{!}{
				\begin{tabular}{|c|c|c|}
					\hline
					Level & $\#$ of invariants & Invariants\\
					\hline
					$\tau^2$ & $0$ &\\
					\hline
					\rule{0pt}{3ex}
					$\tau^4$ & $3$ &$ \begin{array}{c}
						\tilde{Z_1}{}^2 \tilde{Z_2}{}^2+Z_1^2 Z_2^2 \\
						Z_1^2 \tilde{Z_1}{}^2+Z_1^2 \tilde{Z_2}{}^2+Z_2^2 \tilde{Z_1}{}^2+Z_2^2 \tilde{Z_2}{}^2 \\
						\tilde{Z_1}{}^4+\tilde{Z_2}{}^4+Z_1^4+Z_2^4 \\
					\end{array}$\\
					\hline
					\rule{0pt}{3ex}
					$\tau^6$ & $2$ &$ \begin{array}{c}
						Z_1^2 \left(-\tilde{Z_2}\right) \tilde{Z_1}{}^3-Z_2^2 \tilde{Z_2} \tilde{Z_1}{}^3+Z_1 Z_2^3 \tilde{Z_1}{}^2-Z_1^3 Z_2 \tilde{Z_1}{}^2+Z_1^2 \tilde{Z_2}{}^3 \tilde{Z_1}+Z_2^2 \tilde{Z_2}{}^3
						\tilde{Z_1}+Z_1 Z_2^3 \tilde{Z_2}{}^2-Z_1^3 Z_2 \tilde{Z_2}{}^2 \\
						\tilde{Z_2} \tilde{Z_1}{}^5-\tilde{Z_2}{}^5 \tilde{Z_1}-Z_1 Z_2^5+Z_1^5 Z_2 \\
					\end{array}$\\
					\hline
					\rule{0pt}{3ex}
					$\tau^8$ & $9$ &$\begin{array}{c}
						Z_1^2 Z_2^2 \tilde{Z_1}{}^2 \tilde{Z_2}{}^2 \\
						Z_2 Z_1^3 \left(-\tilde{Z_1}\right) \tilde{Z_2}{}^3+Z_2 Z_1^3 \tilde{Z_1}{}^3 \tilde{Z_2}+Z_2^3 Z_1 \tilde{Z_1} \tilde{Z_2}{}^3-Z_2^3 Z_1 \tilde{Z_1}{}^3 \tilde{Z_2} \\
						\tilde{Z_1}{}^4 \tilde{Z_2}{}^4+Z_1^4 Z_2^4 \\
						Z_1^2 Z_2^2 \tilde{Z_1}{}^4+Z_1^4 \tilde{Z_2}{}^2 \tilde{Z_1}{}^2+Z_2^4 \tilde{Z_2}{}^2 \tilde{Z_1}{}^2+Z_1^2 Z_2^2 \tilde{Z_2}{}^4 \\
						Z_1^2 \tilde{Z_2}{}^2 \tilde{Z_1}{}^4+Z_2^2 \tilde{Z_2}{}^2 \tilde{Z_1}{}^4+Z_1^2 Z_2^4 \tilde{Z_1}{}^2+Z_1^2 \tilde{Z_2}{}^4 \tilde{Z_1}{}^2+Z_2^2 \tilde{Z_2}{}^4 \tilde{Z_1}{}^2+Z_1^4 Z_2^2
						\tilde{Z_1}{}^2+Z_1^2 Z_2^4 \tilde{Z_2}{}^2+Z_1^4 Z_2^2 \tilde{Z_2}{}^2 \\
						Z_1^4 \tilde{Z_1}{}^4+Z_1^4 \tilde{Z_2}{}^4+Z_2^4 \tilde{Z_1}{}^4+Z_2^4 \tilde{Z_2}{}^4 \\
						\tilde{Z_2}{}^2 \tilde{Z_1}{}^6+\tilde{Z_2}{}^6 \tilde{Z_1}{}^2+Z_1^2 Z_2^6+Z_1^6 Z_2^2 \\
						Z_1^2 \tilde{Z_1}{}^6+Z_2^2 \tilde{Z_1}{}^6+Z_1^6 \tilde{Z_1}{}^2+Z_2^6 \tilde{Z_1}{}^2+Z_1^6 \tilde{Z_2}{}^2+Z_2^6 \tilde{Z_2}{}^2+Z_1^2 \tilde{Z_2}{}^6+Z_2^2 \tilde{Z_2}{}^6 \\
						\tilde{Z_1}{}^8+\tilde{Z_2}{}^8+Z_1^8+Z_2^8 \\
					\end{array}$\\
					\hline
					\rule{0pt}{3ex}
					$\tau^{10}$ & $8$ &$\begin{array}{c}
						Z_1 Z_2^3 \left(-\tilde{Z_2}{}^2\right) \tilde{Z_1}{}^4+Z_1^3 Z_2 \tilde{Z_2}{}^2 \tilde{Z_1}{}^4+Z_1^2 Z_2^4 \tilde{Z_2} \tilde{Z_1}{}^3+Z_1^4 Z_2^2 \tilde{Z_2} \tilde{Z_1}{}^3-Z_1 Z_2^3
						\tilde{Z_2}{}^4 \tilde{Z_1}{}^2+Z_1^3 Z_2 \tilde{Z_2}{}^4 \tilde{Z_1}{}^2-Z_1^2 Z_2^4 \tilde{Z_2}{}^3 \tilde{Z_1}-Z_1^4 Z_2^2 \tilde{Z_2}{}^3 \tilde{Z_1} \\
						Z_1^2 Z_2^2 \tilde{Z_2} \tilde{Z_1}{}^5-Z_1 Z_2^5 \tilde{Z_2}{}^2 \tilde{Z_1}{}^2+Z_1^5 Z_2 \tilde{Z_2}{}^2 \tilde{Z_1}{}^2-Z_1^2 Z_2^2 \tilde{Z_2}{}^5 \tilde{Z_1} \\
						Z_1^2 \left(-\tilde{Z_2}{}^3\right) \tilde{Z_1}{}^5-Z_2^2 \tilde{Z_2}{}^3 \tilde{Z_1}{}^5+Z_1^2 \tilde{Z_2}{}^5 \tilde{Z_1}{}^3+Z_2^2 \tilde{Z_2}{}^5 \tilde{Z_1}{}^3+Z_1^3 Z_2^5
						\tilde{Z_1}{}^2-Z_1^5 Z_2^3 \tilde{Z_1}{}^2+Z_1^3 Z_2^5 \tilde{Z_2}{}^2-Z_1^5 Z_2^3 \tilde{Z_2}{}^2 \\
						Z_1^4 \left(-\tilde{Z_2}\right) \tilde{Z_1}{}^5-Z_2^4 \tilde{Z_2} \tilde{Z_1}{}^5+Z_1 Z_2^5 \tilde{Z_1}{}^4-Z_1^5 Z_2 \tilde{Z_1}{}^4+Z_1^4 \tilde{Z_2}{}^5 \tilde{Z_1}+Z_2^4 \tilde{Z_2}{}^5
						\tilde{Z_1}+Z_1 Z_2^5 \tilde{Z_2}{}^4-Z_1^5 Z_2 \tilde{Z_2}{}^4 \\
						-Z_1 Z_2^3 \tilde{Z_1}{}^6+Z_1^3 Z_2 \tilde{Z_1}{}^6+Z_1^6 \tilde{Z_2} \tilde{Z_1}{}^3+Z_2^6 \tilde{Z_2} \tilde{Z_1}{}^3-Z_1^6 \tilde{Z_2}{}^3 \tilde{Z_1}-Z_2^6 \tilde{Z_2}{}^3 \tilde{Z_1}-Z_1
						Z_2^3 \tilde{Z_2}{}^6+Z_1^3 Z_2 \tilde{Z_2}{}^6 \\
						\tilde{Z_2}{}^3 \tilde{Z_1}{}^7-\tilde{Z_2}{}^7 \tilde{Z_1}{}^3-Z_1^3 Z_2^7+Z_1^7 Z_2^3 \\
						Z_1^2 \left(-\tilde{Z_2}\right) \tilde{Z_1}{}^7-Z_2^2 \tilde{Z_2} \tilde{Z_1}{}^7+Z_1 Z_2^7 \tilde{Z_1}{}^2-Z_1^7 Z_2 \tilde{Z_1}{}^2+Z_1^2 \tilde{Z_2}{}^7 \tilde{Z_1}+Z_2^2 \tilde{Z_2}{}^7
						\tilde{Z_1}+Z_1 Z_2^7 \tilde{Z_2}{}^2-Z_1^7 Z_2 \tilde{Z_2}{}^2 \\
						\tilde{Z_2} \tilde{Z_1}{}^9-\tilde{Z_2}{}^9 \tilde{Z_1}-Z_1 Z_2^9+Z_1^9 Z_2 \\
					\end{array}$\\
					\hline
				\end{tabular}
			}
		\end{adjustbox}
		\caption{Invariants for $\mathbb{H}^2/G_2(\hat{D}_k,\hat{D}_{\frac{k}{2}})$ at $k=2$}
		\label{tab:invsDkDkb2}
		
	\end{table}

	\begin{table}[htpb]
		\begin{adjustbox}{center}
			\centering
			\resizebox{\textwidth}{!}{
				\begin{tabular}{|c|c|c|}
					\hline
					Level & $\#$ of invariants & Invariants\\
					\hline
					$\tau^2$ & $0$ &\\
					\hline
					\rule{0pt}{3ex}
					$\tau^4$ & $1$ &$ \begin{array}{c}
						\tilde{Z_1}{}^2 \tilde{Z_2}{}^2+Z_1^2 Z_2^2 \\
					\end{array}$\\
					\hline
					$\tau^6$ & $0$ & \\
					\hline
					\rule{0pt}{3ex}
					$\tau^8$ & $4$ &$\begin{array}{c}
						Z_1^2 Z_2^2 \tilde{Z_1}{}^2 \tilde{Z_2}{}^2 \\
						\tilde{Z_1}{}^4 \tilde{Z_2}{}^4+Z_1^4 Z_2^4 \\
						Z_1^4 \tilde{Z_1}{}^4+Z_1^4 \tilde{Z_2}{}^4+Z_2^4 \tilde{Z_1}{}^4+Z_2^4 \tilde{Z_2}{}^4 \\
						\tilde{Z_1}{}^8+\tilde{Z_2}{}^8+Z_1^8+Z_2^8 \\
					\end{array}$\\
					\hline
					\rule{0pt}{3ex}				
					$\tau^{10}$ & $2$ &$\begin{array}{c}
						Z_1^4 \tilde{Z_2} \tilde{Z_1}{}^5+Z_2^4 \tilde{Z_2} \tilde{Z_1}{}^5-Z_1 Z_2^5 \tilde{Z_1}{}^4+Z_1^5 Z_2
						\tilde{Z_1}{}^4-Z_1^4 \tilde{Z_2}{}^5 \tilde{Z_1}-Z_2^4 \tilde{Z_2}{}^5 \tilde{Z_1}-Z_1 Z_2^5 \tilde{Z_2}{}^4+Z_1^5
						Z_2 \tilde{Z_2}{}^4 \\
						\tilde{Z_2} \tilde{Z_1}{}^9-\tilde{Z_2}{}^9 \tilde{Z_1}-Z_1 Z_2^9+Z_1^9 Z_2 \\
					\end{array}$\\
					\hline
				\end{tabular}
			}
		\end{adjustbox}
		\caption{Invariants for $\mathbb{H}^2/G_2(\hat{D}_k,\hat{D}_{\frac{k}{2}})$ at $k=4$}
		\label{tab:invsDkDkb2k4}
	\end{table}
	\clearpage
	
	
	\section{Constrains on Chern-Simons Levels}
	\label{app:CSLEVcons}
	Here we explain how to obtain the constraint on the Chern-Simons levels using the condition for supersymmetry enhancement. We first review the SUSY enhancement condition following \cite{Gaiotto:2008sd,Schnabl:2008wj}.

	We restrict to the discussion of enhancement to $\mathcal{N}=5$ SUSY. Starting with the two hypermutliplets $Q^A_{1}$, $Q^A_{2}$  in the pseduoreal representation $R$ of some gauge group $G$. Let us denote the matrices of this representation as $T^p_{AB}$, where $p=1,2,...,\text{dim}(G)$ and $A,B=1,2,...,\text{dim}(R)$. The superpotential is given as 
	\begin{equation}
		W=f_{ABCD}(Q_1^A Q_2^B + Q_2^A Q_2^B)(Q_1^C Q_2^D + Q_2^C Q_2^D)
	\end{equation}
	where 
	\begin{equation}
		f_{ABCD}=\kappa_{pq}T^p_{AB}T^q_{CD}
	\end{equation}
	and $\kappa_{pq}$ denotes the inverse Chern-Simons coupling. If
	\begin{equation}
		\label{eq:Jacobequivalent}
		f_{A(BCD)}=0
	\end{equation}
	then the Superpotential can be written as
	\begin{equation}
		W=2f_{ABCD}\epsilon^{\alpha\gamma}\epsilon^{\beta\delta}(Q_\alpha^A Q_\beta^B Q_\gamma^C Q_\delta^D) 
	\end{equation}
	This shows that the flavour symmetry is enhanced to $SU(2)_F$, which then combines with the $SU(2)_R$ to form the $SO(5)_R$ symmetry of $\mathcal{N}=5$.
	
	Now, let us explain how this condition is related to the Lie superalgebra structure. A Lie superalgebra consists of even generators $M^a$ which form a Lie algebra and odd generators $\lambda_A$ which are in some representation $R$ and they obey the following commutation/anticommutation relations:
	\begin{equation}
		[M^a,M^b]=f^{ab}_c M^c ,
	\end{equation}
	\begin{equation}
		[M^a,\lambda_A]=T^a_{AB} \lambda^B ,
	\end{equation}
	\begin{equation}
		\{\lambda_A,\lambda_B\}=\kappa_{ab}T^{a}_{AB} M^b .
	\end{equation}
	Furthermore, the $\lambda$'s also obey the Jacobi identity 
	\begin{equation}
		[\lambda_A,\{\lambda_B,\lambda_C\}]+[\lambda_C,\{\lambda_A,\lambda_B\}]+[\lambda_B,\{\lambda_C,\lambda_A\}]=0 ,
	\end{equation}
	which is equivalent to the condition $\ref{eq:Jacobequivalent}$. It restricts the allowed representation of $\lambda$ and also constrains $\kappa_{ab}$ which is the inverse of the Chern-Simons coupling matrix. For a given Lie superalgebra, if the gauge algebra is chosen to be the even part of the algebra and the chirals are chosen to be in a representation of the algebra, we have an enhanced supersymmetry. 
	For the purpose of this paper we are concerned with $OSp(M|2N)$, $G(3)$, $F(4)$ and $D(2,1|\alpha)$ Lie Superalgebras whose gauge groups are $SO(M)\times USp(2N)$, $G_2\times SU(2)$, $SO(7)\times SU(2)$ and $SU(2)\times SU(2)\times SU(2)$ and the hypermultiplets are in the representation $(\mathbf{M},\mathbf{2N})$, $(\mathbf{7},\mathbf{2})$, $(\mathbf{8},\mathbf{2})$  and $(\mathbf{2},\mathbf{2},\mathbf{2})$ respectively of the appropriate gauge algebra.
	
	Now we show how to find the ratio of the Chern Simon levels for the above cases using equation \ref{eq:Jacobequivalent}. Given the highest weight of a representation $\lambda$. Since $f_{A(BCD)}=0$
	\begin{equation}
		\label{eq:Jacobequispeccase}
		f_{\lambda\lambda\lambda\lambda}=0
	\end{equation}
	It is straightforward to derive \cite{Schnabl:2008wj}
	\begin{equation}
		\label{eq:gencond}
		f_{\lambda\lambda\lambda\lambda}=\sum_{l=1}^L\frac{4\pi}{k_l}\left[\sum_{i=1}^{\text{rank}(G_l)}\lambda_{(l)}^i\lambda^i_{(l)}\right]=\sum_{i=1}^l\frac{4\pi}{k_l} \left(\lambda,\lambda\right)_{(l)} ,
	\end{equation}
	where $(\, ,\,)$ denotes the positive definite scalar product \cite{francesco2012conformal}. Here $l$ labels the gauge algebra and runs from $1$ to the number of gauge algebras $L$. Note that \ref{eq:gencond} is a necessary condition but is not sufficient and one needs to impose the full constraint which is equation \ref{eq:Jacobequivalent}. But since we already started with Lie superalgebra, we are guaranteed that equation \ref{eq:Jacobequivalent} is obeyed, and we can simply use  \ref{eq:Jacobequispeccase} to get the constraint on the Chern-Simons levels. We summarize the conditions for the Lie superalgebras which give rise to $\mathcal{N}=5$ theories in table \ref{tab:SUSYenchanceCSlevel}. We use the notations of LieArt \cite{Feger_2020} in which the length of the highest root is normalized to $2$.

	\begin{table}
		\centering
		\resizebox{\textwidth}{!}{
			\begin{tabular}{|c|c|c|c|c|}
				\hline
				Lie-Superalgebra&\makecell{Gauge Algebra and\\ Representation} & Highest Weight vector & Inner Products & SUSY enhancement condition \\
				\hline
				$OSp(M|2N)$&\makecell{$SO(M)_{k_1}\times USp(2N)_{k_2}$ \\$(\mathbf{M},\mathbf{2N})$} & \makecell{$\lambda_{SO(M)}=(1,0,...,0)$ \\ $\lambda_{USp(2N)}=(1,0,....0)$} & 	\makecell{$(\lambda_{SO(M)},\lambda_{SO(M)})=1$\\$ 	(\lambda_{USp(2N)},\lambda_{USp(2N)})=\frac{1}{2}$ }& $	\frac{1}{2k_1}+\frac{1}{k_2}=0\;\implies\; k_1+2k_2=0$ \\
				\hline
				$G(3)$&\makecell{$(G_2)_{k_1}\times SU(2)_{k_2}$ \\ $(\mathbf{7},\mathbf{2})$} & \makecell{$\lambda_{G_2}=(1,0) $ \\ $\lambda_{SU(2)}=(1)$} & 	\makecell{$	(\lambda_{G_2},\lambda_{G_2})=\frac{2}{3}$\\$ 		(\lambda_{SU(2)},\lambda_{SU(2))})=\frac{1}{2}$ }&$		\frac{2}{3k_1}+\frac{1}{2k_2}=0\;\implies\; 3k_1+4k_2=0$  \\
				
				\hline
				$F(4)$&\makecell{$SO(7)_{k_1}\times SU(2)_{k_2}$\\  $(\mathbf{8},\mathbf{2})$  }& \makecell{$\lambda_{SO(7)}=(0,0,1) $ \\ $\lambda_{SU(2)}=(1)$} & 	\makecell{$	(\lambda_{SO(7)},\lambda_{SO(7)})=\frac{3}{4}$\\$ 		(\lambda_{SU(2)},\lambda_{SU(2))})=\frac{1}{2}$ }&  $	\frac{3}{4k_1}+\frac{1}{2k_2}=0\;\implies\; 2k_1+3k_2=0$\\
				\hline
				$D(2|1;\alpha)$&\makecell{$SU(2)_{k_1}\times SU(2)_{k_2}\times SU(2)_{k_2} $\\ $(\mathbf{2},\mathbf{2},\mathbf{2})$} & $\lambda_{SU(2)}=(1)$ & 		$(\lambda_{SU(2)},\lambda_{SU(2))})=\frac{1}{2}$ & $			\frac{1}{2k_1}+\frac{1}{2k_2}+\frac{1}{2k_3}=0$ \\
				\hline
		\end{tabular}}
		\caption{Summary of different constraints on the Chern-Simons levels for different theories.}
		\label{tab:SUSYenchanceCSlevel}
	\end{table}

 \section{Moduli spaces in cases with unequal ranks}
 \label{app:MS}

Here we consider the moduli space of $\mathcal{N}=5$ SCFTs described by a product gauge group, consisting of two groups with ranks $r_1$ and $r_2$, where we assume $r_1<r_2$. The idea is again to go on the moduli space by giving a vev to the bifundamental hyper. On a generic point in the moduli space the gauge group is broken to $r_1$ copies of a $U(1)_k \times U(1)_{-k}$ gauge theory with two bifundamental hypers and a decoupled gauge group of rank $r_2-r_1$ with a Chern-Simons level proportional to $k$. For simplicity, let us first consider the case of $r_1=1$, before remarking on the general case. The moduli space is then spanned by the $U(1)_k \times U(1)_{-k}$ monopole operators properly dressed by the bifundamental in order to be gauge invariant. If the basic monopole has charge $(1;1)$ under the $U(1) \times U(1)$ then the moduli is $\mathbb{C}^4/\mathbb{Z}_k$ \cite{Aharony:2008ug,Tachikawa:2019dvq}. This follows as one must dress this monopole with $k$ copies of the bifundamental, which is then mapped to the order $k$ invariant of $\mathbb{C}^4/\mathbb{Z}_k$. However, if there exist a basic monopole with smaller charges, $(\frac{1}{p};\frac{1}{p})$, then we can use this, properly dressed by $\frac{k}{p}$ copies of the bifundamental, to build a more basic operator spanning the moduli space. The moduli space then becomes $\mathbb{C}^4/\mathbb{Z}_{\frac{k}{p}}$, with this operator playing the role of its order $\frac{k}{p}$ invariant. 

As such, to determine the moduli space one must properly understand what are the basic monopoles that can contribute. This in turn is described by the monopole charge lattice of the gauge group of the SCFT. A subtle issue that arises in cases with unequal ranks is that the basic monopole might involve magnetic charges also in the decoupled gauge group. For instance, consider the case where the minimal monopole with charges only under the $U(1) \times U(1)$ part of the gauge group is $(1;1,0,0,...,0)$, with the additional entries standing for the magnetic charges of the remaining $r_2-r_1$ ranks, but there are also monopoles with charges $(\frac{1}{p};\frac{1}{p},\frac{1}{p},\frac{1}{p},...,\frac{1}{p})$. This would lead to a monopole of charge $(\frac{1}{p};\frac{1}{p})$ under the $U(1) \times U(1)$, but now it is also accompanied by some monopole of the decoupled gauge group. Due to the Chern-Simons level, this monopole will acquire gauge charges under the $U(1) \times U(1)$ and the decoupled gauge group. The former can be canceled by $\frac{k}{p}$ copies of the bifundamental, but what do we do with the latter? In general, there are no hypermultiplets in the decoupled gauge theory so one cannot cancel the gauge charges using hypers. Nevertheless, the decoupled gauge group contains gauginos, and in certain cases, these can be used to cancel the gauge charges. In these cases, a basic monopole invariant of smaller charges exist and the moduli space is $\mathbb{C}^4/\mathbb{Z}_{\frac{k}{p}}$. However, if it is not possible to cancel the gauge charges using the gauginos, then this monopole is not gauge invariant, and the moduli space is $\mathbb{C}^4/\mathbb{Z}_k$. As such the form of the moduli space in these cases depends also on the spectrum of dressed monopoles of the decoupled gauge group.

The discussion so far also carries over the the case of $r_1>1$. The analysis in this case remains the same, with $\mathbb{Z}_{k}$ being replaced with $G(k,1,r_1)$ and $\mathbb{Z}_{\frac{k}{p}}$ being replaced with $G(k,p,r_1)$. Next we shall illustrate this with some examples.

\subsection{$U(N+x)_k \times U(N)_{-k}$}        
\label{sec:UUuneqrank}
As a simple example, consider the case of the ABJ theory $U(1)_k \times U(3)_{-k}$. We can again go on the moduli space, where the gauge group would be broken to the ABJM $U(1)_k \times U(1)_{-k}$ theory and a decoupled $U(2)_{-k}$ gauge theory. In this case, the monopole lattice consists of all integer charges and in particular, contains the monopole $(1;1,0,0)$. This is the minimal possible monopole and so the moduli space is $\mathbb{C}^4/\mathbb{Z}_{k}$.

Next we consider taking a $\mathbb{Z}_p$ quotient, that is we change the gauge group to $[ U(1) \times U(3) ]/\mathbb{Z}_p$, with $p$ a divisor of $k$. Note that for this to be consistent the anomaly found in \cite{Tachikawa:2019dvq} must be trivial, which puts the condition $\frac{2 k}{p^2} = integer$. We assume that this condition is satisfied. Note, that in the quotient gauge group we now allow monopoles with charges that are fractions of $\frac{1}{p}$, but only if all charges carry the fractional charge. As such, $(\frac{1}{p};\frac{1}{p},0,0)$ is not allowed, but $(\frac{1}{p};\frac{1}{p},\frac{1}{p},\frac{1}{p})$ is. The exact form of the moduli space of the $[ U(1) \times U(3) ]/\mathbb{Z}_p$ theory now depends on whether such monopole can be made gauge invariant. 

We can cancel the charges under the $U(1) \times U(1)$ part by using the bifundamental. This leaves the part in the $U(2)_{-k}$ gauge theory. Notice that all bifundamental hypers connecting it to the $U(1) \times U(1)$ have become massive so the only thing we have left to use are the gauginos\footnote{We also remind the reader that this is an $\mathcal{N}=5$ Chern-Simons theory, so there is no Yang-Mills term. As such the scalar in the vector multiplet has no kinetic term so it can be integrated out, leading to a quartic superpotential for the bifundamentals \cite{Schnabl:2008wj}. As such, the vector multiplet only contains the gauge connection and the gauginos.}. However, the gauginos are fermions so we cannot use more than one copy of each gaugino. For $U(2)$, the gauginos carry the charges $\frac{q_1}{q_2}$ and $\frac{q_2}{q_1}$, where we use $q_1$ and $q_2$ for the fugacities of $U(2)$. Therefore, only monoples with charges $\frac{q_2}{q_1}$ or $\frac{q_1}{q_2}$ can be made gauge invariant as these can be canceled by a gaugino. A $U(2)_{k}$ monopole with magnetic charge $(\frac{n_1}{p},\frac{n_1}{p})$ carry the gauge charges $q^{\frac{k n_1}{p}}_1 q^{\frac{k n_2}{p}}_2$, and so one can see that this can only work if $k=p$ and $n_1=-n_2=1$. The condition for the anomaly to vanish, $\frac{2 k}{p^2} = integer$, then further forces $k=p=2$.

We then conclude that the moduli space of $[ U(1)_k \times U(3)_{-k} ]/\mathbb{Z}_p$ is $\mathbb{C}^4/\mathbb{Z}_{k}$ unless $k=p=2$, in which case it becomes $\mathbb{C}^4/\mathbb{Z}_{\frac{k}{2}} = \mathbb{C}^4$. The reason is that on the moduli space, the minimal gauge invariant monopole in almost all cases is $(1;1,0,0)$. We noted that even though monopoles with smaller charges exist, these cannot be made gauge invariant due to their $U(2)_{k}$ charges. The only exceptional case is $[ U(1)_2 \times U(3)_{-2} ]/\mathbb{Z}_2$, as in this case the monopole $(\frac{1}{2};\frac{1}{2},\frac{1}{2},-\frac{1}{2})$ can be made gauge invariant, with the $U(2)_k$ charges canceled by the gaugino.      

It is straightforward to generalize the above discussion. The story unfolds similarly for $U(N)_k \times U(N+2)_{-k}$. Here when going on the moduli space we now have $N$ copies of $U(1)_k \times U(1)_{-k}$ in addition to a decoupled $U(2)_{-k}$. We now have the basic monopoles $(1,0,...,0;1,0,...,0)$, $(1,1,0,...,0;1,1,0,...,0)$ up to $(1,1,...,1;1,1,...1,0,0)$. These can be made gauge invariant using $k$ copies of the bifundamental for the first one, $2k$ copies for the second one and so forth up to $Nk$ copies for the last one. The resulting moduli space is $\mathbb{C}^{4N}/G(k,1,N)$ with these being the associated invariants of dimension $k$, $2k$, up to $Nk$. If we consider $[ U(N)_k \times U(N+2)_{-k} ]/\mathbb{Z}_p$ instead, then we now also allow monopoles where all charges are fractions of $p$. These allow smaller monopole charges, but require the monopole to also carry charges under $U(2)_{-k}$. As we have seen before, the latter can be canceled only if $p=k=2$. As such the moduli space of $[ U(N)_k \times U(N+2)_{-k} ]/\mathbb{Z}_p$ remains $\mathbb{C}^{4N}/G(k,1,N)$ unless $p=k=2$. In this case, the monopole $\frac{1}{2}(1,1,...,1;1,1,...1,1,-1)$ can be made gauge invariant, replacing the invariant of order $Nk$ with one of order $\frac{Nk}{2}$. The moduli space is then $\mathbb{C}^{4N}/G(k,2,N)$.

Similarly we can generalize to the $U(N)_k \times U(N+x)_{-k}$ ABJ theory. We again go on the moduli space where we now have $N$ copies of $U(1)_k \times U(1)_{-k}$ and a decoupled $U(x)_{-k}$ theory. If no quotient is involved than the minimal monopole charge is $1$, and the previous analysis for $x=2$ still holds. We then get that the moduli space is $\mathbb{C}^{4N}/G(k,1,N)$. However, if we instead consider $[ U(N)_k \times U(N+x)_{-k} ]/\mathbb{Z}_p$ then we must determine if new gauge invariant monopoles can appear on the moduli space. We note that as before, the anomaly must vanish for the quotient to make sense, which put the condition $\frac{x k}{p^2} = integer$, and $p$ is still a divisor of $k$. Again, since the quotient is in the diagonal $U(1)$, the monopole carrying fractional charges must carry them in all the $U(1)$'s including those in $U(x)_{-k}$. Due to the Chern-Simons term the monopole would then carry gauge charges under $U(x)_{-k}$, which can only be canceled by the gauginos. This is expected to be quite limiting as each gaugino can only be used once. The problem then reduces to under what conditions can a gauge invariant be formed.

We begin by showing that this is possible for any $p$ dividing $k$ if $x=k$. The idea is to consider the $U(k)_{-k}$ monopole $(\frac{1}{p},\frac{1}{p},...,\frac{1}{p},-\frac{(p-1)}{p},-\frac{(p-1)}{p},...,-\frac{(p-1)}{p})$, where $\frac{1}{p}$ appears $k-q$ times, with $k=pq$. Note that this monopole carries the gauge charges $\frac{\prod_i q^{\frac{k}{p}}_i}{\prod_j q^{\frac{k(p-1)}{p}}_j}$, with $i=1,2,...,k-q$ and $j=k-q+1,k-q+2,...,k$. Next we take all the gauginos with charges $\frac{q_j}{q_i}$ for $i$ and $j$ in the ranges given above. Note that for a chosen $j$ there are $k-q$ choices of $i$, so the total $q_j$ charge would be $k-q = q(p-1)= \frac{k(p-1)}{p}$. Similarly for a chosen $i$ there are $q$ choices of $j$, so the total $q_i$ charge would be $q = \frac{k}{p}$. As such we see that the gauge charges of the gauginos precisely cancel those given from the Chern-Simons term in this case\footnote{One can also show that such a monopole contributes to the superconformal index exactly as the one with no $U(x)$ magnetic charges. Specifically, the powers of the fugacity $x$ coming from the gauginos gets precisely canceled against the term $x^{\epsilon_0}$, where $\epsilon_0$ is negative due to the contribution of the $U(x)$ vector. As such the power of the fugacity $x$ associated with the contribution of this monopole comes solely from the bifundamentals.}.    

What happens for other cases? First, we can consider the case when $k>x$. Increasing $k$ will increase the gauge charges the monopole has, so we would now need to add more gauginos to compensate for it. However, there do not appear to be any more gauginos we can add to increase the charge. As such, it appears that there is no gauge invariant monopole with fractional charges when $k>x$. If we instead take $x>k$, then we now have the same monopole charge but more gauginos. As such we expect to be able to find gauge invariant monopoles with fractional charges in this case, though we would not be interested much in this case.

So to summarize, we seem to find that the moduli space of $[ U(N)_k \times U(N+x)_{-k} ]/\mathbb{Z}_p$ theories, for $x\leq k$, is $\mathbb{C}^{4N}/G(k,1,N)$. The only exceptional cases are $x=0,k$ for which the moduli space is $\mathbb{C}^{4N}/G(k,p,N)$. We note that \cite{Aharony:2008gk} suggested that the $U(N)_k \times U(N+x)_{-k}$ and $U(N)_k \times U(N+k-x)_{-k}$ theories should be dual, and our results for the moduli space are consistent with this duality. The same reference also suggsted that cases with $x>k$ break SUSY spontaneously, and hence our reduced interest in these cases. 

\subsection{$OSp$ cases}
\label{sec:OSpuneqrank}
The behavior of the $OSp$ cases is similar to that of the unitary ones. Here we would be mainly interested in the cases of $USp(2N)_k\times O(2N+2x)_{-2k}$ and $USp(2N+2x)_k\times O(2N)_{-2k}$. The analysis follows in the steps taken in the previous example. We again go on the moduli space where the gauge group is broken to $N$ copies of $U(1)_{2k}\times U(1)_{-2k}$ and a decoupled $O(2x)_{-2k}$ or $USp(2x)_k$ theory, depending on the case. As before, the basic monopoles of the form $(1,0,...,0;1,0,...,0)$, $(1,1,0,...,0;1,1,0,...,0)$, et cetera, can be made gauge invariant using copies of the bifundamental to absorb the gauge charges. However, now we also need to quotient by charge conjugation, which is part of the gauge symmetry (its the parity element of $O$ and part of the Weyl group of $USp$). As in the equal ranks case, this enlarges the group we need to quotient by from $G(2k,1,N)$ to $G_N (\hat{D}_{k},\hat{D}_{k})$. The moduli space is then $\mathbb{H}^{2N}/G_N (\hat{D}_{k},\hat{D}_{k})$.

Next we consider the case where we take the $\mathbb{Z}_2$ quotient, that is when the gauge group becomes $[ USp(2N)_k\times O(2N+2x)_{-2k} ]/\mathbb{Z}_2$ or $[ USp(2N+2x)_k\times O(2N)_{-2k} ]/\mathbb{Z}_2$. In the equal ranks case, this entails the addition of a new minimal monopole, which lead to the change in the moduli space. However, the situation in the differing rank case is more subtle. This subtlety originates from the fact that the new monopole needs to have fractional charges in all the $U(1)$'s, including those that are part of the decoupled gauge group. As before, the problem reduces to whether or not the $O(2x)$ or $USp(2x)$ gauge charges they acquire can be canceled using the gauginos. If so, then there are new minimal monopoles, signaling a change to the moduli space, while if not, then the moduli space remains the same. We shall consider when this can be achieved for each case.

We shall begin with case of $O(2x)_{-2k}$. Here the gaugino carry the charges $q^{\pm 1}_i q^{\pm 1}_j$ for $i\neq j$. We wish to determine what is the highest charge one can build from the gauginos. This is achieved, for instance, by taking $q_i q_j$ for all possible choices of $i$ and $j$. The resulting charge is $\prod_i q^{x-1}_i$. Note that in a $O(2x)$ monopole background of charges $(\frac{1}{2},\frac{1}{2},...,\frac{1}{2})$, the ground state carry the charges $\prod_i q^{k}_i$. We see that the two match when $k=x-1$. For $k>x-1$ then there are not enough gauginos to saturate the charge and the monopole cannot be made gauge invariant. However, for $k\leq x-1$ it is possible to cancel the gauge charges using the gauging and the monopole can be made gauge invariant. 

The analysis for $USp(2x)_k$ follows similarly to that of $O(2x)_{-2k}$. The main difference is that we now have more gauginos. Specifically, we also have gauginos with charges $q^{\pm 2}_i$ in addition to the ones with charge $q^{\pm 1}_i q^{\pm 1}_j$. Adding those of the form $q^{2}_i$, on top of the others, results in the total charge $\prod_i q^{x+1}_i$. As such, in this case the gauginos are sufficient to cancel the charge as long as $k\leq x+1$.

Like the unitary versions, the orthosymplectic ABJ theories are related by dualities \cite{Aharony:2008gk}. Of particular interest are the ones relating the equal ranks case to a case with differing ranks:

\bea
& & O(2N+2k+2)_{2k} \times USp(2N)_{-k} \leftrightarrow USp(2N)_k \times O(2N)_{-2k} , \\ \nonumber & & USp(2N+2k-2)_k \times O(2N)_{-2k} \leftrightarrow O(2N)_{2k} \times USp(2N)_{-k} .
\eea

One can see that the cases with differing ranks that appear are precisely the minimal ones for which the fractional monopole can be made gauge invariant. As such, taking the quotient should modify the moduli space exactly as in the equal rank case, as expected from the duality. As before cases with higher rank difference are expected to break SUSY spontaneously, so these are the highest rank difference we expect to encounter. As such, we see that taking the $\mathbb{Z}_2$ would not change the moduli space, unless the ranks are equal or the difference is such that we are in one of their dual versions.

\subsection{$G(3)$}

Next we consider the case of the $\mathcal{N}=5$ SCFT associated with $G(3)$. As pointed in the main text, the matter content in this case consists of an $SU(2)_{3k}\times (G_2)_{-4k}$ gauge theory with a bifundamental hyper. We can use the previous method to determine the moduli space of the theory. First, we will need to consider what happens to the gauge group once the bifundamental gets a vev. In general, we expect the product group to be broken to some diagonal subgroup, which will be preserved along with a decoupled part, with the rest getting Higgsed. Said Higgsing requires the broken currents to eat components of the hypermultiplets and thus acquire mass. Note that this is only possible if the broken currents carry the same charges as the components of the hypermultiplets under the unbroken gauge group.

We can use the previous observation to determine the low-energy theory on the moduli space. Specifically, we consider the charges of the vector and hyper matter, conveniently expressed in terms of fugacities:

\bea
& & V = 3 + \alpha^2 + \frac{1}{\alpha^2} + a^2 + \frac{1}{a^2} + b^2 + \frac{1}{b^2} + (a + \frac{1}{a})(b^3 + b + \frac{1}{b} + \frac{1}{b^3}) , \\ \nonumber
& & H = (\alpha + \frac{1}{\alpha})\left(1 + b^2 + \frac{1}{b^2} + (a + \frac{1}{a})(b + \frac{1}{b}) \right) ,
\eea       
here we used the fugacity $\alpha$ for the $SU(2)$ group and $a$, $b$ for $G_2$ (using its $SU(2)\times SU(2)$ subgroup). All charges are normalized such that the minimal charge is 1. We also note that the theory is invariant under the charge conjugation of each fugacity individually. In fact, this is part of the Weyl group of $SU(2)\times G_2$ and suggests that each fugacity belong to $O(2)$ rather than just $U(1)$.  

Next note that if we set $\alpha=b^2$, the charges become:

\bea
& & V = 3 + a^2 + \frac{1}{a^2} + b^4 + \frac{1}{b^4} + b^2 + \frac{1}{b^2} + (a + \frac{1}{a})(b^3 + b + \frac{1}{b} + \frac{1}{b^3}) , \\ \nonumber
& & H = 2 +  b^4 + \frac{1}{b^4} + b^2 + \frac{1}{b^2} + (a + \frac{1}{a})(b^3 + b + \frac{1}{b} + \frac{1}{b^3}) .
\eea

We note that the last terms are equal. This suggests that following the above identification, all vector components charged under $b$ can eat the corresponding hyper components and become massive. We are then left with $5$ vectors, forming the gauge group $U(1)\times U(1)\times SU(2)$ and two hypers, which we can interpret as two bifundamentals for $U(1)\times U(1)$. These results suggest that the bifundamental vev leads to the identification of $\alpha=b^2$, breaking the group to $U(1)\times U(1)\times SU(2)$. At low-energies on a generic point in the moduli space, the theory can be effectively described by a decoupled $SU(2)$, associated with the $SU(2)$ with fugacity $a$, a $U(1)\times U(1)$ gauge theory associated with the fugacities $\alpha$ and $b$, and two bifundamentals. Note that the bifundamentals must carry the charges $\frac{\alpha}{b^2} + \frac{b^2}{\alpha}$ such that their vevs would implement the required breaking.       

We can next determine the Chern-Simons levels using the embedding index:

\bea
& & I_{U(1)_{\alpha}\hookrightarrow SU(2)} = \frac{1}{2}\frac{1^2+(-1)^2}{\frac{1}{2}} = 2 , \\ \nonumber & & I_{U(1)_{b}\hookrightarrow G_2} = \frac{1}{2}\frac{2^2+(-2)^2+2\times 1^2+2\times(-1)^2}{1} = 6  , \\ \nonumber & & I_{SU(2)_{a}\hookrightarrow G_2} =\frac{\frac{1}{2}+\frac{1}{2}}{1} = 1 .
\eea 

The result is then that we have the ABJM theory $U(1)_{6k}\times U(1)_{-24k}$ and a decoupled $SU(2)_{-4k}$. Note that here we have used the original charge convention, even though the $U(1)_b$ charge of the bifundamental is $2$ rather than $1$. This results in the absolute values of the Chern-Simons level being unequal. Note that we would indeed get exactly opposite Chern-Simons levels if the charges were properly normalize, though we find it easier to just work with the original charges.

All that is left to determine the moduli space is to understand what is the minimal contributing monopole. Naively, that should be the minimal possible monopole, which carry magnetic charge $(m_{\alpha}=1; m_{a}=\frac{1}{2} , m_{b}=\frac{1}{2})$. Note that the minimal electric charge under $U(1)_b$ in the low-energy theory is $b^2$, and so the minimal magnetic charge consisting with Dirac quantization would be $\frac{1}{2}$. However, that charge is not consistent by itself, due to the presence of doublets in the above decomposition, but must be accompanied with a fractional $U(1)_a$ magnetic charge. As such it is impossible to dress the monopole so as to build a gauge invariant under the $SU(2)$. This follows as the monopole carry the charge $a^{4k}$ due to CS term. This can only be canceled by the gaugino, which carry the charge $a^{\pm 2}$, but as we can only use each gaugino once, it is not possible to cancel the gauge charges in this way.

This leads us to consider the minimal monopole with no $U(1)_a$ magnetic charge, and for which a $U(1)\times U(1)$ invariant can be build. This turns out to be the one with charges $(2; 0 , 1)$. To see this note that said monopole acquire the charges $\frac{\alpha^{12k}}{b^{24k}}$, which can now be canceled using $12k$ copies of the bifundamental, as each copy caries the charges $\frac{b^2}{a}$. note that even though $(1; 0 , 1)$ is a valid monopole, it is not possible to make it gauge invariant using the bifundamental. As such the minimal gauge invariant monopole is $(2; 0 , 1)$ properly dressed by $12k$ copies of the bifundamental. This gives a dimension $12k$ invariant. As such, the moduli space needs to be quotiented by $\mathbb{Z}_{12k}$. Additionally, as the Weyl group includes the charge conjugation operation, one must quotient by it as well, turning $\mathbb{Z}_{12k}$ to $\hat{D}_{6k}$. Overall, we see that the moduli space in this case is $\mathbb{C}^4/\hat{D}_{6k}$.   

\subsection{$F(4)$}    
\label{app:f4}
We can determine the moduli space of the $\mathcal{N}=5$ SCFT associated with $F(4)$ in a similar manner. First, recall that the matter content in this case consists of an $SU(2)_{2k}\times Spin(7)_{-3k}$ gauge theory with a hyper in the $({\bf 2},{\bf 8})$. As before, we begin by considering what happens to the gauge group once the bifundamental gets a vev. This can again be determined by examining the charges of the vector and hyper:

\bea
& & V = 2 + \alpha^2 + \frac{1}{\alpha^2} + {\bf 8} + a^2 \overline{\bf 3} + \frac{1}{a^2}{\bf 3} + a {\bf 3} + \frac{1}{a} \overline{\bf 3} , \\ \nonumber
& & H = (\alpha + \frac{1}{\alpha})\left(a^{\frac{3}{2}} + \frac{1}{a^{\frac{3}{2}}} + \frac{1}{a^{\frac{1}{2}}} {\bf 3} + a^{\frac{1}{2}} \overline{\bf 3} \right) .
\eea       

Here we again use the fugacity $\alpha$ for the $SU(2)$ group and employ the $U(1)\times SU(3)$ subgroup of $Spin(7)$, using the fugacity $a$ for the $U(1)$. As before, we note that the theory is invariant under the charge conjugation of each fugacity individually, which is again part of the Weyl group of $SU(2)\times Spin(7)$.  

Next note that if we set $\alpha=a^{\frac{3}{2}}$, the charges become:

\bea
& & V = 2 + {\bf 8} + a^3 + \frac{1}{a^3} + a^2 \overline{\bf 3} + \frac{1}{a^2}{\bf 3} + a {\bf 3} + \frac{1}{a} \overline{\bf 3} , \\ \nonumber
& & H = 2 + a^3 + \frac{1}{a^3} + a^2 \overline{\bf 3} + \frac{1}{a^2}{\bf 3} + a {\bf 3} + \frac{1}{a} \overline{\bf 3} .
\eea

Again, the last terms are equal, suggesting that following the above identification, all the vector components charged under $a$ can eat the corresponding hyper components and become massive. We are then left with $10$ vectors, forming the gauge group $U(1)\times U(1)\times SU(3)$ and two hypers, which we can interpret as two bifundamentals for $U(1)\times U(1)$. These results suggest that the bifundamental vev leads to the identification of $\alpha=a^{\frac{3}{2}}$, breaking the group to $U(1)\times U(1)\times SU(3)$. At low-energies on a generic point in the moduli space, the theory can be effectively described by a decoupled $SU(3)$, a $U(1)\times U(1)$ gauge theory associated with the fugacities $\alpha$ and $a$, and two bifundamentals. Note that the bifundamentals must carry the charges $\frac{\alpha}{a^{\frac{3}{2}}} + \frac{a^{\frac{3}{2}}}{\alpha}$ such that their vevs would implement the required breaking.       

We can next determine the Chern-Simons levels using the embedding index:

\bea
& & I_{U(1)_{\alpha}\hookrightarrow SU(2)} = \frac{1}{2}\frac{1^2+(-1)^2}{\frac{1}{2}} = 2  , \\ \nonumber & & I_{U(1)_{a}\hookrightarrow Spin(7)} = \frac{1}{2}\frac{(\frac{3}{2})^2+(-\frac{3}{2})^2+3 (\frac{1}{2})^2+3(-\frac{1}{2})^2}{1} = 3 , \\ \nonumber & & I_{SU(3)\hookrightarrow Spin(7)} =\frac{\frac{1}{2}+\frac{1}{2}}{1} = 1 .
\eea 

The result is then that we have the ABJM theory $U(1)_{4k}\times U(1)_{-9k}$ and a decoupled $SU(3)_{-3k}$. As before, the CS level are not exactly opposite due to our chosen charge normalization, giving the bifundamental different charges under the two groups. 

To determine the moduli space we need to find the minimal contributing monopole. It is convenient to first consider monopoles with no $SU(3)$ charges. As the $U(1)_a$ charges are half-integer, the minimal $U(1)_a$ magnetic charge is $2$. In order for the resulting charge to be cancelable by the bifundamental, we must take $m_{\alpha}=3$. Indeed, in this case the monopole charges are $\frac{\alpha^{12k}}{a^{18k}}$, which can be canceled by dressing it with $12k$ copies of the bifundamental (whose charges are $\frac{a^{\frac{3}{2}}}{\alpha}$). As before this implies that the moduli space needs to be quotiented by at least $\mathbb{Z}_{12k}$, which becomes $\hat{D}_{6k}$ once we further quotient by charge conjugation.

However, that is not necessarily the end of the story, as it is possible to find smaller monopoles if we also allow for monopoles charges inside the $SU(3)$. Specifically, consider the monopole $(m_{\alpha}=1, m_a = \frac{2}{3}; m_s = \frac{1}{3} , m_r = \frac{1}{3})$, where we use the fugacities $s$ and $r$ for the Cartan of $SU(3)$, defined as ${\bf 3} = s + r +\frac{1}{rs}$. This is a valid $Spin(7)$ monopole, and note that the monopole charges under $U(1)_{\alpha}$ and $U(1)_{a}$ can now be canceled by merely $4k$ copies of the bifundamental. However, we now also have charges under the $SU(3)$. In fact this monopole is equivalent to the $U(3)_{k'}$ monopole $(\frac{1}{3},\frac{1}{3},-\frac{2}{3})$. We have previously seen that said monopole charge can be canceled by the gauginos if $k'=3$. As such we see that this $U(1)\times U(1)\times SU(3)$ monopole can be made gauge invariant if $k=1$, but not for $k>1$.

To summarize, we conclude that the moduli space of the $\mathcal{N}=5$ SCFT associated with $F(4)$ is $\mathbb{C}^4/\hat{D}_{6k}$ for $k>1$, and $\mathbb{C}^4/\hat{D}_{2}$ for $k=1$.

\bibliographystyle{JHEP}
\bibliography{ref}

\end{document}